\documentclass[twocolumn]{aastex701}

\usepackage{graphicx}
\usepackage{amsmath}
\usepackage{multirow}
\usepackage{color}
\usepackage{icomma}
\usepackage{etipopsbl}

\def\meter{\textrm{m}}

\def\Watt{\textrm{W}}

\def\Hz{\textrm{Hz}}
\def\kHz{\textrm{kHz}}
\def\MHz{\textrm{MHz}}
\def\GHz{\textrm{GHz}}

\def\uJy{\mu\textrm{Jy}}
\def\mJy{\textrm{mJy}}
\def\Jy{\textrm{Jy}}

\def\sec{\textrm{s}}
\def\minute{\textrm{min}}

\def\sr{\textrm{sr}}

\def\Msun{\textrm{M}_{\odot}}
\def\Lsun{\textrm{L}_{\odot}}

\def\gcm3{\textrm{g}\,\textrm{cm}^{-3}}

\def\km{\textrm{km}}

\def\Mpc{\textrm{Mpc}}

\def\sec{\textrm{s}}

\def\ga{\gtrsim}
\def\la{\lesssim}

\newcommand{\editOne}[1]{{#1}}

\shorttitle{ETI Broadcast Populations III.}
\shortauthors{Lacki}

\begin{document}

\title{Artificial Broadcasts as Galactic Populations: III. Constraints on Radio Broadcasts from the Cosmic Population of Inhabited Galaxies}

\author[0000-0003-1515-4857]{Brian C. Lacki}
\email{astrobrianlacki@gmail.com}
\affiliation{Breakthrough Listen, Department of Physics, Denys Wilkinson Building, Keble Road, Oxford OX1 3RH, UK }
\correspondingauthor{Brian C. Lacki}

\begin{abstract}
Any population of artificial radio broadcasts in a galaxy contributes to its integrated radio luminosity. If this radio emission is bright enough, inhabited galaxies themselves form a cosmic population of artificial radio galaxies. We can detect these broadcasts individually or set constraints from their collective emission. Using the formalism in Paper I and II, I set bounds on the artificial radio galaxy population using both of these methodologies. Measured radio source counts set limits on radio broadcasts across the radio spectrum, including the first  Search for Extraterrestrial Intelligence (SETI) constraints at $\sim$250 GHz. I compare these with commensal limits from background galaxies in the fields of large SETI surveys. The field limits are more powerful, but generally only over a limited luminosity range and for frequencies with dedicated SETI surveys. The limits are weaker when broadcasts clump into discrete hosts that are themselves extremely rare. I find that the abundance of Kardashev Type III radio broadcast populations is less than one in $10^{17}$ stars, about one in a million large galaxies. I also examine limits for a power-law distribution in broadcast luminosity.
\end{abstract}


\section{Introduction}

At every wavelength, the extragalactic universe is observed to contain populations of discrete sources.  These sources are generally galaxies and can themselves be a collection of numerous faint emitters -- like the optical emission of a galaxy resolving into stars --  but nonetheless the galaxies can be treated as individual objects. Despite being large assemblages in themselves, galaxies are not all the same. Often, we can discern several distinct populations in a survey, representing different emission processes, luminosities, and abundances.

The Search for Extraterrestrial Intelligence (SETI; \citealt{Tarter01,Worden17}) postulates another cosmic population, the technosignatures of extraterrestrial intelligences (ETIs), in particular their artificial broadcasts.  Most SETI searches have sought individual technosignatures, usually within the Milky Way. But if there is one ETI besides humanity in the present Galaxy, almost certainly there are more. The totality of all the broadcasts in a galaxy forms an intragalactic population, which can be treated statistically. The population properties could plausibly vary enormously between otherwise identical galaxies, as starfaring ETIs evolve and spread across some, establishing billions of transmitters, while others remain uninhabited, the result of the vagaries of their ETI populations.  These collections of ETIs are dubbed metasocieties in Paper I \citep{Lacki24-ETIPop1}, which outlined a point process formalism for their observables.

There are at least two basic kinds of constraints on broadcasts\footnote{In addition to constraints on non-broadcast technosignatures, like probes sent to the Solar System and exoplanet atmospheric changes \citep[e.g.,][]{Freitas85,Lin14}.}, as discussed in Paper II \citep{Lacki24-ETIPop2}. Dedicated SETI surveys derive individualist constraints, searching for extreme features that stick out of a background, like a bright pulse in a time series or line in a spectrogram. Paper II introduces the collective bound, deriving limits on broadcast populations from the total observed emission of a galaxy. Generally, individualist constraints are far more powerful limits on broadcast abundance above a threshold luminosity or energy, because a null result is invalidated by even a single detection. The collective bound, by contrast, requires the broadcast population to dominate the emission of a galaxy, which in turn implies very many broadcasts unless they are individually extremely luminous. Nonetheless, collective bounds are useful in several cases: (1) they provide limits on extremely numerous populations of very faint broadcasts; (2) they provide limits on broadcasts that would be ``missed'' by line or pulse searches, including continuum sources; (3) they are not subject to confusion, in which a large population of broadcasts blends together to make any single one undetectable; and (4) they merely require a flux measurement of the target system.

In this paper, I consider the inhabited galaxies themselves as source populations. There are now many SETI programs that have covered various regions in the Milky Way \citep[e.g.,][]{Tarter85,Enriquez17,Tremblay20,Gajjar21}, and a relative few dedicated to surveying external galaxies, mainly in the Local Group \citep{Shostak96,Gray17,Isaacson17}, and recently, ninety-seven galaxies at distances up to that of the Virgo Cluster \citep{Choza24}. But what if ETIs are so rare that they only exist in one per thousand galaxies, say? Individualist broadcast searches still have something to tell us. There have been a few all-sky surveys, among them Big Ear \citep{Dixon85} and the Megachannel ExtraTerrestrial Assay (META; \citealt{Horowitz93}), which could detect a bright enough narrowband transmission over much of the sky. Recent years have seen the development of another route to extragalactic SETI, setting limits on broadcasts from background objects in the fields of targeted objects. \citet{WlodarczykSroka20} pioneered this method by considering catalogued background Galactic stars in the beams of Breakthrough Listen observations of nearby stars. That effort has now been extended to background galaxies \citep{Garrett23,Uno23,Tremblay24}.

The collective bound can also be applied to cosmic populations through source counts. Source counts are vital tools for understanding cosmic populations. Source count distributions specify the number of sources on the sky within a given flux range (or flux density range). Of course there are many more fainter sources than bright ones because most objects are very far away, but additional features in the distribution reveal the existence of multiple emitter populations and the cosmic evolution of each. In radio and X-rays, extragalactic source counts find two broad classes of galaxies: bright but rare active galactic nuclei (AGNs) and faint but common normal galaxies \citep{Condon84-Evol,Condon02,Sadler02,Bauer04,Lehmer12}. Thus, (suitably normalized) source count distributions are expected to have two ``bumps'' or ``steps'' when probed deeply enough.  Any other major population of galaxies, dominated by some other emission mechanism, could manifest as a distortion of these features or as a third bump entirely.  Indeed, the lack of such features has been used to constrain the existence of new populations of sources, particularly the mysterious radio background excess found by the ARCADE2 experiment \citep{Condon12,Vernstrom14,Matthews21}.

\subsection{Outline of the paper}
A review of relevant mathematics, concepts, and notation in Papers I and II is presented in~section~\ref{sec:Background}. Section~\ref{sec:Individualist} presents a mathematical derivation of how to set constraints on broadcasts from galaxies in the fields of surveys. Section~\ref{sec:Collective} discusses how to derive constraints on broadcasts from observed source counts of galaxies. I then move on to applying these methods. Section~\ref{sec:Methods} presents more details on the methods and data employed. Limits on broadcasts in the base model set, corresponding to the usual assumptions in the SETI literature (which ignore the discreteness of societies and metasocieties), are discussed in section~\ref{sec:Base}. The effects of changing certain assumptions in variant model sets are considered in section~\ref{sec:Variants}. Section~\ref{sec:Discussion} includes additional discussion on how the limits could be strengthened and implications for Kardashev Type III ETIs. The conclusions are summarized in section~\ref{sec:Conclusions}.

\section{Background}
\label{sec:Background}
Papers I and II presented a framework for treating the ETI populations and their observables, including notation for various kinds of selections. In this section, I review only the basics needed for this work; refer to \citet{Lacki24-ETIPop1} and \citet{Lacki24-ETIPop2} for a full discussion.

Refer to Table~\ref{table:Notation} for a quick summary of the most important variables and notation.

\begin{deluxetable*}{cp{13cm}}
\tabletypesize{\footnotesize}
\tablecolumns{2}
\tablewidth{0pt}
\tablecaption{Summary of notation used\label{table:Notation}}
\tablehead{\colhead{Notation} & \colhead{Explanation}}
\startdata
\cutinhead{Statistical notation used}
$\PDF{X}(x)$    					& Probability density function (PDF) of random variable $X$ at value $x$       	\\
$\jkPDFGen{\jjSingle}(x)$ & PDF of random variable $\jjSingle$ over random $\JMark$-objects drawn from the population of $\jjkSampleGen$, instead of for a fixed $\JMark$-object\\
$\jkMeanGen{\jjSingle}$   & Mean of $\jjSingle$ for $\JMark$-objects drawn from the population of $\jjkSampleGen$\\
$\CF{X}(\tilde{x})$ 			& Characteristic function of $X$ at value $\tilde{x}$  													\\
$\jkCFGen{\jjSingle}(\tilde{x})$ & Characteristic function of $\jjSingle$ for $\JMark$-objects drawn from the population of $\jjkSampleGen$\\
$\CDF{X}(x)$							& Cumulative distribution function of $X$, equal to $\fpP(X \le x)$               \\
$\CCDF{X}(x)$        	  	& Complementary cumulative distribution function of $X$, equal to $\fpP(X > x)$	\\
$\IndicatorOf{\fpEvent}$	& Indicator function: equals $1$ if $\fpEvent$ is true, 0 otherwise							\\
$\fDirac(x)$              & Dirac delta distribution in $x$                                               \\
\cutinhead{Framework quantities}
$\jjHaystack$							& Haystack (parameter space) of $\JMark$-type objects																																							\\
$\jjTuple$								& Tuple of a $\JMark$-type object, describing all relevant parameters																															\\
$\jjkSampleGen$						& Point process describing $\JMark$-type objects hosted by $\KMark$ and selected by window $\GenLabel$; each realization is a sample. When no window is given, a trivial window selecting all hosted $\JMark$-type objects is implied.\\
$\jjkDist$                & Number distribution function for $\JMark$-type objects hosted by $\KMark$; the intensity of $\jjkSample$.												\\
$\oBandwidthGen$					& Bandwidth of window $\GenLabel$ in source-frame													         																								\\
$\oeBandwidthGen$         & Bandwidth of window $\GenLabel$ in observer-frame                                                                               \\
$\oeNuMidGen$             & Central frequency of window $\GenLabel$ in observer-frame                                                                       \\                          
$\oSkyFieldGen$						& Sky area covered by window $\GenLabel$																																													\\
\cutinhead{Broadcast and ETI quantities}
$\bLisoBAR$								& Characteristic luminosity of broadcasts in model																																								\\
$\alpha$									& Power-law exponent of broadcast luminosity function in model set D																															\\
$\baAbundnuTotal$					& Instantaneous abundance of broadcasts per unit (source-frame) frequency hosted by society $\SocMark$													  \\
$\bzAbundnu$           		& Instantaneous abundance of broadcasts per unit (source-frame) frequency per star hosted by metasociety $\MetaMark$ 							\\
$\azAbund$ 								& Instantaneous abundance of societies per star hosted by metasociety $\MetaMark$     												  									\\
$\zgAbundEMPTY$						& Effective instantaneous abundance of metasocieties per star hosted by galaxy $\GalMark$ 																				\\
$q$											  & Power-law exponent of evolution of $\zgAbundEMPTY$ with cosmic age in model set A                             									\\
$\bNuMid$									& Source-frame frequency of broadcast $\BcMark$																																										\\
$\hjjNTime$								& Instantaneous number of stars in host $\JMark$																																									\\
$\bjjNGen$    						& Number of broadcasts in host $\JMark$ selected by window $\GenLabel$																														\\
$\azNGen$   	  				 	& Number of societies hosted by metasociety $\MetaMark$ selected by window $\GenLabel$																						\\
$\zgNGen$    	  					& Number of metasocieties in galaxy $\GalMark$ selected by window $\GenLabel$																											\\
$\gNTime$      					 	& Instantaneous number of galaxies                                            																										\\
$\bgAggLnuisoObs$					& Total (effective isotropic) spectral luminosity per unit frequency from broadcasts in galaxy $\GalMark$ and selected by window $\ObsLabel$ \\
$\kgFluxENu$              & Received energy flux per unit frequency from galaxy $\GalMark$ resultant from natural processes                                 \\
$\mgFluxENuObs$						& Received energy flux per unit frequency from broadcasts hosted by galaxy $\GalMark$ within observation window $\ObsLabel$				\\
$\qgFluxENuObs$						& Total received energy flux per unit frequency from galaxy $\GalMark$ within observation window $\ObsLabel$											\\
\enddata
\tablecomments{See Paper I for a comprehensive explanation of the notation. All italicized and san-serif quantities are in source-frame, unless modified by a $^{\oplus}$ superscript; all Fraktur variables (e.g., flux) are observer-frame.}
\end{deluxetable*}

\subsection{Trees and haystacks}
The motivation for this framework is the search for a broadcast, a technosignature consisting of emission released from a specific position over a limited time range. Broadcasts and their hosts are hierarchically organized in a tree structure. Each object on the tree is a parent node which may host zero or more children. At the root of the tree is the model universe (type $\UnivMark$). It contains galaxies (type $\GalMark$). In this work, ETIs in each galaxy are isolated from those in the others, with no significant intergalactic migration. Within each galaxy are metasocieties (type $\MetaMark$), which are collections of societies with coordinated properties (see Section~\ref{sec:MetaScenarios} for further discussion). These in turn contain communicative societies (type $\SocMark$), each of which is a localized site that can finally host broadcasts (type $\BcMark$), the technosignatures we seek.   

Each object $\JMark$ on the tree is described by a tuple of parameters $\jjTuple$. For example, a broadcast may be described by its duration, bandwidth, central frequency, amount of energy released, polarization, among other quantities; a society by its location and lifespan, and so on. For each type of object, there is an entire tuple space $\jjHaystack$, referred to as a ``haystack'' \citep[cf.][and references therein]{Wright18}.

The tuple of an object -- its location within its respective haystack -- completely specifies it within the model. The probability distribution of every observable is assumed to depend solely on the tuple and has no further dependence on any other object's properties. The tuple also specifies the statistical distribution of descendent objects within their haystack. Thus, the properties of the objects contained in a host are assumed to be conditionally independent on the host's properties. Dependent properties resulting from a shared history, like a common broadcast frequency, are thus interpreted instead as emergent properties of the higher-level host, with no further information provided by knowledge of any of the descendents. This lets us employ assumptions of independence in calculations (the use of Poisson point processes in particular) while allowing for mutual influences.

\subsection{Point processes}
Some of the objects may have fixed properties, in particular specific galaxies like the Milky Way, the Magellanic Cloud, or M31; these are realized objects. In general, though, most are treated as random objects, with the object as a whole considered as a kind of random variable. This reflects our uncertainty in the parameters; we may not know exactly which galaxies are in a field, and we certainly do not know the properties of any broadcasts coming from them. A random object has a random tuple indicating a random point in the haystack. Point process theory provides the tools to model the population as a whole \citep{Kingman93,Daley03,Daley08,Baddeley07,Chiu13,Haenggi13,Last17}.\footnote{Paper I treats the subject more rigorously and points out some cavaets.}

The population of $\JMark$-type objects descended from a host $\KMark$ is treated as a point process $\jjkSample$ over the haystack $\jjHaystack$. Point processes are random collections of points in a space, with only a finite number allowed in a finite region of the space. An important property of a point process is its \emph{intensity}, which describes the mean density distribution of points within the space. Given an intensity $\jjkDist(\jjTuple)$ for $\jjkSample$ at $\jjTuple$, the mean number of points, representing objects, occurring in $\jjkSample$ within a subset $\fpSubset \subset \jjHaystack$ is
\begin{equation}
\Mean{\jjkN(\fpSubset)} = \int_{\fpSubset} \jjkDist(\jjTuple) d\jjTuple ,
\end{equation}
with $\jjkDist$ itself a function of $\jkTuple$.

Point processes can be manipulated in several ways. In \emph{superposition}, the points of a collection of point processes are combined, with the intensities adding. If an object $\LMark$ has a descendant population of $\KMark$-type objects, each of which hosts $\JMark$-type objects, the population of $\JMark$-type objects in $\LMark$ is the superposition point process
\begin{equation}
\jjlSample = \bigcup_{\jkTuple \in \jklSample} \jjkSample,
\end{equation}
where $\jklSample$ is the point process representing the $\KMark$-type objects. 

\emph{Thinning} involves selecting points from a point process according to some criteria, resulting in another point process. In the framework, thinning occurs according to the action of \emph{windows}. When a window $\GenLabel$ is applied to a haystack, each point has a probability of being retained according to its position, resulting in a thinned point process $\jjkSampleGen$ with an intensity modulated by this probability function.

A Poisson point process is one in which 1.) the number of points in each subset of the haystack is a Poisson random variable\footnote{Subject to certain reasonable conditions on the subset; see Paper I.} and 2.) the number of points in multiple disjoint subsets are independent. They have many desirable properties, remaining Poissonian under superposition and thinning. In this work, Poisson point processes are used to describe the broadcast population of a \emph{fixed} society, the society population of a \emph{fixed} metasociety or galaxy (depending on scenario), and the galaxy population of a \emph{fixed} model universe. When we take into account the hosts being random, the intensity itself becomes a random function.\footnote{Formally, the random intensities results in a Cox point process \citep{Kingman93}.}

\subsection{Random variables and observables}
Each object has a number of random variables associated with it. Sometimes the randomness is inherent in the object (sample variance) -- the number of ETIs in a galaxy, for one, which depends on unknown evolutionary histories. In other cases, the randomness is a result of noise in our observations, like the measured flux from an object.

Singleton random variables are quantities describing a single object. Singleton random variables describing the emission of an object are typically filtered by an object window, which denote the bounds of integration over which emission is collected (a limited bandpass and integration time, for example). The general notation for singleton random variables in this series is $\jjSingleGen$, which describes the amount of $\jSingle$ associated with object $\JMark$ integrated over the quantity window $\GenLabel$. 

Aggregate random variables describe the sum of singleton random variables attached to all the objects modeled in a point process. Each aggregate variable describes a host object that contains the objects and a window which both selects the objects and filters the quantities:
\begin{equation}
\jjkAggGen \equiv \sum_{\jjTuple \in \jjkSampleGen} \jjSingleGen .
\end{equation}
The aggregate variable may then be interpreted as a singleton variable of the host. For example, the sum total of all broadcast emission from one society may also just be considered to be the society's brightness; the stellar mass of a galaxy can be thought of as a property of the galaxy itself.

Particularly important windows considered in this paper are the observation window $\ObsLabel$, which corresponds to a single resolution element and single channel; $\SurvLabel$, which covers the entire bandwidth and all pointings in a survey; and $\TimeLabel$, which denotes an instantaneous selection, picking only objects active at a single moment. Thus $\hgNObs$ is the number of stars in galaxy $\GalMark$ that happen to fall in a single resolution element, which could be much smaller than $\hgNSurv$, the number that fall within the survey footprint, and $\hgNTime$, the number of stars in the galaxy at any one time.

\subsection{Narrowband radio lines: flux and number}
Most radio SETI searches are concerned with finding narrowband radio broadcasts, the archetypal technosignature since \citet{Cocconi59}, and so this paper will treat that important case. Papers I and II presented the box and chord models for calculating the observables of broadcasts, and low-drift lines are a limit of both of them. Observables are measured within an observation window $\ObsLabel$, which is treated as a ``box'' in time-frequency space. It covers the time interval $[\oTStartObs, \oTStartObs + \oDurationObs]$ and a frequency interval $[\oNuMidObs - \oBandwidth/2, \oNuMidObs + \oBandwidth/2]$; anything that touches those intervals in time and frequency coming from within a sky field $\oSkyFieldObs$ is selected.\footnote{These variables are in source-frame. Their Earth-frame counterparts are marked with a $\oplus$ superscript.} The observations in this paper observe both polarizations. A standard radio observation integrates all of the energy collected within this window, the mean of which is proportional to the energy fluence.

Each narrowband line has an effective luminosity (effective isotropic radiated power, EIRP) $\bLiso$, which is assumed to be constant during the observation. Thus, a line with transverse comoving distance $\yDistanceM$ (luminosity distance $\yDistanceL = (1 + \yRedshift) \yDistanceM$) has a flux of $\lFluxE = \bLiso / [4 \pi (1 + \yRedshift)^2 \yDistanceMSQ] = \bLiso / (4 \pi \yDistanceLSQ)$. The baseline assumption is that the spectral lines have a degenerate luminosity distribution:
\begin{equation}
\PDF{\bLiso}(\bLactualCore) = \fDirac(\bLactualCore - \bLisoBAR) .
\end{equation} 
For this paper, I also ignore scintillation, which will smear out a single-valued distribution into an exponential distribution if it is strong enough \citep{Cordes97,Brzycki23}.

The line is assumed to have zero bandwidth, and at time $\TimeVar$, it is at frequency $\bNuMidTime(\TimeVar)$. Then, the effective flux from the broadcast is $\lFluxEObs = \lFluxE \bDurationObs/\oDurationObs$, where $\bDurationObs$ is the time that the line spends within the bandwidth covered by the observation. If the drift rate $\bDriftRate$ is fast ($|\bDriftRate| \ga \oBandwidthObs/\oDurationObs$), a detected line has $\bDurationObs \approx \oBandwidthObs / |\bDriftRate|$ as the line veers in and out of the frequency window of $\ObsLabel$, and the effective flux is reduced, making it harder to detect individually.\footnote{Dedrifting procedures further complicates the calculation (Paper II).} But in the low drift rate case ($|\bDriftRate| \la \oBandwidthObs/\oDurationObs$), a line is usually either outside the observation's bandwidth, or within it for the entire observation's duration: $\lFluxEObs \approx \lFluxE$. This is the regime I will focus on in this paper, because it applies to the collective constraints, while considering drift rate greatly complicates the analysis of individualist searches. We can also define an effective \editOne{flux density} for the broadcast, which is useful for aggregate flux calculations:
\begin{equation}
\lFluxENuObs = \frac{\lFluxE}{\oeBandwidthObs} \left(\frac{\bDurationObs}{\oDurationObs}\right) \to \frac{\bLiso}{4 \pi \yDistanceLSQ \oeBandwidthObs} = \frac{\bLiso}{4 \pi (1 + \yRedshift) \yDistanceMSQ \oBandwidthObs} .
\end{equation}
The effective spectral luminosity and flux of the galactic population likewise follows from the bandwidth:
\begin{equation}
\bgAggLnuisoObs \equiv \frac{1}{\oBandwidthObs} \sum_{\bTuple \in \bgSampleObs} \bLiso = 4 \pi (1 + \yRedshift) \yDistanceMSQ \mgFluxENuObs .
\end{equation}

The abundance of lines is the other key parameter of broadcasts. The mean number of lines emitted by a society, in the absence of frequency drift, is given by:
\begin{equation}
\Mean{\baNObs} = \baAbundnuTotal \oBandwidthObs,
\end{equation}
where $\baAbundnuTotal$ is the instantaneous frequency abundance of lines. A metasociety likewise has a frequency abundance of line broadcasts per star, $\bzAbundnu$, so that
\begin{equation}
\Mean{\bzNObs} = \bzAbundnu \hzNObs \oBandwidthObs = \bzAbundnu \hgNTime \oBandwidthObs,
\end{equation}
with $\hzNObs$ being the number of stars covered by the observations. Observations are much briefer than stellar lifetimes, and distant galaxies are unresolved by the wide beams of radio surveys, so under these circumstances, $\hzNObs$ is just the total instantaneous number of stars in the galaxy, $\hgNTime$. The average effective spectral luminosity from broadcasts in the metasociety is then:
\begin{equation}
\Mean{\bzAggLnuisoObs} = \frac{1}{\oBandwidthObs} \Mean{\bzNObs} \zMean{\bLiso} = \bzAbundnu \hgNTime \zMean{\bLiso},
\end{equation}
where $\zMean{\bLiso}$ is the mean EIRP of broadcasts within the metasociety.

\subsection{ETIs and metasocieties}
\label{sec:MetaScenarios}
The spread and interconnection of ETI societies has immense implications for their evolution and observable technosignatures. If ETIs are sessile, bound to their world or planetary system of origin, then most likely only a small fraction of stars will have broadcasting societies. This is because technological societies (or series of such societies) are expected to last only a small fraction of their host stars' lifespans. If ETIs are expansive, spreading from one star to multiple others in an initially exponential process, then perhaps nearly all stars have broadcasting societies at any given time once the settling process is complete. Although individual societies come and go, the ``dead'' worlds they once occupied can be resettled \citep{Kuiper77}. Either way, societies in contact with each other may influence each other, through communication or shared history.

The metasociety concept addresses these issues. A metasociety is a collection of societies described with a single distribution function, which are assumed here to be contained within a single galaxy. Paper I presents a few different ways of conceptualizing metasocieties. This paper focuses on two different conceptions.

\editOne{In a \emph{galactic club}, societies arise completely independently of each other. Once they appear, they are inducted into the club and conform to a distribution function that regulates their broadcasts (e.g., rules about how to transmit). To simplify matters, I adopt the ``single metasociety'' assumption in this paper: in galactic club models, every galaxy always has one metasociety ($\zgNTime = 1$).\footnote{Metasocieties, as galactic-level phenomena, are always much longer lived than the observations or surveys we carry out, so $\zgNTime = \zgNObs = \zgNSurv$ is a valid assumption.} I further assume that all the metasocieties converge on the same societal and broadcast distributions in every galaxy. The practical effect of this is that the broadcast population directly traces the number of stars. Broadcasts and societies may still be so rare that there is fewer than one per galaxy, but if one galaxy has a large ETI populations, others of the same size should as well.

An \emph{expansive metasociety} is the end result of an expansion front of interstellar settlement, after the entire galaxy undergoes a kind of phase transition \citep{Cirkovic08}. Expansive metasocieties are conceived as being as in an equilibrium. The main relevant property of expansive metasocieties is that all of the societies are dependent on a single origin event The key difference with the galactic club scenario is that this initial growth is a bottleneck, which may be arbitrarily unlikely. It is thus possible that some galaxies host billions of broadcasts -- in contradiction to what is seen in the Milky Way -- while most are practically empty. This allows an additional mechanism for a population of rare but bright artificial radio galaxies. According to Paper I’s treatment of expansive metasocieties, the instantaneous number of metasocieties in a galaxy $\GalMark$, $\zgNTime$ is either $0$ or $1$. It is an indicator variable for whether there has been a seed technological society in the history of the galaxy (or the lifespan of the metasociety if shorter). Then the probability of a galaxy having a metasociety is regarded as
\begin{equation}
\fpP(\zgNTime = 1) = \Mean{\zgNTime} = \exp(-\zgAbundEMPTY \hgNTime)
\end{equation}
where $\zgAbundEMPTY$ is an effective abundance of metasocieties per star, roughly the probability of a star being the origin site of an expansive metasociety (see Paper I).
}

\subsubsection{The abundance of societies and broadcasts}
If there is a metasociety $\MetaMark$, the abundance of communicative societies within it is in turn given by $\azAbund$, the mean number of societies per star in the metasociety:
\begin{equation}
\Mean{\azNGen} = \azAbund \Mean{\hzNGen} ,
\end{equation}
where $\GenLabel$ stands for a generic window like a survey with a duration much shorter than the societies. Communicative societies are regarded as a Poisson point process within the metasociety once one exists.\footnote{\vphantom{T}\editOne{T}he number of societies is not Poissonian when the number of habitats is saturated \citep{Kipping24}. Nonetheless, the habitats themselves have unknown (from our perspective, random) locations and thus can be treated as a random Poisson point process within the galaxy. Additionally, if only a small fraction of societies are communicative, then a Poisson point process can be a good approximation for that subset.} Finally, each society has some abundance of broadcasts, which is $\baAbundnuTotal$ for the instantaneous number of narrowband lines per unit frequency. It can be converted to a metasocietal abundance just by multiplying:
\begin{equation}
\bzAbundnu = \azAbund \zMean{\baAbundnuTotal} .
\end{equation}
The $\MetaMark$ subscript on the average indicates to take the average over all societies in the metasociety, according to their distribution function. In principle, there could be a whole range of different broadcasting rates in different societies, but in this work, they all are interchangeable ($\zMean{\baAbundnuTotal} \to \baAbundnuTotal$).

To simplify matters further, in some model sets, I adopt the diffuse approximation. This treats the broadcast population of the metasociety itself as a Poisson point process, totally ignoring their clumping into societies. The diffuse approximation applies when the number of broadcasts selected by a window from each society is typically $\ll 1$.

\section{Individualist field constraints from SETI surveys}
\label{sec:Individualist}

Most SETI broadcast constraints are individualist searches for single bright transmitters not subject to confusion in a single target (Paper II). These targets are individual galaxies, if not regions or stars within the Milky Way itself. This can be a very powerful method of setting constraints, because they can detect very rare transmitters as long as they are sufficiently bright. Moreover, the targets are generally nearby on a cosmic scale, allowing for sensitivity to broadcasts of lower luminosity. The downside is that they are unable to constrain very large populations of faint transmitters, they are specialized for specific broadcast types instead of generalized energy release, and they are vulnerable to confusion blending when there are very many broadcasts of similar brightnesses.

But a few SETI surveys have covered much of the sky, in particular, META \citep{Horowitz93}. Additionally, even targeted SETI surveys cover fields sure to include a number of background galaxies. Together, they yield individualist \emph{field constraints} where no particular galaxy is targeted. While some studies have begun to set constraints by finding cataloged galaxies within targeted fields \citep{Garrett23,Tremblay24}, a more general approach is to use galaxy distribution functions to estimate the number of galaxies, and their stellar populations, in each field \citep[see also the ``statistical method'' of][]{Uno23}. This section presents the calculations needed to set constraints on rare but bright broadcasts from background galaxies.

\subsection{The probability of a null detection: the galaxy-level view}
A model is considered incompatible with a null detection if the probability of there being zero detectable broadcasts in the survey sample is less than some threshold $p$, here taken to be $0.05$: $\fpP(\bNDetect = 0) < p$. This probability depends on the selection cuts we impose on broadcasts. We have some survey $\SurvLabel$ which observes a window consisting of a limited part of the sky over some limited frequency and time spans. Not all broadcasts within the window are necessarily detected -- some may be too faint to see; others may be in confused systems, blending with others in its population; some field galaxies may simply be in fields with observations that are not included in the analysis. We can make cuts on the broadcast, society, metasociety, and galaxy haystacks beyond the simple facts about which fields and frequency ranges are covered by the survey to account for these additional conditions, to get the new restricted survey window $\DetectLabel$; that is, $\DetectLabel$ selects broadcasts that are likely to be detectable in the survey $\SurvLabel$, plus all their hosts.

\begin{figure*}
\centerline{\includegraphics[width=12cm]{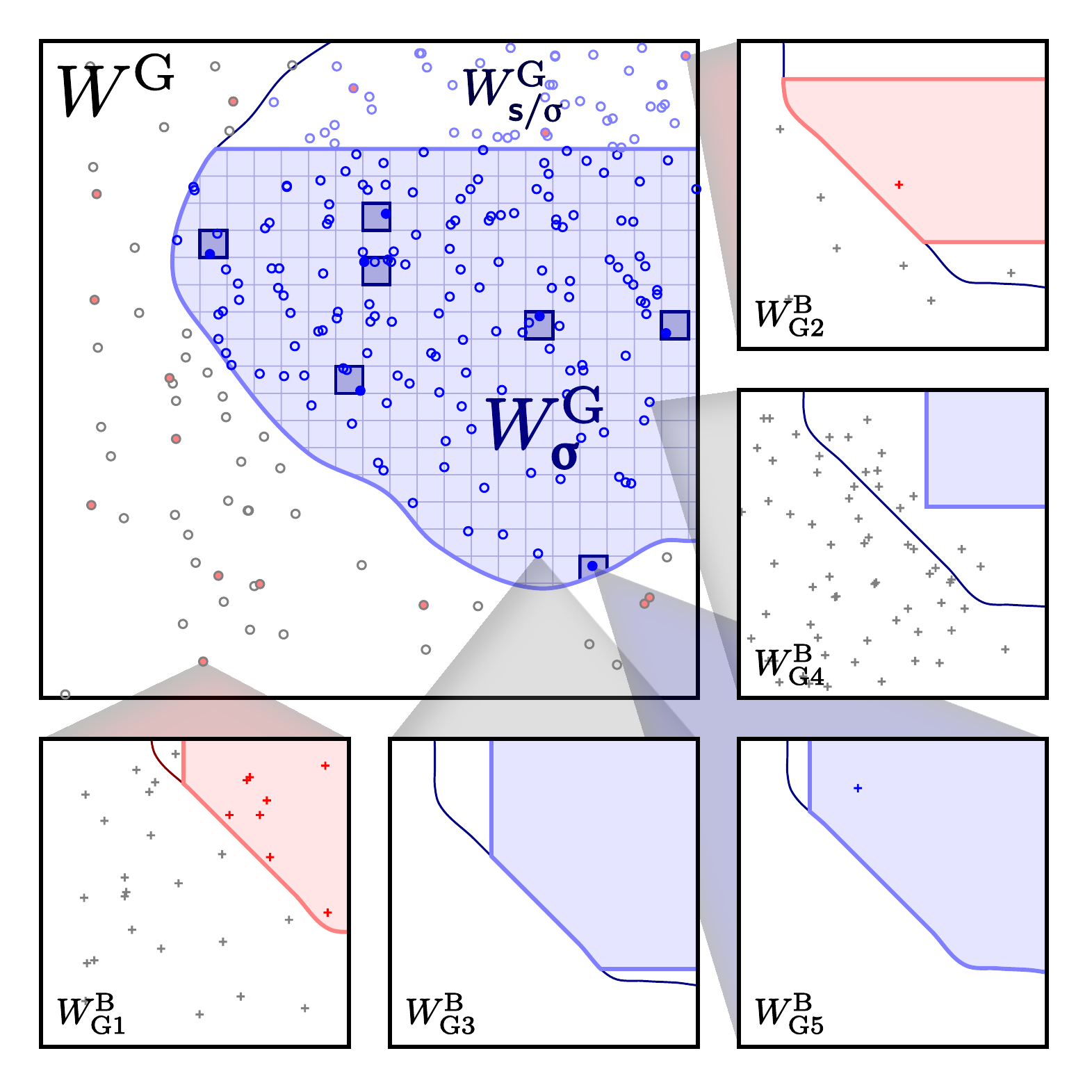}}
\figcaption{Galaxies are scattered in a haystack $\gHaystack$ (upper left), a subset $\gHaystackDetect$ of which is selected by survey $\SurvLabel$ (dark blue outline), in turn restricted to the detection window $\DetectLabel$ (blue shading). Each galaxy in turn has its own broadcast haystack, $\bgHaystack$, with detectable broadcasts falling into a subset selected by $\DetectLabel$. To calculate the probability of a null detection, we can split $\gHaystackDetect$ into many subwindows, only some of which may contain galaxies with detectable broadcasts (highlighted blue squares), and multiply the probabilities that none of the subwindows have one. A null result may happen because the galaxy is outside the survey ($\GalMark 1$; lower left), is in the survey but outside $\DetectLabel$ ($\GalMark 2$; upper right), in $\DetectLabel$ but has no broadcasts ($\GalMark 3$; lower center) or all the broadcasts are undetectable ($\GalMark 4$; middle right). On lower right, $\GalMark 5$ has a detectable broadcast. Not depicted are the metasocietal and societal haystacks, which regulate confusion.\label{fig:FieldDiagram}}
\end{figure*}

The ultimate hosts of broadcasts are galaxies, which are regarded as point sources for this derivation since we are considering galaxies in the background. A null result, with no detected broadcasts in the survey, means that there are no broadcasts from any galaxy within any part of the galaxy haystack sampled by $\DetectLabel$ (Figure~\ref{fig:FieldDiagram}). We can partition $\gHaystackDetect$ into disjoint subsets $\gHaystackSubJ$, each corresponding to a subwindow $\SubLabelJ$ that picks only galaxies of certain properties (in a small piece of the galaxy haystack) but otherwise makes the same selections on broadcasts, societies, and metasocieties that $\DetectLabel$ does. A broadcast is picked by $\SubLabelJ$ if it and its host society and metasociety would be picked by $\DetectLabel$ and its host galaxy specifically is in $\SubLabelJ$.

By assumption in the framework, each galaxy's properties -- including its location and stellar mass -- is determined by its location in the galaxy haystack, its tuple $\gTuple$. Furthermore, by assumption, the subpopulations of different galaxies are independent, conditionalized on the galaxy parameters. The broadcast populations picked by each $\SubLabelJ$ are mutually independent under the same conditions, since the different $\SubLabelJ$ necessarily select different possible host galaxies. Since a null detection means that no broadcast is picked by any $\SubLabelJ$, we can multiply probabilities to get the condition where a model is ruled out by a null detection, or equivalently sum their logarithms:
\begin{equation}
\label{eqn:NullDetectSum}
\sum_{\SubLabelJ \in \oSubOfDetect} \ln \fpP(\bNSubJ = 0) < \ln p ,
\end{equation}
with $\oSubOfDetect$ being the set of subwindows that make up $\DetectLabel$. If we are considering a known set of targeted galaxies, a model is ruled out when the sum of the $\ln \fpP(\bNGal)$ for each galaxy $\GalIndex$ in the sample is less than $\ln p$.

When we consider unidentified background field galaxies, however, we do not know the properties of these galaxies \emph{a priori}. Instead, we must work with a galaxy distribution, with the properties of the galaxies constrained by $\SubLabelJ$. Now, if a (sub)window fails to catch any broadcasts, that could mean either that there were no galaxies picked by that window in the first place, or that galaxies were picked, but none of them had any broadcasts:
\begin{multline}
\fpP(\bNSubJ = 0) = \fpP(\gNSubJ = 0) \\
+ \fpP(\gNSubJ > 0) \cdot \prod_{\gTuple \in \gSampleSubJ} \fpP(\bgNSubJ = 0 | \gTuple) .
\end{multline}
This condition applies as we make the individual $\SubLabelJ$ smaller and smaller, so small that the mean number of galaxies in each $\SubLabelJ$ is much smaller than $1$. We could subdivide physical space into cubic parsecs, for example, and only consider galaxies whose center is in each cubic parsec: most will be empty, a very few will have one, and practically none will have more. In this limit,\footnote{This assumes that the distribution describes a Poissonian point process without ``atoms'' of discrete probability. Thus it is not valid for a set of targeted galaxies with known properties, which would be modeled by a collection of fixed points in the haystack. Even if we considered each galaxy to have uncertain properties and thus a resolvable (if narrow) distribution in the haystack, the fact that there is exactly one and only one such galaxy invalidates the assumption of independence between different subwindows used in equation~\ref{eqn:NullDetectSum}.}
\begin{equation}
\ln \fpP(\bNSubJ = 0) \approx -\fpP(\gNSubJ > 0) \fpP(\bgNSubJ > 0 | \gTuple \in \gHaystackSubJ) . 
\end{equation}
For a Poissonian point process in this limit, $\fpP(\gNSubJ > 0) \approx \Mean{\gNSubJ}$. From Taylor series approximation, $\fpP(\gNSubJ > 0) \approx \gDist(\gTuple | \gTuple \in \gHaystackSubJ) \cdot \gHayvolumeSubJ$, where $\gHayvolumeSubJ$ is the hyper-volume of $\gHaystackSubJ$. An individual $\SubLabelJ$ carves out only a tiny parcel of the haystack, so tiny that variation from one $\gTuple$ to the next is negligible and any of them is representative. Applying these to equation~\ref{eqn:NullDetectSum}, we find that a model is inconsistent with a null detection when:
\begin{equation}
\sum_{\SubLabelJ \in \oSubOfDetect} -\gDist(\gTuple) \fpP(\bgNSubJ > 0 | \gTuple) \gHayvolumeSubJ \bigg|_{\gTuple \in \gHaystackSubJ} \la \ln p .
\end{equation}
In the infinitesimal limit, these approximations become exact, giving us the integral
\begin{equation}
\label{eqn:NullDetectIntegral}
\int_{\gHaystackDetect} -\gDist(\gTuple) \fpP(\bgNDetect > 0 | \gTuple) d\gTuple < \ln p.
\end{equation}

The galaxy distribution represents the astrophysical properties of host galaxies. Probably the two most important are location (which also specificies the cosmological epoch it is observed at), and $\hgNTime$, the instantaneous number of stars it hosts. The space volume element can be considered to be a parcel of comoving volume: $d\gPosition = d\yVolC = \yDistanceH \yDistanceCSQ / \sqrt{\ucOmegaM (1 + \yRedshift)^3 + \ucOmegaL} \cdot d\SkyVar d\yRedshift$ in a flat cosmology, with $\yDistanceH$ being the Hubble distance, $\yDistanceC$ the comoving distance, and $\yRedshift$ the redshift \citep{Hogg99}. So, in practical terms, equation~\ref{eqn:NullDetectIntegral} becomes:
\begin{multline}
\ln p > -\int_{\oSkyFieldSurv} \int_0^{\infty} \int_0^{\infty} \frac{d^2 \Mean{\gNTime}}{d\yVolC d\hgNTime} \fpP(\bgNDetect > 0 | \yRedshift, \SkyVar, \hgNTime) \\
\cdot \frac{\yDistanceH \yDistanceCSQ}{\sqrt{\ucOmegaM (1 + \yRedshift)^3 + \ucOmegaL}} \cdot d\hgNTime d\yRedshift d\SkyVar .
\end{multline}
Observed galaxy mass distributions can be used for the first term, while the broadcast detection probability depends on the details of the model.

\subsection{Detecting broadcasts in a galaxy: the case without confusion}
The conditions for there to be a detectable broadcast in a galaxy $\GalMark$ in survey $\SurvLabel$ are that 1.) there must be a bright enough broadcast in the right frequency range at the right time, also necessitating the existence of both a host society and host metasociety and 2.) neither the host galaxy, nor the metasociety, nor the society can be confused. The first condition is a statement mainly about the broadcasts themselves, and thus can be treated with the broadcast haystack; the latter condition is about the hosts and is left to the next section.

The assumptions of our models are that broadcasts form a Poisson point process for their host society, likewise, societies for their host metasociety, and that there is at most one metasociety per galaxy. Let us start with a single society $\SocMark$. The probability that no broadcasts are detected is
\begin{equation}
\label{eqn:PDetectSoc}
\fpP(\baNDetect = 0) = \exp\left[-\int_{\baHaystack} \baDistDetect(\bTuple) d\bTuple\right],
\end{equation}
where $\baDistDetect$ is the distribution of broadcasts that are selected by $\DetectLabel$ (the unbiased distribution multiplied by the detection probability). Since we are considering long-lived narrowband radio lines without frequency drift in this work, the main parameters defining each broadcast are its EIRP $\bLiso$, frequency $\bNuMid$, and physical position $\bPosition$. A detection is claimed only if some observable is above a minimum threshold, which for radio emission is generally energy collected, or measured energy flux $\dlFluxEObs$, with a minimum detectable flux $\lFluxEObsMIN$. Together, these considerations give us
\begin{align}
\nonumber \baDistDetect(\bTuple) & = \frac{d^3 \Mean{\baNDetect}}{d\bLiso d\bNuMid d\bPosition} \\
                                 & = \baAbundnuTotal \PDF{\bLiso} \fDirac(\bPosition - \aPosition) \IndicatorOf{\dlFluxEObs \ge \lFluxEObsMIN} .
\end{align}
Of course, the measured energy flux is subject to random noise, but at the level of approximation we are working, we can translate it to a luminosity threshold:
\begin{equation}
\dlFluxEObs \approx \iResponseESurv(\lSkyLocation, \bNuMid) \frac{\bLiso}{4 \pi \yDistanceL(\yRedshift)^2} ,
\end{equation}
where $\iResponseESurv$ is the best energy response of the survey beams at the broadcast's location on the sky $\lSkyLocation$ and frequency over all observations and $\yDistanceL$ is the broadcast's luminosity distance. If there are non-zero drift rates, corrections also must be made for drift smearing (Paper II). 
\begin{widetext}
Equation~\ref{eqn:PDetectSoc} reduces to an integral over broadcast frequency:
\begin{equation}
\fpP(\baNDetect = 0) = \exp\left[-\int \CCDF{\bLiso}\left(\ge \frac{4 \pi \yDistanceL(\yRedshift)^2 \lFluxEObsMIN}{\iResponseESurv(\lSkyLocation, \bNuMid)} \right) \cdot \baAbundnuTotal(\bNuMid) d\bNuMid \right] .
\end{equation}
Note that the frequency is in \emph{source-frame}.

Supposing that all of the societies are interchangeable, with similar broadcast properties (e.g., not varying widely in broadcast abundances or brightnesses), and that galaxies are compact enough to be treated as point sources (with similar beam responses for all the societies in a galaxy), our model assumptions give us
\begin{equation}
\label{eqn:PDetectSocA}
\fpP(\bgNDetect > 0) = \fpP(\zgNDetect = 1) \cdot \left[1 - \exp\left[-\azAbundDetect \hgNTime \cdot \left(1 - \exp\left[-\int \CCDF{\bLiso}\left(\ge \frac{4 \pi \yDistanceL(\yRedshift)^2 \lFluxEObsMIN}{\iResponseESurv(\gSkyLocation, \bNuMid)} \right) \cdot \baAbundnuTotal  d\bNuMid \right]\right)\right]\right] .
\end{equation}
The societal abundance $\azAbundDetect$ depends on the selection $\DetectLabel$ to account for confusion (next subsection).

Each individual beam has a response that falls off with distance from its center, and the characteristic beam width itself falls with frequency. At each point within the survey volume, and each possible broadcast luminosity, there is an effective bandwidth $\oBandwidthDetect(\gPosition, \bLiso)$ over which the response is great enough to allow a detection. The frequency integral in equation~\ref{eqn:PDetectSocA} weights this bandwidth with the luminosity distribution. Of course, the response is limited to $1$, so there is a maximum redshift for each luminosity, beyond which the bandwidth is $0$ and $\fpP(\bgNDetect > 0) = 0$ as well. 

All-sky and large field surveys cover the sky with beams packed densely enough that $\iResponseESurv$ can be considered to be $1$ throughout the entire survey sky area and survey (Earth-frame) bandwidth $\oeBandwidthSurv$. If the frequency abundance of broadcasts is constant, and all broadcasts have the same luminosity $\bLisoBAR$,
\begin{equation}
\fpP(\bgNDetect > 0) = \fpP(\zgNDetect = 1) \cdot \left[1 - \exp\left[-\azAbundDetect \hgNTime \cdot (1 - \exp[-\baAbundnuTotal \oeBandwidthSurv (1 + \yRedshift) ])\right]\right] \cdot \IndicatorOf{\yDistanceLSQ \le \bLisoBAR / (4 \pi \lFluxEObsMIN)} ,
\end{equation}
which makes calculations easy: we simply have to integrate the galaxy distribution function over survey volume. Under the usual assumptions of SETI, which effectively employs the single metasociety assumption and the diffuse approximation (ignoring societal discreteness) while ignoring confusion, this expression reduces to $1 - \exp(-\bzAbundnu \hgNTime \oeBandwidthSurv (1 + \yRedshift))$ when the broadcast's flux is detectable.
\end{widetext}

The general case with non-uniform beam patterns that vary with frequency is much more complicated, requiring a calculation of effective bandwidth over all the locations in the survey volume. For this work, however, I adopt $\iResponseESurv = 1$ anyway, because the vast range of parameter space being covered and approximations used do not merit such a precise and lengthy computation for this work. The greatest inaccuracies will occur for the background fields of \citet{Price20}, where the bandwidth is about 44\% of the central frequency, and thus the beam shrinkage is significant although not many orders of magnitude.

\subsection{The effects of confusion}
Populations of broadcasts can be confused, with so much overlap that no single one can be picked out. Paper II showed that in a host $\JMark$ observed by a survey with signal-to-noise threshold $\qSNThreshSurv$, the broadcasts in $\JMark$ are expected to become confused when $\Mean{\bjjNObs} > 1/\qSNThreshSurv^2$, if the broadcasts all have the same observed fluence. Beyond that threshold, the signal-to-noise of an individual broadcast drops low enough that none are detectable. Confusion can set on any ``level'' above the broadcast: the host society, metasociety, or galaxy can be confused. In this paper's models, each galaxy has at most one metasociety, so we can collapse galactic confusion into metasocietal confusion. Furthemore, if we evaluate according to these averages, it is possible that $\Mean{\bzNObs} \ll 1/\qSNThreshSurv^2$ while $\Mean{\baNObs} \gg 1/\qSNThreshSurv^2$ if societies are few in number but packed with broadcasts.

Because confusion is a property of populations, we account for its effects by making selections on the metasocietal and societal haystacks with $\DetectLabel$. Metasocietal confusion is treated by lowering the probability that a galaxy has a ``detectable'' metasociety when the flux distribution is narrow:
\begin{equation}
\fpP(\zgNDetect = 1) = \fpP(\zgNSurv = 1) \cdot \IndicatorOf{\MinSurv{\Mean{\bzNObs}} \le 1/\qSNThreshSurv^2},
\end{equation}
where $\MinSurv{\Mean{\bzNObs}}$ takes the minimum value over all the observations of the survey (e.g., if different observations have different bandwidths or broadcast abundances). If all the societies in a metasociety are interchangeable, with similar broadcast populations, then we can define an abundance
\begin{equation}
\azAbundDetect = \azAbund \cdot \IndicatorOf{\MinSurv{\Mean{\baNObs}} \le 1/\qSNThreshSurv^2} .
\end{equation}
If some societies have more broadcasts than others, then the possibility that some are confused while others are not must be taken into account.

Confusion only applies if the flux distribution is narrow. An intrinsic spread in luminosity (as with a power-law luminosity distribution $\PDF{\bLiso} \propto \bLiso^{-\alpha}$ with $1 < \alpha < 3$), or even a wide spread in drift rate, prevents confusion: as the number of broadcasts detected increases, so does the flux of the brightest single broadcast, allowing it to rise above the rest (Paper II). For this reason, we do not expect confusion to occur when many galaxies, each individually unconfused, are blended together: one of them will be the closest, and the flux distribution is already broadened to $\propto \lFluxEObs^{-5/2}$. A detection is possible as long as the \emph{brightest} broadcast rises above the mean, not the \emph{mean} broadcast -- just as Cygnus A is an easily detectable radio galaxy in GHz surveys even if it is observed with a wide beam that covers thousands of distant background sources.

\section{Collective bounds from source counts}
\label{sec:Collective}
Collective bounds result from the simple observation that the broadcasts of a metasociety contribute to the host galaxy's luminosity. Thus, the host's observed flux necessarily sets an upper bound to the flux from broadcasts within it. Paper II applied the concept to individual galaxies, with known radio fluxes. But in population terms, if a subset of galaxies host metasocieties that shine with bright radio broadcasts, then there should be a population of artificial radio galaxies, all with observed radio luminosities boosted by the technosignatures. We already know how many radio galaxies there are in the Universe, however, from source counts studies. This section deals with the method of using source counts to constrain metasocieties according to the collective bound.

\subsection{The aggregate flux distribution function}
Each galaxy is a (potential) radio source, with possible contributions from both natural and artificial emission. Generally, source counts are given in terms of \editOne{flux density}. A galaxy $\GalMark$ hosts a population of metasocieties represented by $\zgSampleObs$ that overlaps with the observation $\ObsLabel$, and
\begin{equation}
\label{eqn:TotalFlux}
\qgFluxENuObs = \sum_{\zTuple \in \zgSampleObs} \mzFluxENuObs(\zTuple) + \kgFluxENu,
\end{equation}
with $\mzFluxENuObs$ being the effective flux density from all the broadcasts sampled in the observation, and $\kgFluxENu$ being the natural emission.

The source count distribution follows from the galaxy distribution, which also regulates the distribution of metasocieties. Now, the relevant dimensions of the galaxy haystack include a galaxy's position (which also sets the redshift), the instantaneous number of stars present during the epoch of observation, and possibly other parameters for natural radio emission and galaxy morphology that are represented as the subtuple $\gParamsOther$:
\begin{equation}
\gDist(\gPosition, \hgNTime, \gParamsOther) = \frac{d^2 \Mean{\gNTime}}{d\yVolC d\hgNTime} \cdot \PDF{\gParamsOther}.
\end{equation}
\begin{widetext}
The source counts are then
\begin{equation}
\frac{d^2 \Mean{\gNTime}}{d\SkyVar d\qgFluxENuObs} (\qFluxENu) = \frac{1}{4\pi} \iiint \frac{d^2 \Mean{\gNTime}}{d\yVolC d\hgNTime} \cdot \PDF{\gParamsOther} \cdot \PDF{\qgFluxENuObs}(\qFluxENu | \gParamsOther) \cdot d\gParamsOther d\hgNTime d\yVolC.
\end{equation}
These fluxes are restricted to the observation $\ObsLabel$ because there will be some sample variance in the number of broadcasts active during the measurements, resulting in fluctuations in the brightness.

In this work, I ignore the natural flux ($\kgFluxENu = 0$). This is a conservative assumption, in that adding emission makes a galaxy brighter. Because there are fewer bright things than faint things, any model where the artificial broadcasts alone cause galaxies to violate source count constraints will only fare worse if the natural emission is added. 

Furthermore, I work under the assumption of at most $1$ metasociety per galaxy. Thus, we have:
\begin{equation}
\qgFluxENuObs = \left\{ \begin{array}{cl}
                         \mzFluxENuObs  & \text{if}~\GalMark~\text{hosts metasociety}~\MetaMark\\
                         0              & \text{if}~\zgNObs = 0
                         \end{array} \right. .
\end{equation}
Now, the metasocietal flux distribution will depend on both the number and brightness of individual broadcasts. These will depend on the number of stars in the metasociety (and its host galaxy), its cosmic epoch and position, and other parameters like a characteristic luminosity scale, broadcast abundance per society, and rate of societies per star that are collected into $\zParamsOther$. The probability density function (PDF) for the emission from metasocieties is weighted by these parameters:
\begin{equation}
\PDF{\qgFluxENuObs} (\qFluxENu | \gTuple) = \fpP(\zgNObs = 0) \fDirac(\qFluxENu) + \int_{\zHaystackSurv} \zgDist(\zTuple | \gTuple) \cdot \PDF{\mzFluxENuObs}(\qFluxENu | \zTuple) d\zTuple. 
\end{equation}
Given a host galaxy's parameters, the metasocietal distribution is $\zgDist(\zTuple | \gTuple) = \fpP(\zgNObs = 1) \fDirac(\zPosition - \gPosition) \fDirac(\hzNObs - \hgNTime) \PDF{\zParamsOther | \gParamsOther}$.\footnote{In this paper, since we observe each galaxy essentially at a single instant, I disregard the parameters for the origin time and longevity of a metasociety, even if the abundance evolves with time.} So, 
\begin{equation}
\PDF{\qgFluxENuObs} (\qFluxENu | \gTuple) =  \fpP(\zgNObs = 0) \cdot \fDirac(\qFluxENu) + \fpP(\zgNObs = 1) \cdot \int \PDF{\zParamsOther | \hgNTime, \gPosition} \cdot \PDF{\mzFluxENuObs}(\qFluxENu | \hgNTime, \gPosition, \zParamsOther) d\zParamsOther .
\end{equation}

We finally calculate the source count distribution:
\begin{equation}
\label{eqn:SourceCountsCalc}
\frac{d^2 \Mean{\gNTime}}{d\SkyVar d\qgFluxENuObs} (\qFluxENu) = \frac{1}{4\pi} \iiint \frac{d^2 \Mean{\gNTime}}{d\yVolC d\hgNTime} \fpP(\zgNObs = 1) \cdot \PDF{\zParamsOther | \hgNTime, \PosVecVar} \cdot \PDF{\mzFluxENuObs} (\qFluxENu | \hgNTime, \PosVecVar, \zParamsOther)  d\zParamsOther d\hgNTime d\yVolC.
\end{equation}
The stellar mass density distribution is well known (Section~\ref{sec:MassFunctions}), leaving the flux distribution.
\end{widetext}

\subsection{Aggregate flux as an observable}

The aggregate broadcast flux from a metasociety is composed of the aggregate broadcast flux from all its active societies:
\begin{equation}
\label{eqn:AggFlux}
\mzFluxENuObs = \sum_{\aTuple \in \azSampleObs} \maFluxENuObs = \sum_{\aTuple \in \azSampleObs} \sum_{\bTuple \in \baSampleObs} \lFluxENuObs .
\end{equation}
The discreteness of both societies and broadcasts add a graininess to the flux distribution function. The broadcasts of a host society are regarded as a Poisson point process on the broadcast haystack, and likewise the societies within the metasociety are a Poisson point process. Now, there is no general formula for the probability density of an aggregate sum like equation~\ref{eqn:AggFlux} where the number of variables is itself Poissonian, although if all the variables are independent and identically distributed, the resulting aggregate is called compound Poisson \citep[e.g.,][]{Adelson66}.

However, convenient numerical expressions are found if we use the characteristic function of the variables. The characteristic function of a variable is essentially the Fourier transform of its probability density: 
\begin{equation}
\CF{\jSingleCore}(x) = \Mean{\exp(i x \jSingleCore)} = \int \exp(i x \jSingleCore) \PDF{\jSingleCore} d\jSingleCore .
\end{equation}
A useful property of these functions is that when independent random variables are added together, the characteristic function of their sum is equal to the product of the characteristic functions of the individual random variables.

As a Poisson point process, the broadcasts in a society can be shown to have aggregate emission with a characteristic function 
\begin{align}
\label{eqn:CFFluxBcInSoc}
\nonumber \CF{\maFluxENuObs}(x) & = e^{\int \baDistObs(\bTuple) \left(\CF{\lFluxENuObs | \bTuple}(x) - 1\right) d\bTuple} \\
                      &  = \exp\left[\Mean{\baNObs} \left(\aCFObs{\lFluxENuObs}(x) - 1\right)\right] 
\end{align}
\citep{Kingman93}. The characteristic function in the second expression, $\aCFObs{\lFluxENuObs}$, is a characteristic function that uses the flux distribution for a broadcast randomly drawn by $\ObsLabel$ from society $\SocMark$, as opposed to a broadcast with fixed parameter tuple $\bTuple$. We generally know the flux distribution of broadcasts within a society already, since it follows from the luminosity distribution, which is a core assumption of the model. Thus, we can save an integration; if we have a power-law flux distribution for individual broadcasts, for example, we can just use the characteristic function for that power law.

The societies of a metasociety are another Poisson point process, so likewise,
\begin{multline}
\CF{\mzFluxENuObs}(x) = \exp\left[\Mean{\azNObs} \left(\zCFObs{\maFluxENuObs}(x) - 1\right)\right] .
\end{multline}
If all the societies are interchangeable, with the same mean number and flux distribution of broadcasts, then $\zCFObs{\maFluxENuObs}$ is simply $\CF{\maFluxENuObs}$. The diffuse approximation lets us skip this extra compounding step, allowing us to plug in the number and distribution of broadcasts in the metasociety in equation~\ref{eqn:CFFluxBcInSoc}.

All that is left is to find the distribution function for $\lFluxENuObs$, from which the characteristic function follows from a numerical Fourier transform. In the box model, the energy fluence from a line is directly proportional to the luminosity. Lines with frequency drift have a more complicated fluence distribution (Paper II), but the observations used for source count measurements are very broadband, so it is unlikely a line enters or leaves a frequency window during the observation. Furthermore, the measured \editOne{flux density} is directly proportional to the energy fluence. Thus, up to a distance-dependent constant of proportionality, the $\lFluxENuObs$ distribution is directly proportional to the (effective isotropic) luminosity $\bLiso$ distribution. When all the lines have the same luminosity, it reduces to simply the number distribution, up to distance-dependent constants of proportionality.

\subsection{The source count constraint}
Source counts reported in the literature come in two forms. Differential source counts report estimates for the distribution of the flux at various reported flux levels $\qFluxENuI$. For convenience's sake, I write
\begin{equation}
\gNFluxESkyI \equiv \left. \frac{d^2 \Mean{\gNTime}}{d\SkyVar d\qgFluxENuObs} \right|_{\qgFluxENuObs = \qFluxENuI} .
\end{equation}
The reported values for this quantity are $\dgNFluxESkyI$ with an uncertainty of $\fDelta \dgNFluxESkyI$.

A subset of galaxies with enhanced radio emission from artificial radio broadcasts is a new population of radio sources, appearing as a shelf in the Euclidean normalized differential source counts. If this shelf sticks out of the observed distribution function according to a model, that model predicts too many radio galaxies at some brightness to be consistent with observations.

To fully evaluate a model's consistency with the source counts, we need a model of natural emission as well as artificial. This requires the distribution of star-formation rate and the amount of AGN activity, which is beyond the scope of this work. Instead, I use a simpler criterion: the artificial radio emission alone should not produce a population of galaxies that violates observed source counts. This is a conservative criterion. Any natural emission will make galaxies brighter, and since brighter things are much rarer than dimmer things (at flux levels observed in source count studies), the violation in the observed source counts would become even worse. This approach does not penalize a model for underproducing the source counts; any discrepancy there can be attributed to the unpredicted natural radio emission (either as a distinct population of galaxies, or as an enhancement of the radio flux of inhabited galaxies), which is essentially a free parameter.

I evaluate consistency with a quasi-$\chi^2$ test. If the observations have Gaussian errors\footnote{An assumption that is not precisely correct because the errors are usually Poissonian.}, then the $\chi^2$ value is closely related to the likelihood of a model, its consistency with observation. When $\dgNFluxESkyI < \gNFluxESkyI$ for a given data point $i$, the contribution to the $\chi^2$ sum is taken to be $0$, which is equivalent to asserting a likelihood of $1$; the model is deemed perfectly consistent with observations and the remaining unaccounted-for radio sources are attributed to the effects of the unmodeled natural contribution. If $\dgNFluxESkyI \ge \gNFluxESkyI$, however, the best possible case is when there is no natural radio emission. Any overproduction of radio sources necessarily must be an overproduction in the artificial radio sources. As a test-statistic, I use
\begin{equation}
\chi^2 = \sum_{i = 1}^{\oNI} \left[\frac{\max(\gNFluxESkyI - \dgNFluxESkyI, 0)}{\fDelta \dgNFluxESkyI}\right]^2 ,
\end{equation}
given $\oNI$ data points. A model is inconsistent with the differential source counts if $\chi^2$ is greater than the degrees of freedom in the case of no excess (and thus no additional free parameters). Assuming these are independent, a model with $N_p$ parameters needs to have
\begin{equation}
\label{eqn:CountConstraint}
\chi^2 \le \oNI - N_p.
\end{equation}
A more rigorous treatment would account for the complicated relationship between the nonlinear parameters and degrees-of-freedom; the goal here is to establish, at a very basic level, what kinds of ETI transmitter populations are grossly inconsistent with source counts.

On occasion, integral source counts are reported, giving the number of sources on the sky above a particular flux (fluence) level. A model may produce too many galaxies brighter than a given threshold. We could again use a $\chi^2$-like statistic to quantify consistency between observations and model predictions. But source counts generally have a $\gNFluxESky \propto {\qgFluxENu}^{-5/2}$ dependence over a wide range of fluxes; in these ranges, integral source counts are greatly dominated by sources with the minimum reported flux. If the typical brightness of an artificial radio galaxy falls within the flux ranges probed by observations, then, the differential source counts provide a sufficient constraint. I therefore mostly ignore integral source counts, especially given their relative paucity, with one important exception: the brightest source constraint. 

In some models with extremely abundant, luminous broadcasts, \emph{every} inhabited galaxy in the observable Universe is absurdly bright (e.g., with megajansky \editOne{flux densities}). But these models are obviously ruled out because the inhabited galaxies would be brighter than any actual radio galaxies on the sky. I define a \emph{brightest source constraint}:
\begin{equation}
\label{eqn:BrightestRatio}
\Mean{\gNTime(\qgFluxENuObs \ge \qgFluxENuObsMAX)} \le n
\end{equation}
where, for this equation, $\qgFluxENuObsMAX$ is the flux of the brightest radio source intercepted by an observation $\ObsLabel$, and thus an upper limit on the brightness of a galaxy. For frequencies near $1\ \GHz$, the brightest source in the radio sky is Cygnus A; the constraint then asks how many sources brighter than Cygnus A are predicted by a model. Of course, it is possible the lack of sources above that brightness is a Poissonian fluke, which is accounted for by using $n > 1$. I set $n = 3.0$, which rules out models that predict too many bright sources at $95\%$ confidence. This constraint turns out to be quite important in constraining a wide range of model space.

\subsection{Miscellaneous issues working with source counts} 
Another problem is that deep surveys are wideband, which they achieve by constructing images that are weighted sums of images in narrower subbands \citep[as in][]{Condon12,Matthews21}. If all the subbands have equal weights, this is no issue.  Nor is it a problem for sources that span the entire bandwidth, like for the continuum sources assumed by most source count derivations. But when narrowband broadcasts are sparse enough that most subbands are expected to be unoccupied, the apparent flux of each broadcast in the final image depends on which subband it appears in. The effective flux is a weighted combination of all the subbands, which can have a nontrivial PDF given all the different possible combinations of subbands. SETI limits span order-of-magnitude in sensitivity to both number and luminosity, however, so simply assuming uniform sensitivity across the survey band is sufficient. Thus, I ignore these complications.

In any case, if we are approaching the sparse limit for narrowband broadcasts, source counts are not the most potent constraints.  The vast majority of sources in radio surveys are continuum-dominated and thus will be detected in all subband images if bright enough.  But if the radio spectrum resolves into a very few narrowband broadcasts, we expect the galaxy to appear in one, or perhaps a few, of the subband images instead of all of them.  In fact, they should stand out more in individual subband images because they are not competing against the noise from all the other subbands.  A search of individual subband images for narrowband searches could be fruitful from a SETI perspective.

\section{Models of extragalactic broadcast populations: Methods}
\label{sec:Methods}
In this section, I describe models to compare extant individualist field and collective source count bounds on narrowband transmitters with zero frequency drift.

To keep the number of variant models to a manageable level, I consider a small number of model sets that test the effects of different basic assumptions (Table~\ref{table:ModelSets}). Within each model set, different models explore the effects of different characteristic luminosities and abundances. The base model set describes a scenario where every galaxy has one metasociety, the diffuse approximation applies, the broadcasts all have a single luminosity, and there is no evolution. It ignores the discreteness of metasocieties and societies. The base model set reflects the usual assumptions made when deriving broadcast abundances in the literature. Other model sets test the effects of rare expansive metasocieties (A), discrete societies (B and C), and a broad (power-law) luminosity distribution (D).

In all model sets, $\bzAbundnu$ (and $\baAbundnuTotal$ in model sets B and C) has no $\bNuMid$ dependence. Thus, there are an equal number of broadcasts expected per unit frequency, as opposed to per decade of frequency. The mean aggregate radio spectrum is then flat, $\Mean{\qgFluxENuObs} \propto \FreqVar^0$. 

\begin{deluxetable*}{ccccccc}
\tabletypesize{\footnotesize}
\tablecolumns{7}
\tablewidth{0pt}
\tablecaption{Model sets \label{table:ModelSets}}
\tablehead{\colhead{Model set} & \colhead{Metasociety} & \colhead{Societies} & \colhead{Evolution} & \colhead{Luminosity Distribution} & \colhead{Bands} & \colhead{$N_p$}}
\startdata 
Base & Single          & Diffuse  & None                                                  & Degenerate                                 & 0.15, 1.4, 16, 250 GHz & 2\\
A    & Expansive       & Diffuse  & $\zgAbundEMPTY \propto \TimeVar^q (q = 0, 1, 2)$ 			& Degenerate                                 & 1.4 GHz                & 4\\
B    & Galactic Club   & Discrete & None                                                  & Degenerate                                 & 1.4 GHz                & 3\\
C    & Expansive       & Discrete & None                                                  & Degenerate                                 & 1.4 GHz                & 4\\
D    & Expansive       & Diffuse  & None                                                  & Pareto $\bLiso$ ($\alpha = 1.0, 1.5, 2.0$) & 1.4 GHz                & 4
\enddata
\end{deluxetable*}

I adopt a $\Lambda$CDM cosmology with $\ucHubbleZero = 67.66\ \km\ \sec^{-1}\ \Mpc^{-1}$, $\ucOmegaM = 0.3111$, and $\ucOmegaL = 0.6889$ \citep{Aghanim20}.

\subsection{Galaxy stellar mass functions}
\label{sec:MassFunctions}
The galaxy stellar mass density function is a key ingredient in calculating both source count constraints and commensal individualist bounds. Ideally, the distribution should cover a wide range of redshift including $\yRedshift = 0$, have a smooth evolution over cosmic time, and cover a wide range in stellar mass. In practice, most works either focus on very low redshifts or redshifts significantly above $0$, without ensuring total consistency between the two regimes. The very low redshift mass function is critical because it determines the entire bright end of the flux distribution, all the way down to the faint-end tail. The nearby galaxies provide all limits on fainter broadcasts from field surveys. The higher redshift mass functions are mainly used for constraining the faint-end tail of the source count function, or in a narrow range of broadcast abundances where the number of observed galaxies must be maximized without being so rare that there are none in the entire Universe. If the high-end mass functions do not match up with those at low redshift, an apparent jump appears in the calculated flux distributions, which is a completely unphysical feature. The stellar mass distributions are thus chosen to ensure that the jump is minimized.

When the redshift is exactly $0$, I apply the \citet{Driver22} mass function derived from GAMA applies. At redshifts between $0.5$ and $4.0$, the \citet{Thorne21} fits to the mass functions are used directly. When $0 < \yRedshift < 0.5$, I interpolate between the \citet{Driver22} $\yRedshift = 0$ mass function and the fit mass function from \citet{Thorne21} at that redshift. The \citet{Driver22} and $\yRedshift = 0.5$ \citet{Thorne21} mass functions have similar shapes, so using this combination results in the fewest artifacts appearing in the resultant source count distributions. I impose an upper redshift limit of $4.0$, which is near the maximum limit for \citet{Thorne21}.

The \citet{Driver22} and lower-redshift \citet{Thorne21} mass functions both go down to very low stellar masses, near $10^7\ \Msun$, but galaxies this small make up only a small fraction of the total stellar mass (and thus sites of habitable planets) in the present Universe, and they are poorly constrained at moderate-to-high redshift. I somewhat arbitrarily impose a lower mass limit of $10^9\ \Msun$, of order the size of the Large Magellanic Cloud. Galaxies this small are constrained out to $\yRedshift \sim 1$ in the observations used for \citet{Thorne21}. 

To convert stellar mass into number of stars, the mean mass of an individual star in a galaxy $\GalMark$, $\gMean{\hmstar}$, is assumed to be $0.2\ \Msun$ from the \citet{Chabrier03} initial mass function.

\subsection{Data for source count constraints}
Observed source counts come from a variety of surveys listed in Table~\ref{table:SourceCounts}. I group them into four wide frequency bands centered at 1.4 GHz, 150 MHz, 16 GHz, and 250 GHz. Each band includes data from several surveys: these vary from shallow, wide-field surveys and deep, pencil beam surveys. The 1.4 GHz source counts are among the most well-studied and cover the most range in flux. Additionally, the frequencies near 1.4 GHz are among the most well-studied in SETI. Thus I emphasize the 1.4 GHz results for most models.

\begin{deluxetable*}{cccccc}
\tabletypesize{\footnotesize}
\tablecolumns{6}
\tablewidth{0pt}
\tablecaption{Source count data\label{table:SourceCounts}}
\tablehead{\colhead{Survey} & \colhead{Instrument} & \colhead{$\oeNuMidObs$} & \colhead{$\oeBandwidthObs$} & \colhead{$\log_{10} [\qFluxENu (\Jy)]$ Range} & \colhead{Notes}}
\startdata
\cutinhead{150 MHz}
GLEAM              & MWA           & $154\ \MHz$    & $30.72\ \MHz$ & $[-1.1, +1.9]$ & a\\
\citet{Williams16} & LOFAR         & $146.5\ \MHz$  & $71.48\ \MHz$ & $[-3.1, -0.1]$ & \\
\cutinhead{1.4 GHz}
NVSS               & VLA           & $1.4\ \GHz$    & $42\ \MHz$    & $[-2.5, +1.3]$ & b\\
DEEP2              & MeerKAT       & $1.266\ \GHz$  & $661\ \MHz$   & $[-4.9, -2.5]$ & \\
\cutinhead{16 GHz}
AT20G              & ATCA          & $19.904\ \GHz$ & $256\ \MHz$   & $[-1.2, +1.3]$ & \\
9C                 & Ryle          & $15.2\ \GHz$   & $350\ \MHz$   & $[-2.2, -0.1]$ & c\\
10C                & AMI           & $15.7\ \GHz$   & $4.5\ \GHz$   & $[-3.8, -2.1]$ & d\\
\cutinhead{250 GHz}
Intermediate       & \emph{Planck} & $217\ \GHz$    & $68\ \GHz$    & $[-0.4, +0.8]$ & \\
SPT-SZ             & SPT           & $220\ \GHz$    & $120\ \MHz$   & $[-1.8, +0.1]$ & \\
\citet{Lindner11}  & MAMBO         & $250\ \GHz$    & $80\ \GHz$    & $[-2.7, -2.4]$ & \\
GOODS						   & ALMA          & $265\ \GHz$    & $7.5\ \GHz$   & $[-3.1, -2.7]$ & \\
\citet{Fujimoto16} & ALMA          & $250\ \GHz$    & $7.5\ \GHz$   & $[-4.7, -2.9]$ & 
\enddata
\tablenotetext{a}{Only includes the source counts for 154 MHz. Additional source counts are available at 88, 118, and 200 MHz \citep{Franzen19}.}
\tablenotetext{b}{Source counts reported in \citet{Matthews21}.}
\tablenotetext{c}{Bandwidth for the Ryle telescope from \citet{Jones91}.}
\tablenotetext{d}{Bandwidth reported as ``usable'' in \citet{Davies11}.}
\tablereferences{NVSS: \citet{Condon98,Matthews21}; DEEP2: \citet{Matthews21}; GLEAM: \citet{Franzen19}; AT20G: \citet{Murphy10,Massardi11}; 9C: \citet{Waldram10,Jones91}; 10C: \citet{Whittam16,Davies11}; MAMBO: \citet{Greve04}; SPT-SZ: \citet{Everett20,Carlstrom11,Vieira10}; GOODS: \citet{Franco18}; \emph{Planck}: \citet{Ade13-SourceCounts,Tauber10,Ade11-Intro}}
\tablecomments{Acronyms -- ALMA: Atacama Large Millimeter/submillimeter Array; AMI: Arcminute Microkelvin Imager; ATCA: Australia Telescope Compact Array; GLEAM: Galactic and Extragalactic All-sky MWA; GOODS: Great Observatories Origins Deep Survey; LOFAR: Low Frequency Array; MAMBO: Max-Planck Millimeter Bolometer; MWA: Murchison Widefield Array; NVSS: NRAO VLA Sky Survey; SPT: South Pole Telescope; VLA: Very Large Array}
\end{deluxetable*}

Cygnus A is used as the standard in the brightest source constraint at 150 MHz, 1.4 GHz, and 16 GHz. Cassiopeia A and the Crab Nebula are brighter, but are intragalactic. Cygnus A is the brightest extragalactic source at these frequencies based on comparison with the fluxes reported for other bright radio galaxies in \citet{Perley17}. I use the \citet{Perley17} fit to the radio flux, which actually only used data from 50 MHz to 12 GHz; comparing the extrapolated flux to that in the Second Planck Catalog of Compact Sources (PCCS2) shows a good match at 30 GHz \citep{Ade16-PCCS2}. According to PCCS2, Centaurus A is brighter at 30 GHz at 52 Jy to Cygnus A's 44 Jy, but the 20 GHz flux for Centaurus A in AT20G (28 Jy; \citealt{Murphy10}) is about half that of Cygnus A (69 Jy). 

At frequencies around 250 GHz, the brightest extragalactic sources in the PCCS2 are flat-spectrum radio galaxies. The brightest at 143 GHz and 217 GHz is 3C454.3 \citep{Ade16-PCCS2}. Unfortunately, no detection is reported for that galaxy in the next \emph{Planck} band at 353 GHz, but the spectrum below that frequency appears to be relatively flat. I adopt Planck Early Release Compact Source Catalog fluxes for 3C454.3, which does include fluxes for frequencies as high as 857 GHz, albeit at levels up to $\sim 50\%$ higher than in the PCCS2 \citep{Ade11-EarlyCSC}. I use a power-law interpolation between these reported fluxes at 217 GHz and 353 GHz to get a conservative estimates of 3C454.3's radio flux at 250 GHz. 

At still higher frequencies, the nearby starbursts NGC 253 and M82 are the brightest sources because of their steeply rising spectra \citep{Ade16-PCCS2}, a result of the far-infrared emission from dust.

\subsection{Data for field constraints from individualist searches}
The best surveys to use for commensal searches of individual broadcasts have wide sky field coverage, are wideband, and are sensitive. I calculate these field constraints for a few representative surveys listed in Table~\ref{table:IndividualistSearches}.

\begin{deluxetable*}{ccccccccc}
\tabletypesize{\footnotesize}
\tablecolumns{9}
\tablewidth{0pt}
\tablecaption{Field broadcast searches considered\label{table:IndividualistSearches}}
\tablehead{\colhead{Survey} & \colhead{Instrument} & \colhead{$\oeNuMidSurv$} & \colhead{$\oeBandwidthSurv$} & \colhead{$\oeBandwidthObs$} & \colhead{$\oSkyFieldSurv$} & \colhead{$\qSNThreshSurv$} & \colhead{Sensitivity} & \colhead{Notes} \\ & & \colhead{($\GHz$)} & \colhead{($\MHz$)} & \colhead{($\Hz$)} & \colhead{($\sr$)} & & \colhead{($\Watt\,\meter^{-2}$)}}
\startdata
META (H93)                & Harvard/Smithsonian 26 m & $1.42$  & $1.2$  & $0.05$   & $8.6$    & 30 & 1.70E-23  & a\\
Breakthrough Listen (P20) & GBT                      & $1.50$  & $660$  & $2.79$   & $0.0041$ & 10 & 6.82E-26  & b\\
\citet{Tremblay20} (T20)  & MWA                      & $0.113$ & $15.4$ & $10,000$ & $0.12$   & 5  & 2.50E-23  & c
\enddata
\tablereferences{H93: \citet{Horowitz93}; P20: \citet{Price20}; T20: \citet{Tremblay20}}
\tablenotetext{a}{Only includes the observations at 1.42 GHz. Additional observations at 2.84 GHz with the same characteristics (but smaller beam size) were also taken, which would double the bandwidth covered, but these are not considered when deriving constraints. Sensitivity is average on sky, but in some sky regions is much worse because of noise.}
\tablenotetext{b}{L-band observations taken with the GBT, reported for 882 stars in \citet{Price20}. Additional observations at S band for a partly overlapping set of stars are also reported, but are not included in the constraints. $\oSkyFieldSurv$ calculated as $882$ times solid angle covered by FWHM at 1.5 GHz from the GBT Observer's Guide. Sensitivity calculated using SEFD of $10\ \Jy$, signal-to-noise of $10$, channel bandwidth $2.79\ \Hz$, and integration time of $5\ \minute$ with both polarizations.}
\tablenotetext{c}{Sensitivity calculated from a reported rms flux sensitivity of $0.05\ \Jy\,\mathrm{beam}^{-1}$ multiplied by a nominal signal-to-noise of $5$ and the reported channel bandwidth of $10\ \kHz$.}
\tablecomments{$\qSNThreshSurv$ is the signal-to-noise threshold of the survey. The total observer-frame frequency range covered by the survey is $\oeBandwidthSurv$, while $\oeBandwidthObs$ is the bandwidth of a single channel.}
\end{deluxetable*}

My calculations do not correct for the drift rates of broadcasts. High drift rates greatly decrease the sensitivity to broadcasts \citep{Margot21} but also reduce confusion (Paper II). The calculations also assume uniform sensitivity on the sky field $\oSkyFieldSurv$ of each survey. In surveys where the primary beams of different pointings do not overlap, the fall-off in sensitivity towards the edge of the beam leads to reduced ability to detect broadcasts \citep{Garrett23}.

\section{Results for the base model set: broadcast luminosity and abundance}
\label{sec:Base}

\begin{figure}
\centerline{\includegraphics[width=9cm]{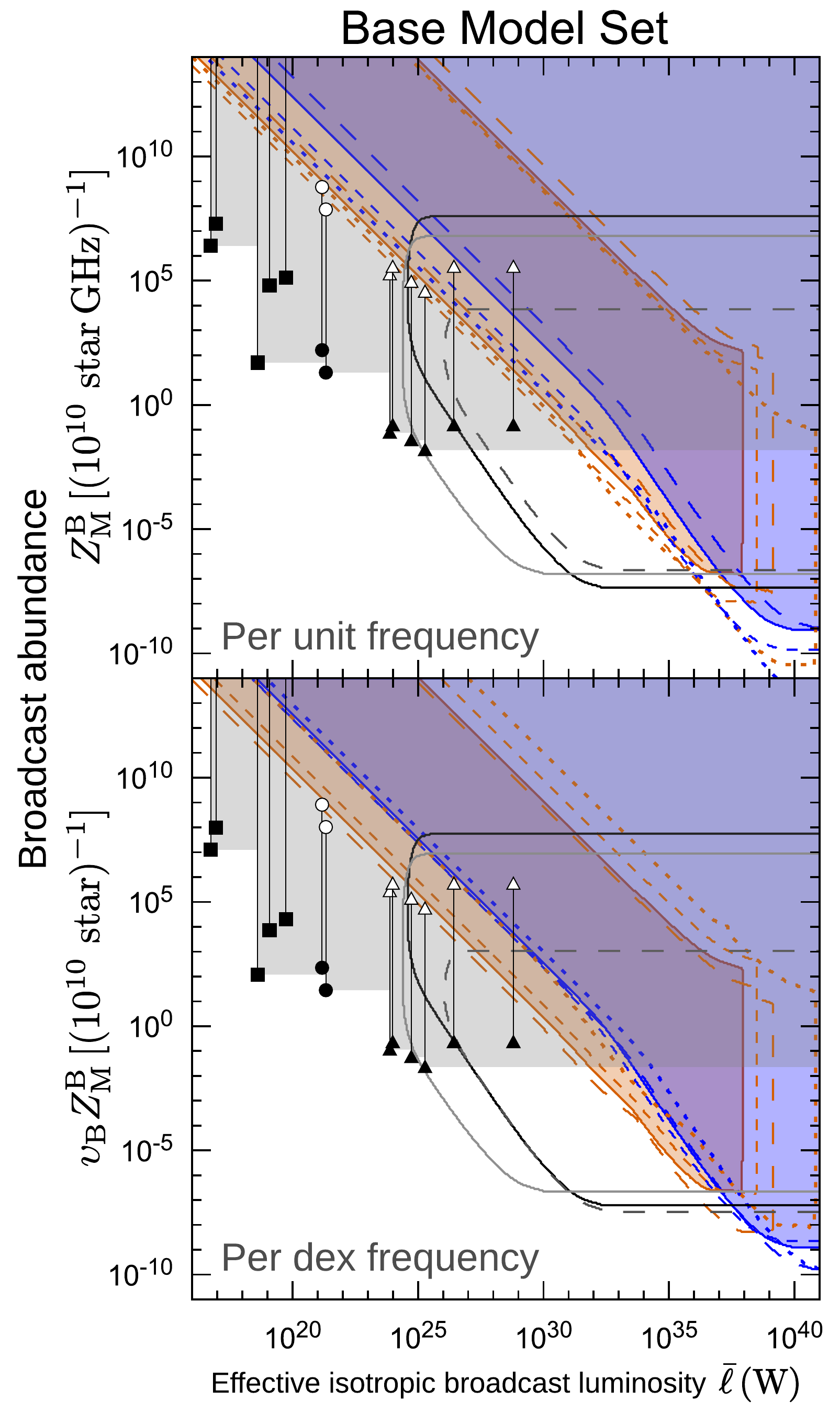}}
\figcaption{Constraints on luminosity and broadcast frequency abundance per metasociety in the base model set. On top, the frequency abundance is per unit frequency, but below, it is per unit log frequency. Orange limits delineate regions excluded by differential source counts, and blue limits for regions excluded by the brightest source limit. Frequency band is given by line style: solid (and shaded) for 1.4 GHz, long-dashed for 150 MHz, short-dashed for 16 GHz, and dotted for 250 GHz. These are compared with constraints from targeted surveys (points) and commensal background galaxies in large-scale surveys (contours: black solid for META, light grey solid for P20, dark grey long-dashed for T20). Each targeted survey has a pair of points, one for when there are too few broadcasts to detect (black, below) and another for when there are too many and they are confused (white, above); it excludes the grey shaded region extending to its right. The data points for targeted surveys are square for Milky Way surveys (from \citealt{Harp16,Tremblay20,Tremblay22,Gajjar21}), circles for M31 (\citealt{Gray17}), and triangles for the galaxies reported in \citet{Garrett23}. \label{fig:ConstraintsBase}}
\end{figure}

To start, I consider models where every galaxy has an active metasociety and the diffuse approximation is applied. The population of broadcasts in each metasociety -- and thus each galaxy -- has the same abundance per star, and all broadcasts everywhere in the Universe have the same luminosity. The diffuse approximation ignores clumping of broadcasts into societies, leaving only the discreteness of the broadcasts themselves. These are the assumptions implicit in most luminosity--abundance plots in the SETI literature. Since we are considering galaxies at cosmological distances, however, we are probing the far end of the luminosity distribution, much further than even targeted observations of nearby galaxies.

\subsection{Individualist constraints from targeted searches and the field}
Targeted searches yield powerful results at relatively low luminosity. This is because they just need a single broadcast to be detected over the background noise in a single observation, rather than requiring an entire population of them to outshine the natural background of a galaxy. Their reach is limited, however, because they only cover a few galaxies at most; thus, they are unable to constrain scenarios where broadcasts are rare (see data points in Figure~\ref{fig:ConstraintsBase}, with grey shading).

In setting constraints on the stellar population of all the galaxies included in these limits, targeted searches are limited by the small number of galaxies observed so far explicitly for SETI purposes. As one pushes into the Kardashev III regime, where broadcasts are enormously powerful but cosmically rare, we need to rely upon field limits. Even some background field results like \citet{Garrett23} and \citet{Uno23} may be limited by the small portion of the sky covered by targeted searches, and are further limited if they only consider galaxies listed in catalogs (the ``cross-matching'' method of \citealt{Uno23} and used in \citealt{Garrett23}).

The field limits shown in Figure~\ref{fig:ConstraintsBase} essentially continue the trend well-known for targeted programs -- probing deeper into abundance as we consider brighter and brighter broadcasts.\footnote{The commensal limits I derive for \citet{Price20} (grey lines) approach the reported limits in \citet{Garrett23} for the nearest background galaxies. These galaxies are actually nearer than might be expected for a typical sample, so the \citet{Garrett23} limits are a bit stronger.} The shape of the constrained region can be understood as coming from the number of galaxies from which a broadcast could be detected, with lower and upper bounds at low and high redshift. The ultimate floor for field constraints, $\sim 10^{17}\ \mathrm{star}\ \GHz$, comes from the limited bandwidth of each survey, and the limited number of galaxies in the survey volume, which can never exceed the total number of galaxies in the Universe. META has a small bandwidth but covered the entire sky, while the other shown surveys cover smaller footprints but with bandwidths of MHz to GHz. Hence, they extend down to roughly the same abundances.

Individualist limits are subject to confusion for high broadcast abundances. These apply to field bounds as well. Confusion imposes a ceiling on the abundance constraints, at which point the broadcasts in any sampled galaxy are confused. Confusion in these models is always worse in big galaxies -- there are more stars and more broadcasts blended together. Therefore, the confusion ceiling occurs when the smallest galaxy expected to be sampled is confused. But it should be noted that to some extent, the confusion limit is arbitrary in this work, being set by the $10^9\ \Msun$ lower limit on the mass functions. Smaller dwarf galaxies dominate the galaxy count, though not the cosmic stellar mass budget, which ETIs might be expected to trace.

\subsection{Collective emission and source counts}
\subsubsection{The luminosity function of artificial radio galaxies}
\begin{figure*}
\centerline{\includegraphics[width=18cm]{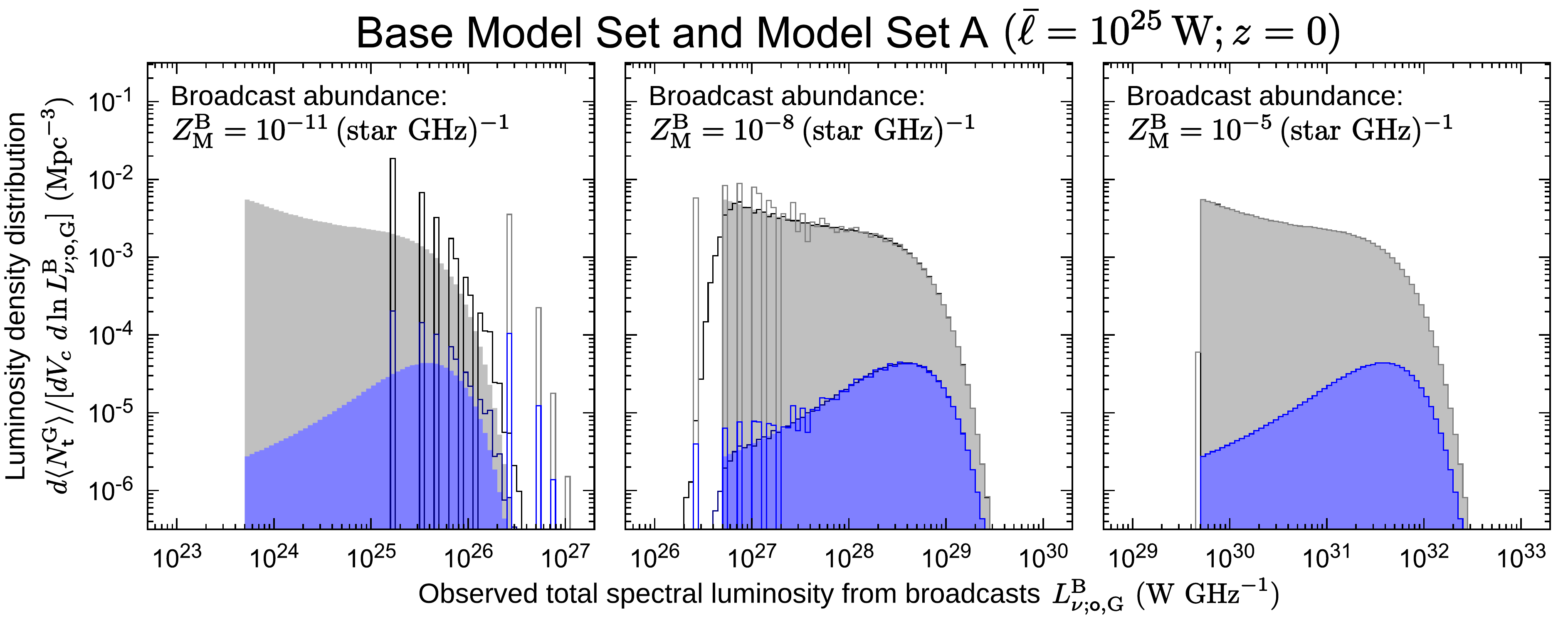}}
\figcaption{Luminosity distributions in the base model set and model set A at $\yRedshift=0$ in the absence of natural emission.  From left to right, $\bzAbundnu$ increases while broadcast luminosity remains fixed. The shaded curves are the naive mean luminosity distributions; dark lines are for DEEP-2; and light lines are for NVSS. These are shown for the base model with $\zgN = 1$ (black/grey) and when metasocieties are rare ($\zgAbundEMPTY = 10^{-13}\ \mathrm{star}^{-1}$; blue). The bin width is $0.05$ in natural logarithm of the luminosity.\label{fig:LumDistA}}
\end{figure*}

If broadcast sampling effects could be ignored, the galaxy luminosity distribution would be a smooth curve, with a shape that can be derived from the galaxy mass function:
\begin{multline}
\frac{d^2 \Mean{\gNTime}}{d\VolVar_C d\bgAggLnuiso} (\Mean{\bgNObs} \gg 1) = \fpP(\zgN = 1 | \hgMAggTime)  \\
    \cdot \frac{\gMean{\hmstar}}{\bLisoBAR \bzAbundnu} \frac{d^2 \Mean{\gNTime}}{d\VolVar_C d\hgMAggTime} \Bigg|_{\hgMAggTime = \bgAggLnuiso \Mean{\hmstar}/(\bLisoBAR \bzAbundnu)} .
\end{multline} 
Some examples of these mean luminosity functions are plotted as the shaded curves in Figure~\ref{fig:LumDistA}. In the base model, these trace the galaxy mass function.

Yet the measurable luminosity function is distorted by discrete sampling effects when $\bzAbundnu$ is small, however. The number of line broadcasts within a frequency window is an integer; we cannot detect $0.01$ broadcasts, for example. This can lead to a series of discrete peaks in the luminosity distribution (Figure~\ref{fig:LumDistA}). For mass ranges where $\Mean{\bgNObs} \ll 1$, most galaxies lack an observable broadcast, but a few do have some leading them to be overluminous compared to the naive mean-value expectation. As a result, the luminosity function is cut off below the threshold luminosity for one broadcast and is a comb above that threshold. Furthermore, $\Mean{\bgNObs}$ depends on the observational window itself, increasing with bandwidth. The discreteness effects are more prominent in line surveys with smaller bandwidths, as seen in Figure~\ref{fig:LumDistA} when comparing NVSS ($42\ \MHz$; light) and DEEP-2 ($661\ \MHz$; dark).

\subsubsection{Broadcast luminosities, abundances, and source count distributions}

\begin{figure*}
\centerline{\includegraphics[width=18cm]{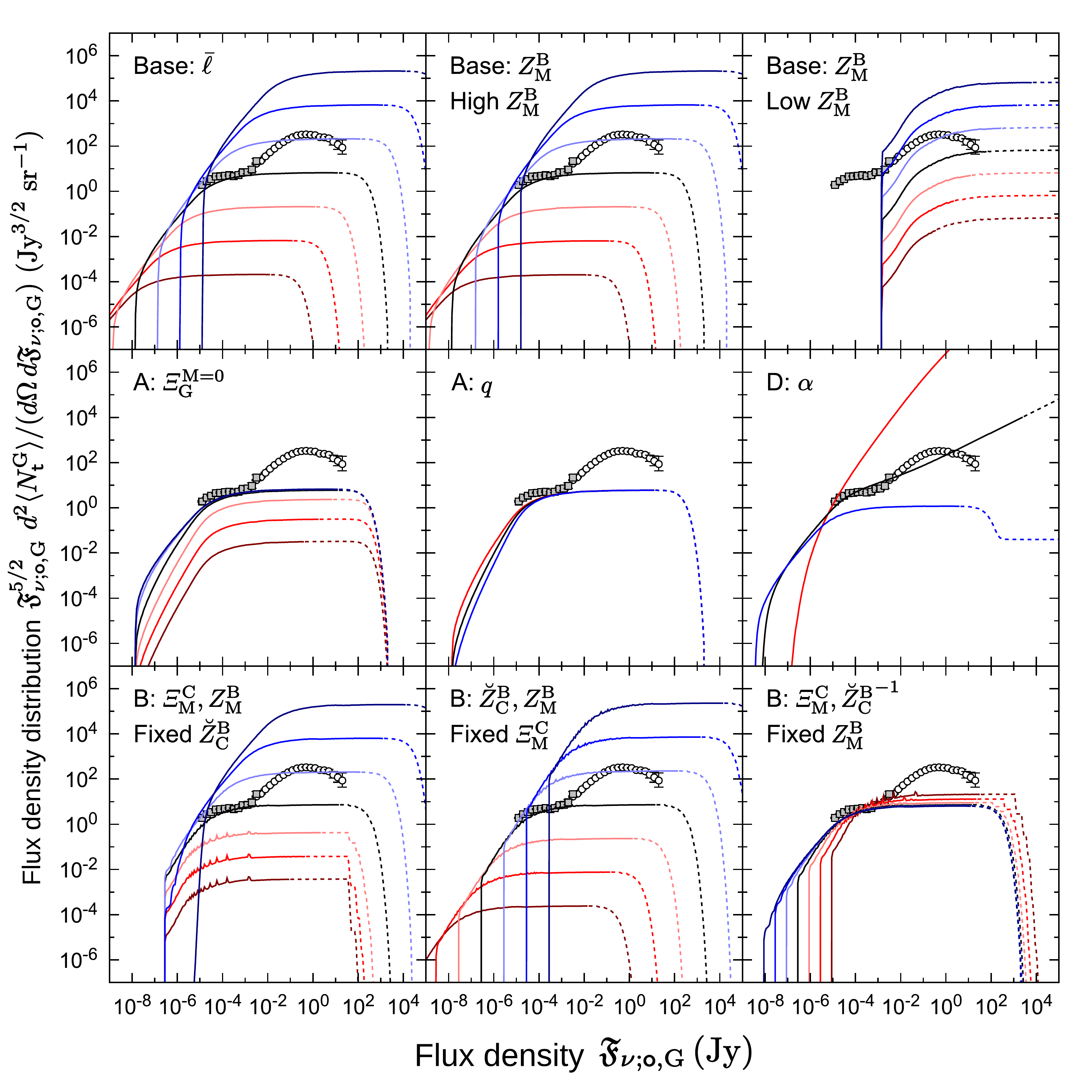}}
\figcaption{Effects of adjusting various model parameters on the predicted Euclidean-normalized flux distributions in NVSS for the base model set. The model set, along with the adjusted parameters are given by the panel labels. Moving from dark red through black to dark blue, the value of the parameter is increased by a factor of 10, except the evolution exponent in model set D, which is incremented by 1, and the luminosity power-law slope $\alpha$ in model set D, which is incremented by 0.5. The NVSS \editOne{(white-filled circles)} and DEEP2 \editOne{(grey-filled squares)} source counts at 1.4 GHz are shown to give a sense of scale. The transition to dashed lines in the predicted source counts indicates fluxes for which $\le 1$ sources of equal or brighter flux are expected on the sky.\label{fig:SourceCountsParams}}
\end{figure*}

The two parameters of the models, $\bLisoBAR$ and $\bzAbundnu$, have somewhat different effects on the source count distribution, as illustrated in Figure~\ref{fig:SourceCountsParams}. The scale luminosity has a very simple relationship with these functions: if all broadcasts are intrinsically brighter by a constant factor, all populations of broadcasts are brighter too by that same factor. This just shifts the source counts over.\footnote{It moves up and to the right in Figure~\ref{fig:SourceCountsParams} because the distribution is multiplied by ${\qgFluxENuObs}^{5/2}$.}

When $\bzAbundnu$ is high, $\Mean{\bgNObs} \gg 1$, and the broadcasts blend into a continuum. In this regime, $\bzAbundnu$ has the same effect on the source counts as luminosity (upper-middle panel in Figure~\ref{fig:SourceCountsParams}): it does not matter whether a galaxy is brighter because it has more sources or each source is more luminous, because there are so many. This behavior breaks down when $\Mean{\bgNObs} \la 1$, though. Then almost all galaxies has either zero or one broadcast, which solely accounts for the artificial radio emission. What $\bzAbundnu$ then controls is the \emph{fraction} of galaxies with a broadcast. The source count function then reaches a limiting shape, and changing $\bzAbundnu$ simply shifts that shape up or down (upper-right panel in Figure~\ref{fig:SourceCountsParams}).

\subsubsection{Collective constraints from source count distributions}

\begin{figure*}
\centerline{\includegraphics[width=18cm]{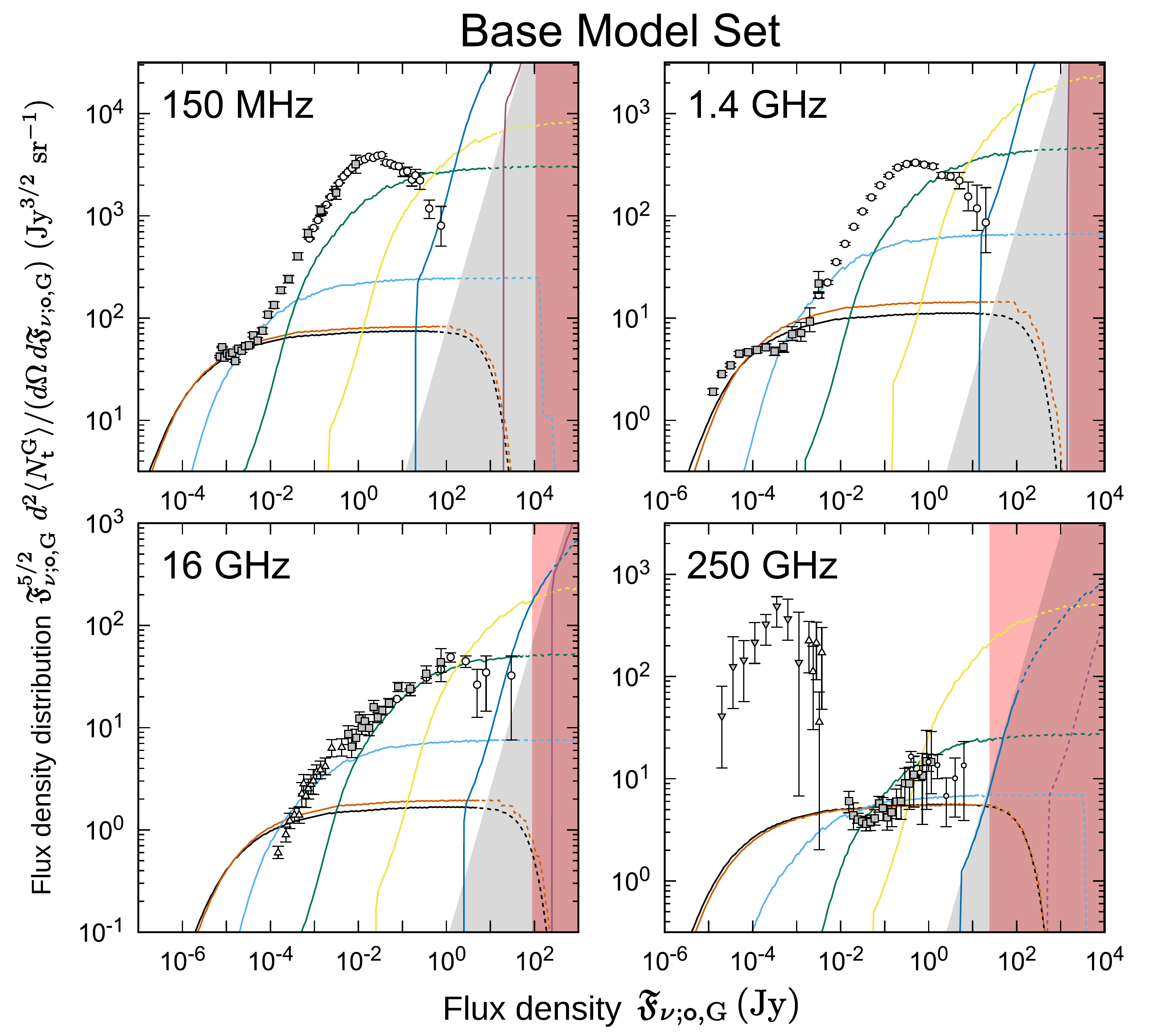}}
\figcaption{Euclidean-normalized flux distributions for the base model set.  
On upper left are distributions for 150 MHz, with source count data from GLEAM (circles) and LOFAR (squares). On upper right, the distribution for 1.4 GHz, with data from NVSS (circles) and DEEP2 (squares). On lower left, flux distributions for 16 GHz are compared with data from AT20G (circles), 9C (squares), and 10C (triangles). Finally, on lower right are the distributions for 250 GHz, compared to counts from \emph{Planck} (circles), SPT-SZ (squares), MAMBO (upward-pointing triangles), and ALMA (downward-pointing triangles).
The predicted flux distributions for the survey most relevant for constraining models (GLEAM, NVSS, AT20G, SPT) are shown as solid lines for fluxes where at least one source is expected on the sky, and dashed at higher fluxes. The distributions are shown at the maximum allowed $\bzAbundnu$ for different $\bLisoBAR$ (along the bottom edge of blue/red bounded regions in Figure~\ref{fig:ConstraintsBase}): $10^{20}$ (black), $10^{30}$ (dark orange), $10^{32}$ (light blue), $10^{34}$ (dark green), $10^{36}$ (yellow), $10^{38}$ (dark blue), and $10^{40}\ \Watt$ (pink-purple). Along this progression, broadcasts go from common but dim to rare but bright.
The pink shading fills fluxes above the brightest source in the radio sky, while the grey shading roughly indicates values for which no sources are expected on the sky ($d\Mean{\gNTime}/d\ln\,\qgFluxENuObs \la 1$). \label{fig:SourceCountsBase}}
\end{figure*}

Like individualist constraints, the source count constraints in the base model set disallow models where broadcasts are bright and numerous (Figure~\ref{fig:ConstraintsBase}). For a wide span of broadcast luminosities, up to around $10^{32}\ \Watt$, the limits all fall along lines of constant $\bLisoBAR \bzAbundnu$, which reflects how they blend into a continuum with $\Mean{\bgNObs} \gg 1$. In this range, the differential constraints are more powerful, setting limits of
\begin{equation}
\label{eqn:ConstraintsBase_Common}
\bLisoBAR \bzAbundnu \la \begin{cases}
			 5.0 \times 10^{20}\ \Watt\,\GHz^{-1}\,\mathrm{star}^{-1}                   & (150\ \MHz)\\
                         1.4 \times 10^{20}\ \Watt\,\GHz^{-1}\,\mathrm{star}^{-1} & (1.4\ \GHz)\\
                         4.0 \times 10^{19}\ \Watt\,\GHz^{-1}\,\mathrm{star}^{-1} & (16\ \GHz)\\
                         8.9 \times 10^{19}\ \Watt\,\GHz^{-1}\,\mathrm{star}^{-1} & (250\ \GHz)
                         \end{cases} .
\end{equation}
Figure~\ref{fig:SourceCountsBase} shows examples of the source count distributions at these limits for different luminosities. The low luminosity limit corresponds to the black lines. We see that at 150 MHz and 1.4 GHz, the constraints are set by a plateau of microjansky to millijansky radio sources. This plateau corresponds to the population of star-forming galaxies along the far-infrared correlation (FRC), glowing in natural synchrotron radiation associated with cosmic rays \citep{Condon92,Yun01}. This becomes clear when the low-luminosity constraints are written in terms of the galaxy's mean radio luminosity,
\begin{multline}
\FreqVar \Mean{\bgAggLnuisoObs} \la \left(\frac{\hgMAggTime}{10^{10}\ \Msun}\right) \\
                   \cdot \begin{cases}
												 9.8 \times 10^{3}\ \Lsun  & (150\ \MHz)\\
                         2.5 \times 10^{4}\ \Lsun  & (1.4\ \GHz)\\
                         8.3 \times 10^{4}\ \Lsun  & (16\ \GHz)\\
                         2.9 \times 10^{6}\ \Lsun  & (250\ \GHz)
                         \end{cases} .
\end{multline} 
At 1.4 GHz, this corresponds to $\sim 10^{-5.5}$ of the bolometric luminosity. The obvious interpretation of this constraint is that if every galaxy had such bright and numerous broadcast populations, the artificial radio emission is on a level with the synchrotron emission of star-forming galaxies according to the FRC (Paper II). The integral (brightest source) constraint also applies at still higher broadcast abundances (blue shading in Figure~\ref{fig:ConstraintsBase}), and at very high abundances, is the primary limit.

The limits on $\bzAbundnu$ itself become more severe with frequency. This is because there is more room in the spectrum at high frequency, basically. Observational bandwidths can be wider:  one cannot meaningfully conduct a survey with 10 GHz bandwidth at 150 MHz, but this is easily possible at 250 GHz. However, in terms of broadcasts per \emph{log frequency}, the limits become weaker (compare between the top and bottom panels of Figure~\ref{fig:ConstraintsBase}). At 16 GHz, the source counts that are most constraining for models with low $\bLisoBAR$ are at the faint flux end. The 250 GHz constraints bear special discussion. This region of the spectrum, at around 1 mm in wavelength, is unexplored in SETI. In local star-forming galaxies, it is roughly the end of the radio region, just before the immense far-infrared peak of dust emission rises out of the spectrum. But at redshift $\ga 1$, the Rayleigh-Jeans tail of the thermal emission is shifted to these frequencies, resulting in an enormous bump in the faint end of the flux distribution. Now the limits are set by the bright end of the distribution, dominated by flat spectrum AGNs (Figure~\ref{fig:SourceCountsBase}).

As the luminosity increases to $\ga 10^{32}\ \Watt$, $\Mean{\bgNObs} \ga 1$ and the constraints deviate from equation~\ref{eqn:ConstraintsBase_Common}. Now the bright end of the radio source count distribution -- essentially the prevalence of nearby radio AGNs -- is what sets limits on the broadcast population. When broadcasts are this luminous, even a single one outshines the natural synchrotron emission associated with star formation. Obviously, this is not observed in most galaxies. At most, we can have a rare population of radio-bright galaxies, which can describe either natural or artificial radio galaxies. The predicted radio source counts for broadcast-dominated galaxies deviates from the continuum limit (non-black curves in Figure~\ref{fig:SourceCountsBase}), and the curves in Figure~\ref{fig:ConstraintsBase} bend as the source count function traces the shape of the observed distribution at different flux levels.

When $\bLisoBAR \ga 10^{38}\ \Watt$, any broadcast anywhere within $\yRedshift < 4$ appears like a $\ga 100\ \Jy$ source in NVSS. The brightest source constraint becomes the sole limit on abundance, though a powerful one. Thus, the predicted source counts are technically consistent with the number of jansky, millijansky, and microjansky radio sources, because every broadcasting galaxy is a lot brighter than that.\footnote{In the same way that a single red supergiant at the distance of Saturn would be consistent with the observed number of magnitude 20 optical sources in the sky, while obviously being an absurd scenario.} Finally, there cannot be any line broadcasts in the observable Universe in the covered frequency ranges with $\bLisoBAR \ga 10^{40}\ \Watt$, which sets a final abundance limit.\footnote{One could legitimately question why we even need constraints at such luminosities, and whether any transmitter could sustain that amount of power. However, $\bLiso$ refers to \emph{isotropic} luminosity. Additionally, the broadcast may be the result from a large number of transmitters operating at a common frequency. A Dyson swarm that captures the light from a red supergiant and beams it into a square arcsecond as radio waves could achieve $\bLiso$ levels of this order.}

The relative strengths and weaknesses of the two approaches are in line with Paper II's discussion. As with individual galaxies (Paper II), the source count limits happen to remain our best constraints on extremely rare but extraordinarily bright broadcasts ($\ga 10^{38}\ \Watt \approx 10^{12}\ \Lsun$).

\section{Results for variant model sets}
\label{sec:Variants}

\subsection{Model set A: When metasocieties are rare}
\subsubsection{Results without evolution}
Model set A expands on the base model set by assuming that only some galaxies are inhabited, as might be expected in the expansive metasociety scenario. The effect of metasocietal abundance, $\zgAbundEMPTY$, is simply to set the fraction of galaxies of a given stellar mass that are inhabited. As such, increasing $\zgAbundEMPTY$ increases the number of galaxies at a given flux level without changing their brightness. When it is high, it becomes indistinguishable from the base model, because $\zgNTime$ is at most $1$ (Figure~\ref{fig:SourceCountsParams}, middle-left panel). 

If metasocieties are rare, the basic picture still holds but with major adjustments. Now only a small fraction of galaxies have any broadcasts. One effect a small $\zgAbundEMPTY$ has is that targeted individualist constraints cease to apply. The Milky Way has $\sim 10^{11.5}$ stars, so Galactic constraints are of negligible use in constraining any model where $\zgAbundEMPTY \la 10^{-12}$. Even constraints on the somewhat larger M31 \citep{Gray17} or the galaxies with reported limits in \citet{Garrett23} will fail to be informative for $\zgAbundEMPTY \la 10^{-14}$, as depicted in the figure. Indeed, even the largest galaxies only have a few trillion stars, so observations of single galaxies considered in isolation can say nothing about cases when $\zgAbundEMPTY \ll 10^{-12}$. Instead, we need to consider constraints from whole populations of galaxies. 

\begin{figure*}
\centerline{\includegraphics[width=18cm]{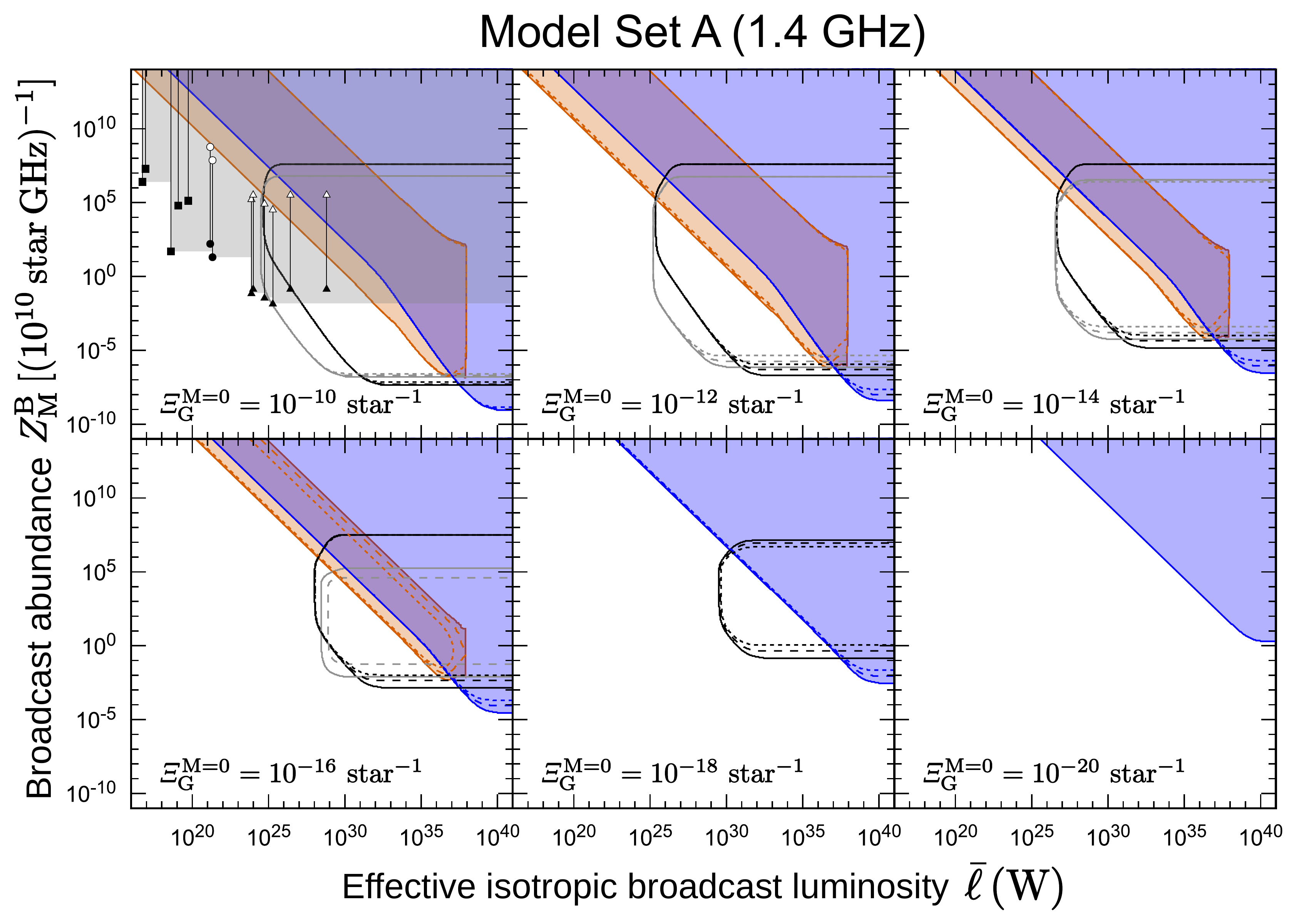}}
\figcaption{Constraints on luminosity and broadcast frequency abundance compared for different metasocietal abundances in model set A at 1.4 GHz. Line colors and shading are the same as in Figure~\ref{fig:ConstraintsBase}. Solid lines are for no evolution in $\zgAbundEMPTY$, long-dashed lines for $\zgAbundEMPTY$ linearly proportional to cosmic age, and short-dashed lines are for $\zgAbundEMPTY$ quadratically proportional to cosmic age.\label{fig:ConstraintsA}}
\end{figure*}

The field constraints from individualist surveys do better than targeted searches, because they necessarily include many galaxies, whether by design or commensally. Nonetheless, they too are limited because they only include a finite number of galaxies. In some cases, the field itself is limited to a small fraction of the sky, as for \citet{Price20} and \citet{Tremblay20}. We thus must rely on all-sky surveys to set meaningful field limits on $\zgAbundEMPTY \sim 10^{-20}$ scenarios. An additional effect is that these field constraints get weaker for small $\zgAbundEMPTY$, in the sense that higher $\bzAbundnu$ are allowed at a given luminosity. This is just because broadcasts are rarer if their host metasocieties are rarer; eventually, even if the field does include many inhabited galaxies, too few of them happen to be broadcasting in the frequency range where we are observing during the survey.

As far as the collective source count constraints are concerned, $\zgAbundEMPTY$ simply increases or decreases their normalization without changing the distribution shape or shifting it to higher of lower fluxes (Figure~\ref{fig:SourceCountsParams}). This is similar behavior as $\bzAbundnu$ in the rare broadcast limit: both set the fraction of galaxies with broadcasts. But while $\bzAbundnu$ directly controls the rate of broadcasts, with small $\zgAbundEMPTY$, metasocieties may be rare even while broadcast are common in them. A few thousand inhabited galaxies in the Universe are always outnumbered by the millions and millions of microjansky radio sources in the sky. Thus, if the collective luminosity of their broadcasts is small, scenarios with rare metasocieties cannot be excluded -- for all we know, there really is a population of artificial radio galaxies buried in the microjansky radio population.

\begin{figure*}
\centerline{\includegraphics[width=15cm]{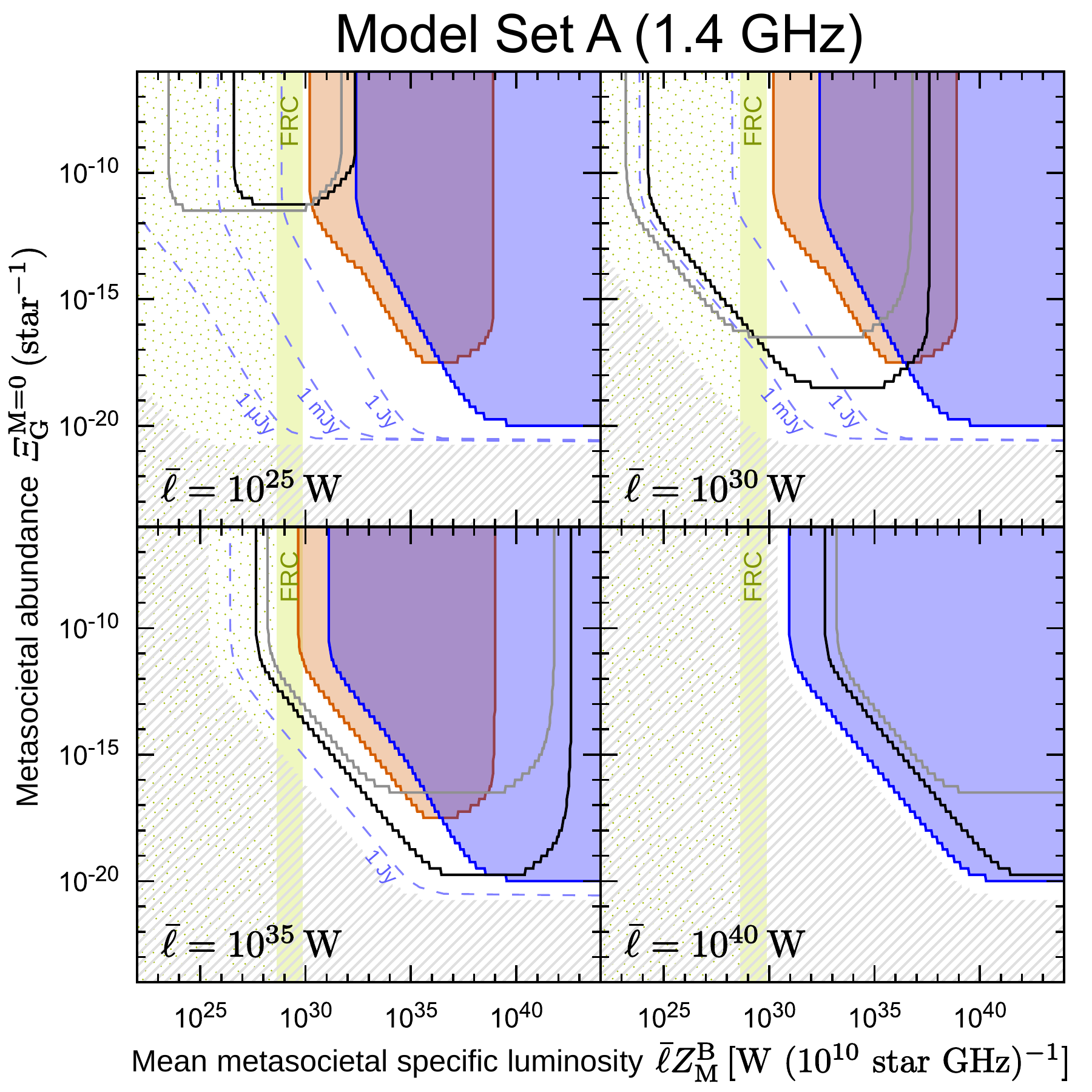}}
\figcaption{Constraints on broadcast and metasocietal abundance compared for different broadcast EIRPs in model set A at 1.4 GHz. Line styles and shading are the same as in Figure~\ref{fig:ConstraintsBase}. The yellow band indicates the natural level of synchrotron emission in star-forming galaxies, according to the FRC. To the left of the band (yellow-dotted region), the collective emission is buried in this glow. Collective constraints cannot rule out models in this region, but individualist constraints still work. In the grey hatched regions on the bottom, no broadcasts are expected anywhere in the Universe in the NVSS survey. The blue dashed lines delineate models where the brightest artificial radio source in NVSS is (from right to left) $1\ \Jy$, $1\ \mJy$, and $1\ \uJy$.\label{fig:AbundConstraintsA}}
\end{figure*}

Another view of the constraints is shown in Figure~\ref{fig:AbundConstraintsA}, which now plots them in terms of broadcast and metasocietal abundances. For all constraints, larger $\zgAbundEMPTY$ never results in the constraints becoming less powerful; the observables saturate for $\zgAbundEMPTY \sim 10^{-10}$. At low luminosities, it is the product of broadcast EIRP and abundance that is constrained, as plotted. When $\bLisoBAR = 10^{25}\ \Watt$, the field constraints are limited, because of the restricted distance to which they are detectable. The source count constraints, on the other hand, run into the FRC (yellow band): they cannot rule out a large fraction of galaxies emitting these low levels of radio emission, because a large fraction of galaxies \emph{do} in fact have that level of radio emission. With luminosities of $\bLisoBAR = 10^{30}\ \Watt$, individualist field constraints are much more effective in comparison because these can be detected out to cosmological distances while collective constraints still have a background of natural radio galaxies. At the highest luminosities, individualist field constraints are limited by frequency range or field covered. However, we also see that, for large luminosities, in most of the parameter space not already ruled out, we expect no broadcasts within the observable Universe in a survey like NVSS with 40 MHz of bandwidth. Indeed, any broadcast that bright would appear as the most luminous radio source on the sky whatever frequency it was observed in, brighter than Cygnus A. Thus, although in a formal sense, it is possible that there are ETIs out there with broadcasts that luminous, functionally, we can say that none exist within the horizon in the NVSS observed frequency range, and that such technosignatures \emph{might as well} not exist \citep[cf.,][]{Wesson90}.

Figure~\ref{fig:AbundConstraintsA} also plots the apparent flux density of the brightest artificial radio galaxy, as would be measured in NVSS. There is a corner of parameter space where transmitters are faint, broadcasts are abundant, and metasocieties are rare, in which these would have brightnesses in the millijansky to jansky range, as implied by Figure~\ref{fig:AbundConstraintsA}. Hypothetically, this means that some sources in the NVSS catalog, for example, could actually represent artificial radio galaxies. 

\subsubsection{Effects of metasocietal evolution}
\label{sec:ModelsAEvol}

ETIs take time to evolve. The formation rate of planets in the early Universe was low, with fewer habitable locations at high redshift although some probably were formed even at $\yRedshift \sim 4$ \citep{Behroozi15,Zackrisson16}. If evolution on Earth is anything to go by, it may take billions of years further for intelligence to develop. In addition, violent phenomena with the potential to inhibit complex life like gamma-ray bursts and supernovae were more prevalent billions of years ago, and galaxies may have been more impassable to interstellar travel \citep{Lineweaver04,Gowanlock16,Lacki21-Traversability}. 

A full accounting of these effects would require assumptions about at least the star-formation histories of all galaxies and a delay-time distribution for ETIs relative to their host stars. That is beyond the scope of this work, so I consider two simpler models, where $\zgAbundEMPTY \propto t^q$ evolves linearly ($q = 1$) or quadratically ($q = 2$) with time since the Big Bang.\footnote{Linear or quadratic evolution is predicted in the \citet{Carter83} model, if there are only one or two ``hard steps'' in the evolution of technological societies \citep[see also][]{Hanson21}.}

Cosmic evolution has little bearing on whether a model is allowed or not, as seen in Figure~\ref{fig:ConstraintsA}. This is because the effect is negligible for small redshifts. It is the galaxies at low redshift that determine the entire bright end of the broadcast flux distribution. If there are many broadcasts in the survey footprint, all of the same luminosity, the nearest ones are both more likely to be individually detected and to be part of a galactic population that masquerades as a bright radio galaxy. The source count distribution, for example, is only affected in the faint-end tail (Figure~\ref{fig:SourceCountsParams}; middle panel). Cosmic evolution thus only matters if broadcasts, or their host societies and metasocieties, are intrinsically rare. Then, only a few broadcasts are present in the observable Universe, mostly in the large comoving volumes at high redshift, and suppressing the number at high redshifts causes these to vanish. Thus, when $\zgAbundEMPTY \ga 10^{-16}$, the only significant differences in the constraints are their boundaries at high luminosities and low abundances. For very low $\zgAbundEMPTY$, the constraints on $\bLisoBAR$ and $\bzAbundnu$ combinations also weaken considerably or vanish, as seen in Figure~\ref{fig:ConstraintsA}.

\subsection{Model set B: Ubiquitous galactic clubs and rare societies}
The assumption of broadcasts being emitted entirely independent of each other only is viable if the transmitters can come from anywhere at any time within a metasociety. Actual metasocieties must have a finite number of transmitting entities, referred to here as societies. This finiteness imposes a clumping in the distribution of broadcasts. Model set B is a treatment of ubiquitous galactic clubs that illustrates this clumping effect. As with model set A, there are two fundamental quantities that regulate how common broadcasts, but with societies playing the role of metasocieties instead: frequency abundance of lines per society $\baAbundnuTotal$ and the stellar abundance of societies in a metasociety/galaxy $\azAbund$. Their product is the frequency abundance of lines per star in a metasociety, $\bzAbundnu$, but it does not behave the same way as it does in model set A.

\subsubsection{The discreteness of societies and the number of broadcasts}
Although one could draw an analogy between metasocieties in model set A and societies in model set B, there is one big difference: a galaxy can have more than one society. In model set A, the number of expansive metasocieties in a galaxy is modeled as a Bernoulli random variable: either there is one or there is not. Beyond some point all galaxies are fully inhabited and there is no further dependence on metasocietal abundance. In contrast, the number of societies in a model set B galaxy is considered to be a Poissonian variable that can increase without limit. For fixed $\Mean{\baNObs}$, increasing $\azAbund$ also increases $\Mean{\bzNObs}$ (and $\Mean{\bgNObs}$) proportionally.

\begin{figure}
\centerline{\includegraphics[width=8.5cm]{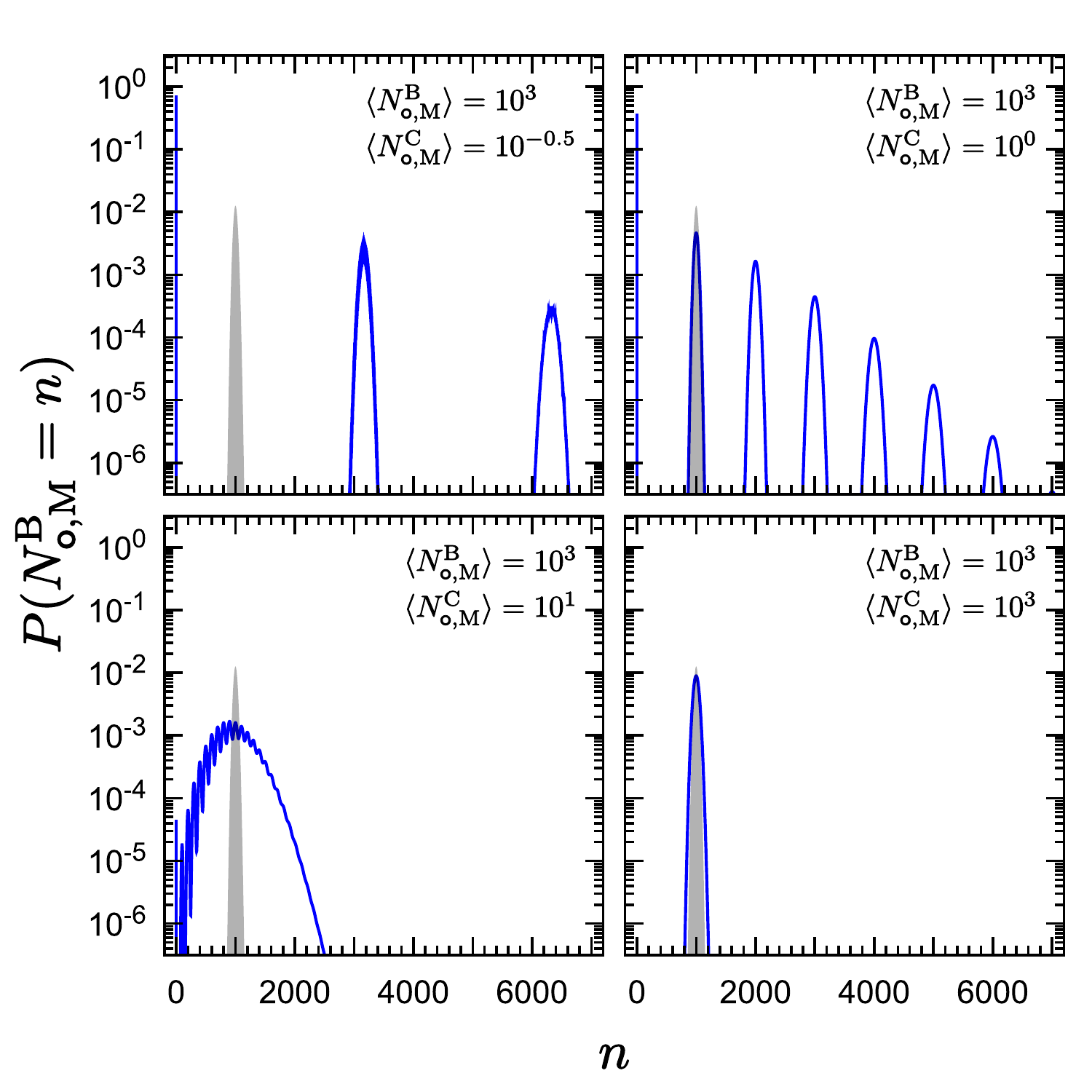}}
\figcaption{Number distributions in Compound Poisson distributions \editOne{(blue lines)}, as used in model set B. In each case plotted in the figure, there are on average 1,000 broadcasts per metasociety (galaxy), but these are grouped into societies. In the diffuse approximation, the clumping into societies is ignored and the distribution is simply Poissonnian (grey shading). As the mean number of societies per metasociety decreases, the broadcasts are increasingly clumped, resulting in a wider scatter in the number of broadcasts\editOne{.}\label{fig:NDistB}}
\end{figure}

The number of intercepted broadcasts from the system is a compound Poisson variable. Figure~\ref{fig:NDistB} illustrates the effects of the clumping when keeping $\Mean{\bzNObs}$ constant at a high value on the number distribution of broadcasts. When the number of societies is very high, the distribution converges to the Poissonian limit, nearly Gaussian (dark blue). Decreasing the number of societies -- dividing the broadcasts into fewer groups -- widens the broadcast number distribution (bright blue), due to the Poissonian fluctuations in $\azNObs$ itself. Clumping even more, the number distribution resolves into a comb (bright red). Each ``tooth'' represents the probability associated with a fixed value of $\azNObs$. But the discreteness of societies imposes a minimum spacing between the teeth, given by $\Mean{\bzNObs | \azNObs = n} = n \Mean{\baNObs} = n \Mean{\bzNObs} / \Mean{\azNObs}$. When $\Mean{\azNObs} \ll 1$ (dark red), the first nonzero peak occurs for $\Mean{\bzNObs | \azNObs = 1} \gg \Mean{\bzNObs}$. There are essentially no systems for which $\Mean{\bzNObs}$ broadcasts are intercepted -- from most, zero are intercepted, while a few are seemingly overpopulated. 

\begin{figure}
\centerline{\includegraphics[width=8.5cm]{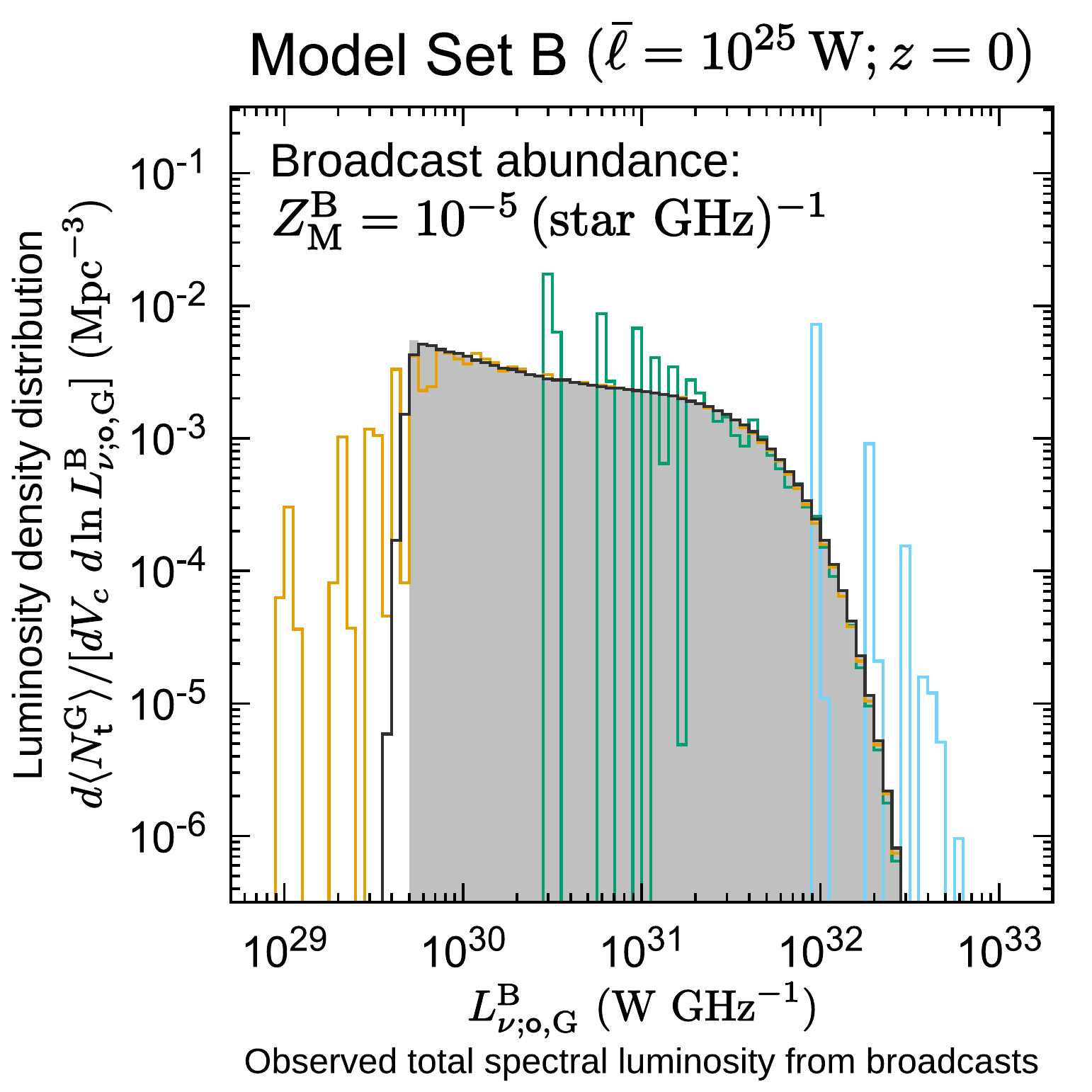}}
\figcaption{Example of a luminosity distributions in model set B, demonstrating the effects clumping of broadcasts into rare societies. The peaks in the distribution correspond to small integer values of $\azNObs$. Luminosity distributions are shown for $\bzAbundnu = 10^{-5}\ \mathrm{star}^{-1}\ \GHz^{-1}$, with $\azAbund$ values of $10^{-12}$ (\editOne{sky }blue), $10^{-10.5}$ (\editOne{dark green}), $10^{-9}\ \mathrm{star}^{-1}$ (\editOne{orange}), and $10^{-7.5}\ \mathrm{star}^{-1}$ (dark grey). Because the number of societies does not depend on which survey is observing, the peaks have the same apparent luminosity in both NVSS and DEEP-2. As in Figure~\ref{fig:LumDistA}, the grey shading is the expected luminosity distribution if all discreteness effects are ignored.  \label{fig:LumDistB}}
\end{figure}

As a result, when societies are very rare but broadcasts are common, the broadcasts are clumped into a few bright galaxies with most appearing empty. The clumping in broadcast number translates to peaks in the luminosity distribution (Figure~\ref{fig:LumDistB}), each corresponding to an integer number of societies. Unlike the sampling of broadcasts, the sampling of societies generally does not depend on the survey when observing distant galaxies -- societies presumably last longer than observing programs and all lie within the survey footprint.

The bottom row of Figure~\ref{fig:SourceCountsParams} illustrates the different effects of $\azAbund$, $\baAbundnuTotal$, and $\bzAbundnu$ on the source count distribution, each centered on the values where the transition between rare and common objects occurs.

\subsubsection{Results for model set B}

\begin{figure*}
\centerline{\includegraphics[width=18cm]{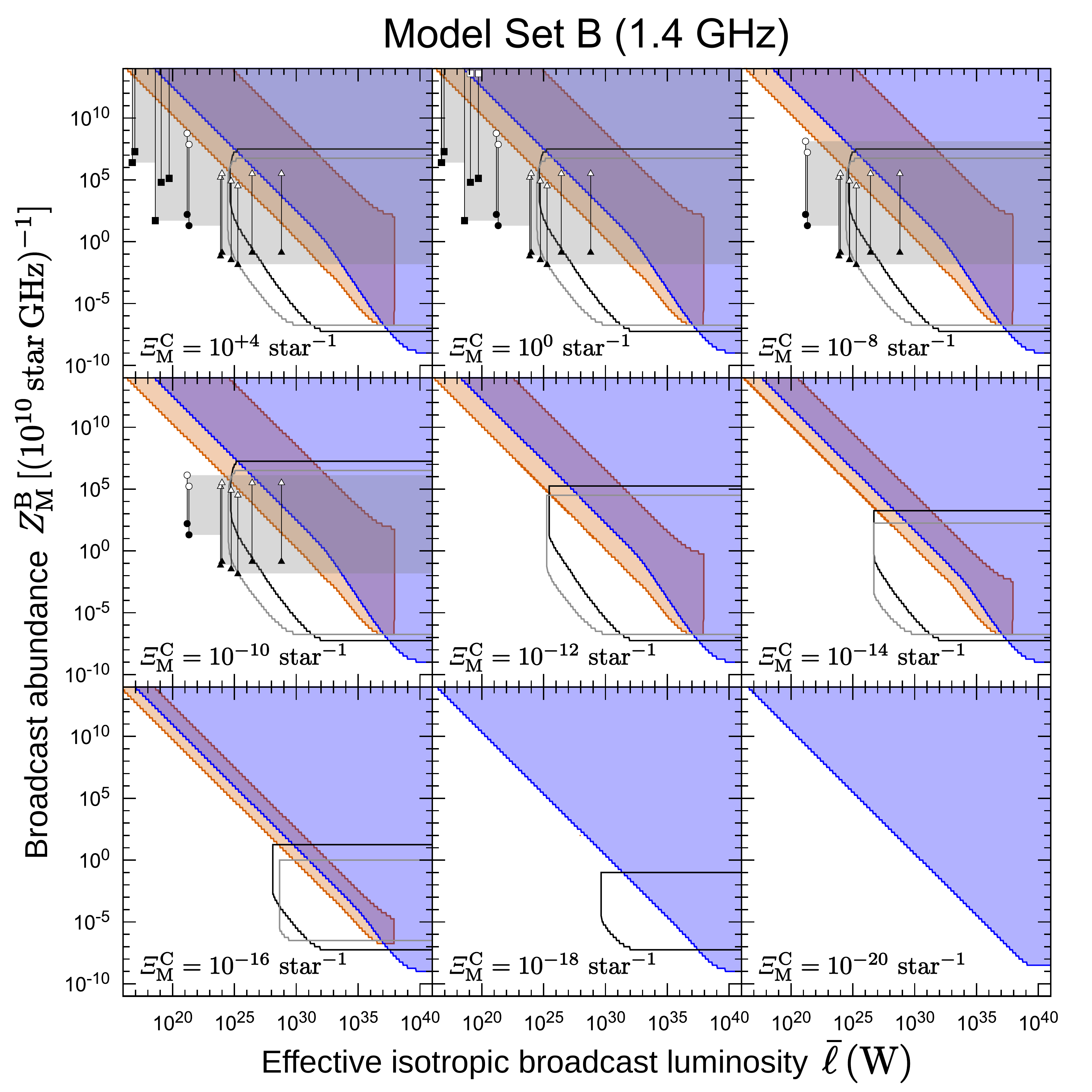}}
\figcaption{Constraints on luminosity and broadcast frequency abundance compared for different metasocietal abundances in model set B at 1.4 GHz. Line colors and shading are the same as in Figure~\ref{fig:ConstraintsBase}.\label{fig:ConstraintsB}}
\end{figure*}

The resulting constraints are shown in Figure~\ref{fig:ConstraintsB}. In the case when societies are common ($\azAbund \ga 10^{-8}$), the limits are largely the same as in the base model set, or model set A with high $\zgAbundEMPTY$. This is to be expected: when this holds, all galaxies are populated to roughly the same degree for a given stellar mass, so all that matters are the properties of the broadcasts themselves.

\begin{figure*}
\centerline{\includegraphics[width=15cm]{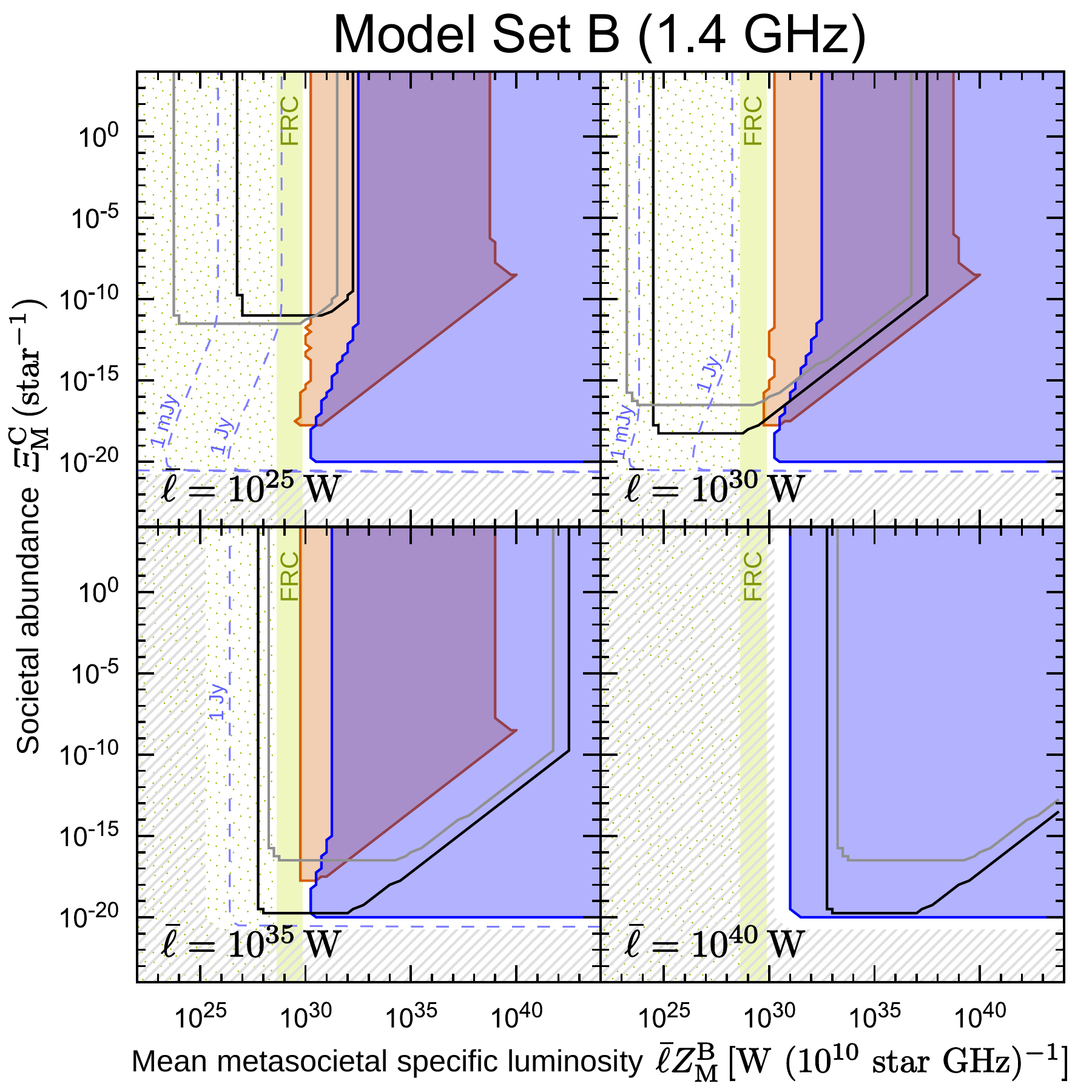}}
\figcaption{Constraints on broadcast and societal abundance compared for different broadcast luminosities in model set B at 1.4 GHz. Line styles and shading are the same as in Figure~\ref{fig:AbundConstraintsA}. \label{fig:AbundConstraintsB}}
\end{figure*}

When societies are rare, however, the results may seem baffling at first glance. Even as societies become very rare, the minimum $\bzAbundnu$ probed remain the same, implying the limits remain as powerful as they were before! A careful examination reveals that the \emph{maximum} $\bzAbundnu$ probed shrinks, narrowing the constrained parameter space. This seemingly paradoxical behavior is the result of the galactic club assumption that $\zgNTime = 1$, and the resulting fact that $\bzAbundnu$ is the mean abundance over \emph{all} galaxies rather than \emph{inhabited} galaxies. The bottom boundary of the field limits, for example, then indicates the abundance of broadcasts in the Universe for which at least one detection is made. Roughly speaking, if a survey covers $10^{18}$ stars with 1 GHz of bandwidth, then a detection is made if there are $\ga 10^{-18}$ broadcast per star-GHz, regardless of whether one in a thousand or one in a million galaxies has a society that might be a host. Similar logic applies to the source count constraints. However, as $\azAbund$ decreases, it takes more broadcasts per society to reach a given level of $\bzAbundnu$; the broadcasts are clumped into fewer and fewer societies. This makes individual societies more prone to confusion and brighter in collective emission (red lines in bottom right panel in Figure~\ref{fig:SourceCountsParams}), the former invalidating individualist constraints, and the latter eventually leading to absurdly bright galaxies that are beyond the reach of the differential source count constraint. Yet this also makes the integral (brightest source) constraint more powerful when societies are rarer. This same behavior is also seen in Figure~\ref{fig:AbundConstraintsB}, with the constrained regions bending the ``wrong'' way compared to Figure~\ref{fig:AbundConstraintsA}. There are cases where increasing $\azAbund$ at fixed $\bzAbundnu$ \emph{eases} the constraints; the model can pass between allowed and disfavored multiple times as $\azAbund$ increases.

To get around this, we could instead plot with respect to abundance of broadcasts per \emph{society}, $\baAbundnuTotal$. Then, when societies are rare, it acts equivalently to $\bzAbundnu$ in model set A, and the constraints behave just like they do in that model set. The catch is that the correspondence would then break down if societies are common, with the limited regions continuing to shift as $\azAbund \la 10^{-10}$.

\subsection{Model set C: Expansive metasocieties and rare societies}

\begin{figure*}
\centerline{\includegraphics[width=18cm]{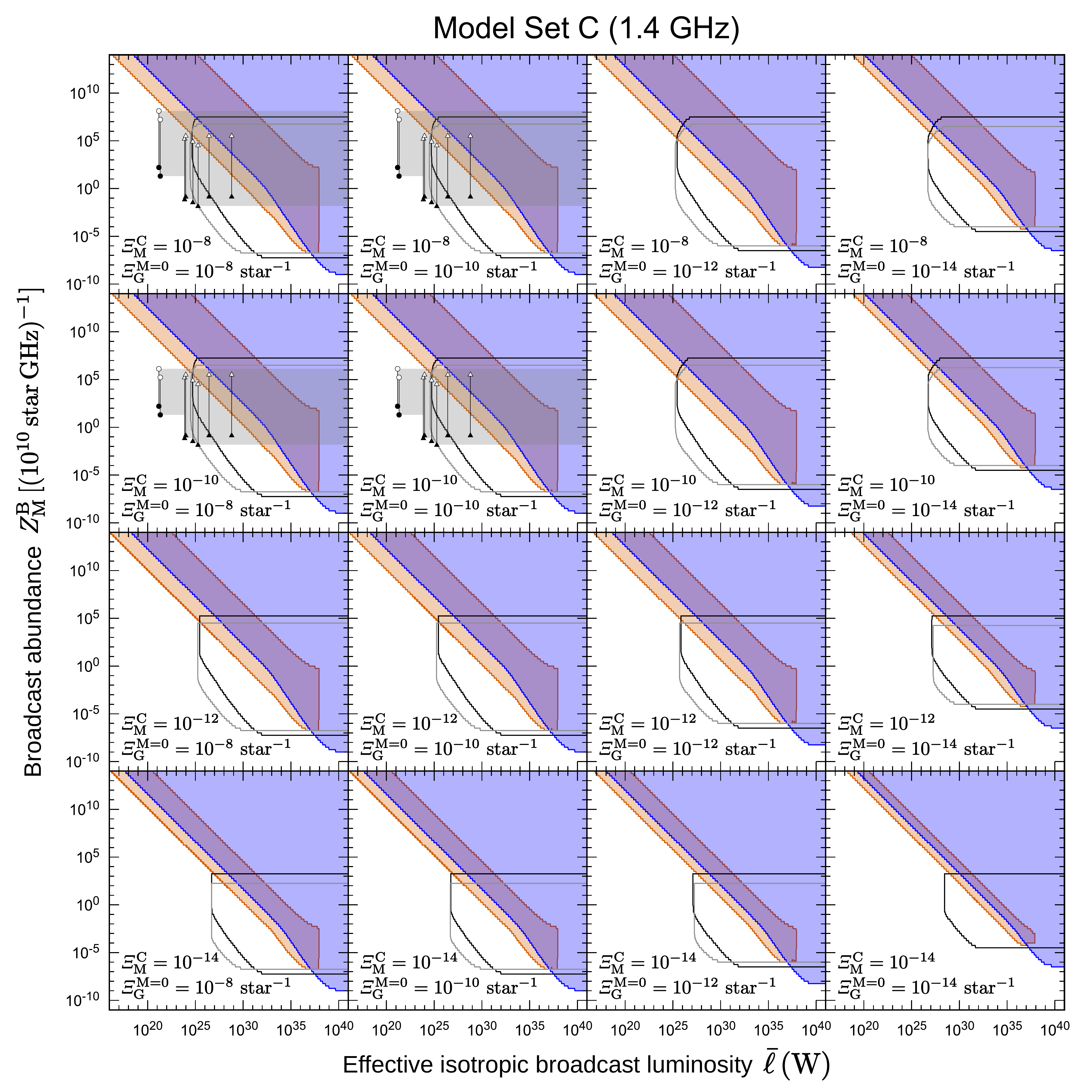}}
\figcaption{Constraints on luminosity and broadcast frequency abundance compared for different metasocietal abundances in model set C at 1.4 GHz. Line colors and shading are the same as in Figure~\ref{fig:ConstraintsBase}. \label{fig:ConstraintsC}}
\end{figure*}

What if we include the discreteness of both metasocieties and societies? Model set C is like model set A, but also accounts for the clumping of broadcasts into societies; equivalently, it is like model set B, but where only a fraction of galaxies host metasocieties. The results are shown in Figure~\ref{fig:ConstraintsC} for a limited range in $\zgAbundEMPTY$ and $\azAbund$. There are no real surprises in the constraints; both parameters behave as they do in model sets A and B.

\subsection{Model set D: A universal power-law luminosity distribution for broadcasts}
\label{sec:ModelsD}

What if ETIs use a variety of broadcasting strategies, some using very bright broadcasts and others dim, for example, or some beaming into tight cones while others radiate into the entire sky? Perhaps the effective isotropic emission from broadcasts span many orders-of-magnitude. The power-law distribution is a popular model for broadcast luminosity. A specific example is the Pareto distribution. The effective isotropic luminosity has a probability density
\begin{equation}
\PDF{\bLiso}(\bLactualCore) = \begin{cases}
                     \displaystyle \frac{\alpha - 1}{\bLisoBAR} \left(\frac{\bLactualCore}{\bLisoBAR}\right)^{-\alpha} & (\bLactualCore \ge \bLisoBAR) \\
										 0 & (\bLactualCore < \bLisoBAR)
										 \end{cases}
\end{equation}
with $\alpha > 1$, allowing for a finite density of transmitters of arbitrarily bright emission. This section considers a universal Pareto luminosity distribution, one where the distribution is identical in all galactic systems, considering $\alpha = 3/2$, $2$, and $5/2$.

\begin{figure*}
\centerline{\includegraphics[width=18cm]{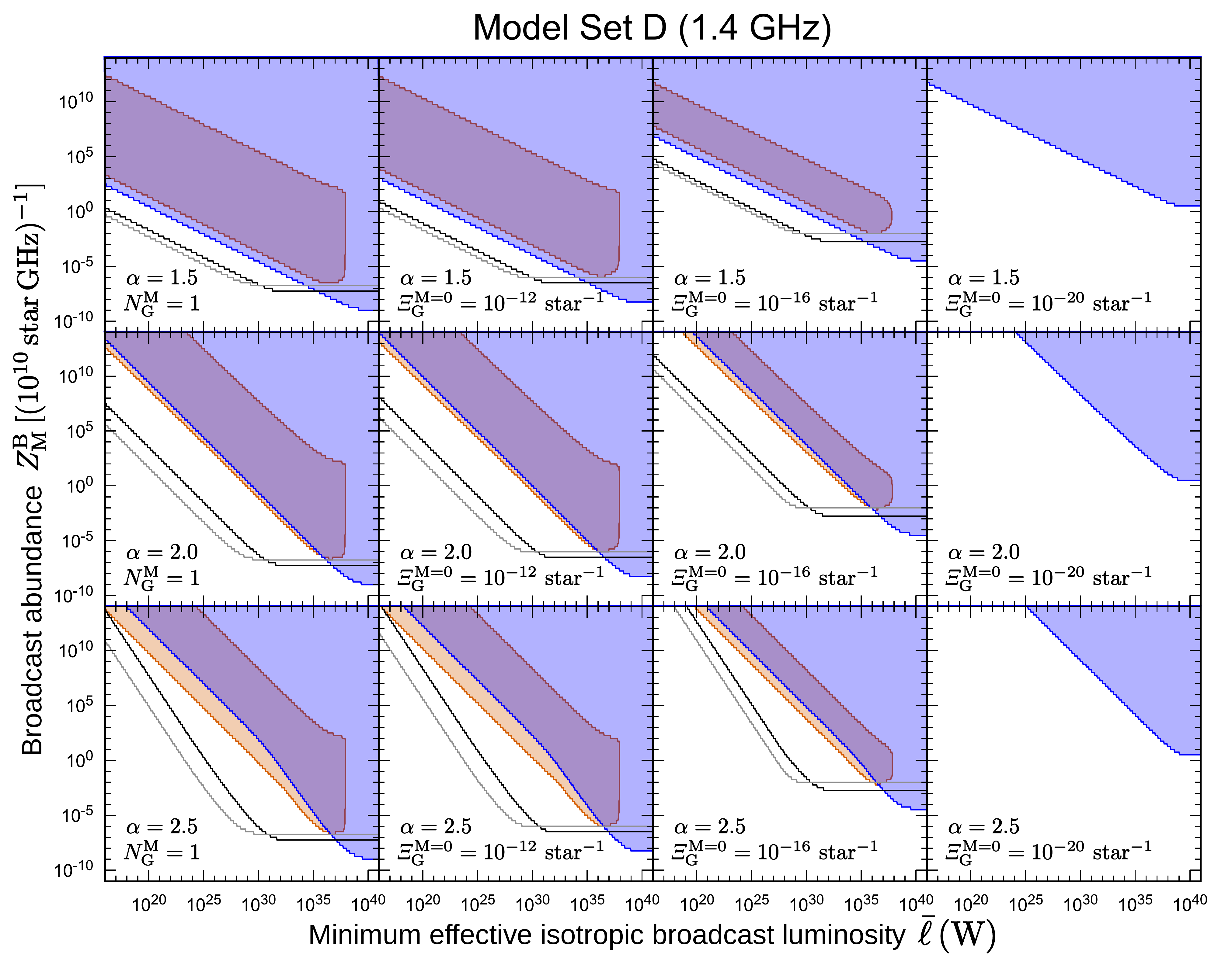}}
\figcaption{Constraints on luminosity and broadcast frequency abundance compared for different metasocietal abundances in model set D at 1.4 GHz. Line colors and shading are the same as in Figure~\ref{fig:ConstraintsBase}. \label{fig:ConstraintsD}}
\end{figure*}

It has been noted that the limits on broadcasts in the literature roughly fall along a power law over a vast range of luminosities \citep{Enriquez17,Tremblay20,Garrett23}. At first glance, this could erroneously imply that the limits themselves are actually made under the \emph{assumption} of a power-law luminosity distribution. This is not the case. Power-law luminosity distributions extending over many orders of magnitude provide a very long lever arm. Given a shallow power law with $\alpha < 2$ and no upper luminosity limit, the lack of bright transmitters are an extremely powerful constraint on faint transmitters; the reverse is true for steep power laws. If we believe the luminosity distribution extends from a gigawatt to infinity with $\alpha \sim 3/2$, for example, then we already know there are no transmitters among the thousands of nearby stars observed in SETI because we should already have seen their bright counterparts in all-sky surveys like META, and there would be kilojansky artificial radio galaxies lighting up the radio sky.

The lever-arm effect is illustrated in Figure~\ref{fig:ConstraintsD}. The limits on abundance for broadcasts when $\alpha = 3/2$ remain quite powerful even when the minimum luminosity is small. Even though most broadcasts are faint in these cases, enough bright ones are present to be detected across the Universe unless the normalization is very low. In detail, we see that the individualist field constraints are more powerful than the collective source count bounds except when broadcasts are extraordinarily rare. Normally, field surveys are limited by a distance: if all broadcasts are faint, then they just cannot go very deep. However, the continuing power law ensures that there are broadcasts bright enough to be seen clear across the Universe, and enough will land in the survey footprint and bandwidth even if the population as a whole is somewhat rare. Another advantage field constraints have is that confusion does not apply to them: even if there are many broadcasts blended together in an observation, we expect one to outshine all the rest combine, sticking out in visibility. The integral brightest-source criterion is also more powerful than the differential constraint because bright galaxies are overproduced (red line in Figure~\ref{fig:SourceCountsParams}, right-middle panel). 

Even for $\alpha$ values of $2$ and $5/2$, the lever-arm effect continues to ensure the continued prominence of individualist field constraints. If $\bLisoBAR$ was the luminosity of every broadcast like in the other model sets, then the vast distances to even the nearest galaxies would impose a floor on how small it could be with detections still possible. But that is not the case; with a wide span in luminosities, much smaller \emph{minimum} luminosities can be explored. In fact, the field survey constraints shown in Figure~\ref{fig:ConstraintsD} follow the luminosity distribution shape until they saturate. The source counts bounds on the other hand show a stable shape for $\alpha \ga 2$, for which the faint broadcasts dominate the aggregate luminosity of a metasociety. Additionally, Euclidean-normalized source constraints already have an effective power-law shape of $\alpha = 5/2$.

So what are the role of targeted individualist results if a power-law distribution holds? Because of the lever-arm effect, there is a nontrivial dependence between the luminosity limit, the minimum luminosity of broadcasts, and total broadcast abundance. What they do provide is a measure of the number of broadcasts around a particular luminosity (and all luminosities above it). A limit on terawatt transmitters around nearby stars applies to terawatt broadcasts \emph{regardless} of the shape of the luminosity distribution -- whether it is degenerate, Gaussian, power-law, bimodal, or anything else. Thus, while the limits in the literature to this point generally do not assume a power-law luminosity distribution, they do probe different pieces of them.

\section{Discussion}
\label{sec:Discussion}
\subsection{How prevalent are Kardashev 3 ETIs?}
The \citet{Kardashev64} scale is a convenient, if sometimes contentious, measure of the power of an ETI. Roughly speaking, a Kardashev I ETI harnesses the energy incident on a planet, Kardashev II uses the power of a whole sun, and Kardashev III the power of an entire galaxy. Variations of it have been used to gauge the energy usage of an entire (meta)society, but here its use is more in line with \citet{Kardashev64}, quantifying the radio emission of an ETI, albeit the apparent luminosity after beaming. I adopt the common conversion to luminosity, $K = (1/10) \log_{10} (\AggLisoCore / 10^6\ \Watt)$ \citep{Wright20}. 

\begin{deluxetable*}{ccccccccc}
\tabletypesize{\footnotesize}
\tablecolumns{9}
\tablewidth{0pt}
\tablecaption{Field and source count limits on individual bright beacons in the base model set\label{table:IndividualKardashev}}
\tablehead{\colhead{Kardashev rating} & \colhead{$\log_{10} [\bLisoBAR\ (\Watt)]$} &  \multicolumn{7}{c}{Maximum allowed $\log_{10} [\bNuMid \bzAbundnu (\mathrm{star}^{-1})]$} \\ & & \multicolumn{3}{c}{Field individualist} & \multicolumn{4}{c}{Source count} \\ & & \colhead{H93} & \colhead{P20} & \colhead{T20} & \colhead{150 MHz} & \colhead{1.4 GHz} & \colhead{16 GHz} & \colhead{250 GHz}}
\startdata
$1$             & $16$   & \nodata & \nodata & \nodata & $+3.9$  & $+4.3$  & $+4.9$  & $+6.4$  \\
$1 \frac{1}{2}$ & $21$   & \nodata & \nodata & \nodata & $-1.1$  & $-0.7$  & $-0.1$  & $+1.4$  \\
$2$             & $26$   & $-9.9$  & $-12.9$ & \nodata & $-6.1$  & $-5.7$  & $-5.1$  & $-3.6$  \\
$2 \frac{1}{4}$ & $28.5$ & $-13.6$ & $-15.9$ & $-13.7$ & $-8.6$  & $-8.2$  & $-7.6$  & $-6.1$  \\
$2 \frac{1}{2}$ & $31$   & $-16.6$ & $-16.6$ & $-16.8$ & $-11.2$ & $-10.7$ & $-10.2$ & $-8.9$  \\
$2 \frac{3}{4}$ & $33.5$ & $-17.2$ & $-16.6$ & $-17.6$ & $-13.6$ & $-13.3$ & $-12.8$ & $-12.2$ \\
$3$             & $36$   & $-17.2$ & $-16.6$ & $-17.6$ & $-16.7$ & $-16.2$ & $-15.7$ & $-14.4$ \\
$3 \frac{1}{4}$ & $38.5$ & $-17.2$ & $-16.6$ & $-17.6$ & $-18.4$ & $-18.2$ & $-18.3$ & $-17.7$\\
$3 \frac{1}{2}$ & $41$   & $-17.2$ & $-16.6$ & $-17.6$ & $-19.8$ & $-18.9$ & $-18.6$ & $-19.8$ \\
$3 \frac{3}{4}$ & $43.5$ & $-17.2$ & $-16.6$ & $-17.6$ & $-19.8$ & $-18.9$ & $-18.6$ & $-19.9$ \\
\enddata
\end{deluxetable*}

Given that a moderately large galaxy has of order $10^{11}$ stars, the field constraints in this paper yield maximum abundances of about one ultranarrowband beacon rated Kardashev 2.5 or higher per million galaxies or so (Table~\ref{table:IndividualKardashev}). The source count constraints are weaker at low luminosities, but they imply similar conclusions for Kardashev 3 beacons. Interestingly, these constraints only vary by two orders of magnitude from 150 MHz to 250 GHz: at most one in two thousand galaxies with $10^{11}$ stars hosts a Kardashev 3 beacon at 250 GHz. Thus, despite how little SETI has been done at high frequencies, existing source counts are sufficient to conclude that individual Kardashev 3 beacons are rare across the radio spectrum.

\begin{deluxetable*}{ccccc}
\tabletypesize{\footnotesize}
\tablecolumns{5}
\tablewidth{0pt}
\tablecaption{Limits on ETIs with bright collective emission from faint broadcasts\label{table:CollectiveKardashev}}
\tablehead{\colhead{Kardashev Rating} & \multicolumn{2}{c}{Metasocieties: model set A (no evolution; 1.4 GHz)} & \multicolumn{2}{c}{Societies: model set B} \\ & \colhead{$\log_{10} [\bLisoBAR \bzAbundnu (\Lsun\ \mathrm{star}^{-1}\ \GHz^{-1})]$} & \colhead{$\log_{10} [\zgAbundEMPTY (\mathrm{star}^{-1})]$} & \colhead{$\log_{10} [\bLisoBAR \baAbundnuTotal (\Lsun\ \GHz^{-1})]$} & \colhead{$\log_{10} [\azAbund (\mathrm{star}^{-1})]$}}
\startdata
$1$             & $-21.6$ & \nodata           & $-10.6$ & $+4 \frac{1}{4}$\\
$1 \frac{1}{2}$ & $-16.6$ & \nodata           & $-5.6$  & $-0 \frac{3}{4}$\\
$2$             & $-11.6$ & \nodata           & $-0.6$  & $-5 \frac{3}{4}$\\
$2 \frac{1}{4}$ & $-9.1$  & \nodata           & $+1.9$  & $-8 \frac{1}{4}$\\
$2 \frac{1}{2}$ & $-6.6$  & \nodata           & $+4.4$  & $-10 \frac{3}{4}$\\
$2 \frac{3}{4}$ & $-4.1$  & $-13 \frac{3}{4}$ & $+6.9$  & $-13 \frac{1}{2}$\\
$3$             & $-1.6$  & $-17$             & $+9.4$  & $-16 \frac{1}{4}$\\
$3 \frac{1}{4}$ & $+0.9$  & $-18 \frac{3}{4}$ & $+11.9$ & $-18 \frac{1}{4}$\\
$3 \frac{1}{2}$ & $+3.4$  & $-20$             & $+14.4$ & $-20$\\
$3 \frac{3}{4}$ & $+5.9$  & $-20$             & $+16.9$ & $-20$
\enddata
\tablecomments{The Kardashev ratings are converted into a luminosity by assuming a bandpass of $1\ \GHz$, and in model set A, a host galaxy containing $10^{11}$ stars.}
\end{deluxetable*}

But we can also use the Kardashev scale to measure the aggregate emission of an ETI, not just an individual broadcast. We now consider host metasocieties or societies with a great many faint broadcasts, but where the host societies or metasocieties are rare. Individualist constraints from external galaxies are of no use, but the collective source count constraints still apply. In model set A, the collective spectral luminosity of an inhabited metasociety with $10^{11}$ stars is $10^{11} \zMean{\bLiso} \bzAbundnu$; in model set B, the collective spectral luminosity of an inhabited society is $\zMean{\bLiso} \baAbundnuTotal$. If we take a generic bandwidth of order the observing frequency, say, $1\ \GHz$, we can convert a Kardashev rating into a broadcast luminosity-abundance product. From Table~\ref{table:CollectiveKardashev}, the limits are weak or nonexistent below Kardashev $\sim 2.5$ -- of course, the majority of galaxies have synchrotron emission at this level. Yet we find that radio-broadcasting Kardashev 3 ETIs, whether metasocieties or societies, can be present around less than one in $\sim 10^{17}$ stars. We thus again find that they are rare: at most one in a million moderately large galaxies.

The paucity of Kardashev level 3 radio emission from ETIs is thus a robust conclusion, implicit in already extant source counts even without any dedicated SETI surveys.

\subsection{Are collective constraints really this weak?}
The collective constraints, as presented in this paper, make the conservative assumption that any or all of the observed radio source population can be artificial. This is extremely unlikely. Cygnus A, the anchor for the brightest radio source constraint, is presumably a natural AGN. All of the brightest radio sources like Centaurus A and M87 have identified counterparts, with clear natural explanations and no signs of ultranarrowband emission. Thus, we arguably should set the radio flux for the brightest artificial radio galaxy much lower -- of the unidentified radio sources in \citet{Maselli16}, the greatest flux is $13.68\ \Jy$ for 3C 409, a hundred times lower than Cygnus A. Doing this would expand the constrained regions of parameter space, approaching the $1\ \Jy$ curves in Figures~\ref{fig:AbundConstraintsA} and~\ref{fig:AbundConstraintsB}.

Even when we cannot resolve it into individual narrowband transmitters, the tell for an artificial radio galaxy may be its morphology. A population of radio transmitters would most likely either trace stellar mass, or be dominated by one or a few sites. In the latter case, these could be off-center, simply because that is where the host society is. Radio-loud AGNs, however, have jets and lobes, as well as emission from their cores, and thus do not match the expected morphology.\footnote{Admittedly, it is conceivable that galactic clubs would build beacons specifically in their galactic centers, perhaps to harness an already extant AGN.} Star-forming galaxies do have radio emission that trace their disks, more like what we might expect from aggregate artificial radio emission. Their distinguishing characteristic may be that natural radio emission follows a close correlation with infrared emission, while there is no reason artificial broadcasts would. Indeed, any Kardashev 3 radio ETI necessarily falls well off the FRC simply because star-forming galaxies just are not that radio-bright. As far as I am aware, no nearby galaxy has the properties expected of an artificial radio galaxy: extended emission tracing stellar mass far exceeding the predicted natural luminosity.

\section{Conclusions}
\label{sec:Conclusions}
This paper discussed constraints on radio broadcasts from entire populations of inhabited galaxies. As has been emphasized in Papers I and II, broadcast populations may differ dramatically from one galaxy to another, with some being empty of ETIs and others blazing with technosignatures. While most SETI surveys hitherto have focused on individual targets -- nearby stars or sometimes nearby galaxies -- extant data can also be applied to set limits on a large number of galaxies. These fall in two basic approaches, according to the division in Paper II. Individualist surveys look for single broadcasts standing out against a background of noise and background, and account for most SETI work. As multiple works over the past few years have noted, there is a large number of background galaxies in targeted observations, and we can set limits on bright but rare objects from the vast number of stars in them. Additionally, there have been a few large-field and all-sky SETI surveys like META. These together form the field constraints in this paper. I demonstrated how one can derive statistical constraints using the framework of Paper I and II, in a way that accounts for the possibility that galaxies may have differing metasocieties, with divergent broadcast populations within them.

The other philosophy is to seek out signs of the aggregate glow of all broadcasts in a galactic metasociety. If there were galaxies with many bright broadcasts, they would essentially be artificial radio galaxies. We know how many radio galaxies at different luminosities exist, as summarized in source counts. Thus the collective bound of Paper II can be generalized to populations of galaxies by predicting source counts in a given model and comparing to observations. If there were many galaxies with many bright broadcasts, the population of radio sources would either appear as an unexpected plateau in the source count function (limited by the differential constraint), or as a population of sources that are brighter than any source on the sky (integral constraint). These limits can be applied at any wavelength for which source counts, or limits on source counts, exist, even without any dedicated SETI work.

My results demonstrated the different tradeoffs between broadcast luminosity, the abundance of ETIs, and the number of broadcasts each ETI hosts. It is well-known that null results are consistent with many faint broadcasts or a few bright broadcasts, but if the host metasocieties or societies are themselves rare, it is possible that each individually has many bright broadcasts. The results show that Kardashev III radio galaxies, in the classical sense of broadcast populations with the luminosity of a galaxy, are extremely rare, less than one in a million galaxies at 1.4 GHz. When they apply, individualist field constraints are usually more powerful than the collective source count limits. However, they are limited by limited sky footprint and bandwidth, having a minimum luminosity sensitivity at extragalactic distances, and confusion when broadcasts are too common, although the last two are not relevant when the broadcasts have a power-law luminosity distribution. Source count constraints, at least as presented here, ultimately are limited by the natural radio emission of galaxies, which would bury any artificial radio emission in the wideband observations of sky surveys. They are thus compatible with $\la 25,000\ \Lsun (\hgMAggTime/10^{10}\ \Msun)$ of artificial radio emission near 1.4 GHz in every galaxy. They can however be used at any frequency with source counts. I used extant millimeter surveys to derive the first SETI constraints around 250 GHz. 

There are of course many possible extensions and variations to study, some explored here. Different luminosity distributions, abundance distributions, population evolution, and variances between systems can all be studied. Perhaps most promising is extending the constraints across the electromagnetic spectrum, even to other messengers. Although collective bounds are expected to be weak in optical and infrared (Paper II), reasonably powerful limits might be set in X-rays and gamma-rays.

\begin{acknowledgments}
{I thank the Breakthrough Listen program for their support. Funding for \emph{Breakthrough Listen} research is sponsored by the Breakthrough Prize Foundation (\url{https://breakthroughprize.org/}).  In addition, I acknowledge the use of NASA's Astrophysics Data System and arXiv for this research. \editOne{I thank the referee for their suggestions on the manuscript.}}
\end{acknowledgments}

\appendix
\section{Calculation of the collective luminosity distributions}
\subsection{Numerical representation of probability distributions}
Numerically, probability distributions are represented as a discrete array of $\iFont{N}_R$ floating point numbers. Although this can be interpreted as the probability density function, it can also be viewed as the probability mass function (PMF) of a discrete distribution that approximates a continuous one,
\begin{equation}
\label{eqn:DiscretePDF}
\PDF{X}(x) \approx \sum_{j \in \mathbb{Z}} \ProbabilityCore_{X;j} \fDirac(x - j \Delta x)
\end{equation}
for a random variable $X$, where $\Delta x$ is the spacing in the array. This simplifies integration, in that probability in bins can be added without transformations. 

I use characteristic functions to numerically evaluate the probability distributions. I use the GNU Scientific Library (GSL) radix-2 Fast Fourier Transform (FFT) functions to calculate characteristic functions of real-valued probability distributions. These require $\iFont{N}_R = 2^{l}$, where $l$ is a positive integer. The resulting complex-valued characteristic functions have $\iFont{N}_C = 2^{l - 1} + 1$ values, with $\CF{X}(s = 0) = 1$. They too can be interpreted as a modulated comb:
\begin{equation}
\CF{X}(s) \approx \sum_{j \in \mathbb{Z}} \CFCore_{X;j} \fDirac(s - j \Delta s) .
\end{equation}
According to the GSL FFT conventions on $\pi$ and sign, $\Delta s = 2 \pi / (\iFont{N}_R \Delta x)$. An important consideration is the scale of the spacing $\Delta x$ and $\Delta s$. If $\Delta x$ is too narrow, the array is too short to include the bulk of the probability mass; if it is too wide, there is insufficient resolution to correctly represent the values of $X$.

Discrete Fourier Transforms (FFT) on finite arrays strictly speaking represent functions that are periodic in $x$ and $s$. This is because a finite array in one domain represents a continuous functions multiplied by a Dirac comb; in the conjugate domain, the (inverse) Fourier transform is the conjugate continuous function convolved with a Dirac comb. The discrete arrays represent only a single period in the $x$ domain ($0 \le x < \iFont{N}_R \Delta x)$, and only a half period of the characteristic function ($0 \le s \le \iFont{N}_C \Delta s$). Of course, the actual probability distributions and characteristic functions are not periodic. Nonetheless, the representations are accurate as long as they have sufficient resolution for all the important features of the distribution and sufficient length to cover almost all of the probability mass.

I use array lengths of $\iFont{N}_R = 2^{15}$ for the base model set and model sets A, B, and C; for the Pareto-distributed luminosity functions of model set D, I use $\iFont{N}_R = 2^{19}$.

\subsection{The stability of Poisson variables and divide-and-conquer algorithms}
\label{sec:DivideAndConquer}
The sum of two independent compound Poisson random variables is itself compound Poisson. If $N \sim \Poisson(\Mean{N_k})$, then $S_k = \sum_{i=1}^{N_k} X_i$ with all $X_i$ independent is equivalent to the sum of two independent random variables $S_{k-1} = \sum_{i=1}^{N_{k-1}} X_i$ with $N_{k-1} \sim \Poisson(\Mean{N_k}/2)$. In terms of characteristic functions, we then have the simple relation
\begin{equation}
\label{eqn:SquaringCF}
\CF{S_k}(s) = \CF{S_{k-1}}(s)^2 .
\end{equation}

This property is the underlying principle of the divide-and-conquer algorithms I use for calculating compound Poisson distributions, allowing their distribution to be computed even when $\Mean{\jNCore}$ is extremely large. I start by factoring $\Mean{\jNCore}$ into $b \cdot 2^a$, where $1/2 \le b < 1$. I then calculate $S_0 = \sum_{i = 1}^{\jNCore_0} X_i$ for $\jNCore_0 \sim \Poisson(b)$, using the methods below. Then it is a matter of iterating: taking the FFT to get $\CF{S_k}$, squaring it, and taking the inverse FFT to find the PMF of $\CF{S_{k+1}}$, until I reach $S_a$, which is the desired compound Poisson variable.

One cavaet is that as the values spanned by $S_k$ grows, the bulk of the probability eventually exceeds the limits of the array. Thus, the resolution must be upscaled at appropriate intervals by a factor that depends on the properties of the variables being summed.  Probability mass for each bin in the old PMF is added to the bin in the new PMF with the $x_{k, j^{\prime}}$ closest to $x_{k-1,j}$.

\subsection{The degenerate luminosity distribution}
When all narrowband broadcasts have zero drift rate and equal luminosity, the total flux from a galaxy is equal to the number of intercepted broadcasts times the flux of a single broadcast. The flux distribution calculation for a galaxy at fixed distance reduces to the calculation of the number distribution of broadcasts, $\ProbabilityCore(\bzNObs = n)$.

\subsubsection{Expansive interstellar metasocieties under the diffuse approximation: simple Poisson distributions}
According to the expansive interstellar metasociety scenario, almost all galaxies are either uninhabited or fully settled (the Milky Way possibly being in a transitory intermediate phase with humanity confined on one planet). Furthermore, fully settled galaxies likely have many communicative societies, suggesting minor Poisson fluctuations in $\azNTime$. Under the diffuse approximation, the number of broadcasts observed from a settled galaxy, $\bzNObs$ has a simple Poisson distribution. 

When $\Mean{\bzNObs} \le \iFont{N}_R / 2 = 16,384$, I simply use a spacing of $\Delta n = 1$ and directly calculate the PMF: $\ProbabilityCore(\bzNObs = n) = \exp(-\Mean{\bzNObs}) \Mean{\bzNObs}^n / n!$. The standard deviation of the Poisson distribution remains $\le 128$, so there is always negligible probability distribution at $n \ge \iFont{N}_R$.

For larger $\Mean{\bzNObs}$, I upscale the resolution to $\Delta n = \Mean{\bzNObs} / (2 \iFont{N}_R)$, so that the peak of the Poisson distribution is at the center of the array. When $\Mean{\bzNObs} \le \iFont{N}_R^2$, I use the CDF to calculate the probability distribution: $\ProbabilityCore_{j} = \CDF{\bzNObs \le j \Delta n} - \CDF{\bzNObs \le (j - 1) \Delta n}$, with $\ProbabilityCore_{0} = \exp(-\Mean{\bzNObs})$. This cumulative distribution can be calculated using the normalized incomplete gamma function $Q(a, x)$: $\CDF{\bzNObs \le n} = Q(\lfloor n + 1 \rfloor, \Mean{\bzNObs})$. 

Finally, when $\Mean{\bzNObs} \ge \iFont{N}_R^2$, the distribution is unresolved. I then model $\bzNObs$ itself with a degenerate distribution, setting $\ProbabilityCore_{\iFont{N}_R/2} = 1$ and all other values to $0$, with $\Delta n = \Mean{\bzNObs} / (2 \iFont{N}_R)$.

\subsubsection{Broadcasts in discrete societies: compound Poisson distributions}

Without the diffuse approximation, the number of broadcasts per society $\baNObs$ is assumed to be Poissonian, which is then compounded by a Poissonian random number of societies, $\azNObs$. The result can be a complicated probability mass function with structure on multiple scales. 

There are several regimes where the calculation of $\bzNObs$ can be simplified:
\begin{itemize}
\item When $\Mean{\baNObs} \ge \iFont{N}_R^2$, the distribution of $\baNObs$ cannot be resolved by the array. For all intents and purposes, it is a degenerate variable with $\baNObs = \Mean{\baNObs}$. Then, $\bzNObs$ can be calculated from the simple Poisson distribution of $\azNObs$, simply by multiplying the resultant values of $\azNObs$ by $\Mean{\baNObs}$.
\item Likewise, when $\Mean{\azNObs} \ge \iFont{N}_R^2$, the variation in $\azNObs$ cannot be resolved, and numerically it might as well be the constant value $\Mean{\azNObs}$. I then calculate $\bzNObs$ as a simple Poisson variable with mean $\Mean{\azNObs}\Mean{\baNObs} = \Mean{\bzNObs}$.
\item When $\Mean{\baNObs} \ll 1$, virtually all societies have either zero or one broadcasts in the observational window. The very rare cases where $\baNObs \ge 2$ are vastly outnumbered by the cases when $\baNObs = 1$, so they have negligible effect on $\bzNObs$. Thus, $\baNObs$ is essentially a Bernoulli variable, and summing a Poissonian number of them yields a simple Poisson variable. I again calculate $\bzNObs \sim \Poisson(\Mean{\bzNObs})$. 
\end{itemize}

Outside of these cases, I proceed with the divide-and-conquer algorithm. I start by calculating the distribution for $\baNObs$, and then convert it to $S_0 = \sum_{i = 0}^{N_0} \baNObs$ with $N_0 \sim \Poisson(b)$, adopting the $b$ value for $\azNObs$ from Appendix~\ref{sec:DivideAndConquer}. For this, I use the characteristic function $\CF{S_0}(s) = \exp(b (\CF{\baNObs} - 1))$. Recursive upscaling and squaring of the characteristic function ultimately gives me the characteristic function and probability distribution of $\bzNObs$ itself.

The resolution of $\Delta n$ must be chosen with care, because the distribution of ${\bzNObs}$ can have significant probability mass far from its mean. The worst case is when $\Mean{\azNObs} \approx 1$: a good many metasocieties will have $\azNObs = 2$, $3$, and the next few integers. It is important to use a resolution such that even when this is true, virtually all of the probability is covered by the array. At all points, I ensure that $\Delta n_k \ge b \cdot 2^k \Mean{\baNObs} / (16.5 \iFont{N}_R)$. When $k = 0$, this means that the distribution covers all peaks associated with $\azNObs$ from $1$ to $16$, above which there is no significant probability mass. The extra half guarantees that even if there is any probability mass at the $\azNObs = 16$ or $17$ peaks, it is never wrapped around to $\azNObs \sim 0$. Without these measures, wrapping effects can lead to artificial ramps in the probability distribution of $\bzNObs$.

\subsection{The Pareto luminosity distribution in the diffuse approximation}

There is no analytical expression for the PDF of a sum of Pareto random variables, much less a Poisson sum of them, necessitating the use of the characteristic function approach. The Pareto luminosity distribution results in a very wide range of broadcast luminosities, spanning many orders of magnitude. 

\subsubsection{Setting the resolution of the distribution} 
The typical aggregate luminosity can grow faster than the number of broadcasts: if there is no upper cutoff, the mean luminosity of an individual broadcast is infinite when $\alpha \le 2$. The typical values of $\bzAggLisoObs$ increase faster than $\Mean{\bzNObs}$ if this is true. The reason is that as the number of sampled objects increases, broadcasts further and further on the luminosity distribution tail are sampled, and these tend to dominate the collective luminosity. This is important when doing numerical calculations because it describes the range and resolution needed for the distribution. Estimates of where the bulk of the probability mass occurs are given by a regularized mean defined in Paper I, integrating over values of $\bLiso$ likely to occur in a sample of $\bzNObs$ broadcasts:
\begin{equation}
\zMeanObs{\bLisoREGzObs} = \int_{\bLisoLO}^{\bLisoHI} \bLactualCore \PDF{\bLiso}(\bLactualCore) d\bLactualCore,
\end{equation}
where $\CDF{\bLiso}(\bLisoLO) = \CCDF{\bLiso}(\bLisoHI) \approx 1/\Mean{\bzNObs}$ when $\bzNObs \gg 1$. For a power-law luminosity distribution in this limit, $\bLisoLO \approx \bLisoBAR$ and $\bLisoHI \approx \bLisoBAR [\Mean{\bzNObs}]^{1/(\alpha - 1)}$. Then,
\begin{equation}
\zMeanObs{\bLisoREGzObs} \sim \begin{cases}
                                  \bLisoBAR \Mean{\bzNObs}^{(2-\alpha)/(\alpha - 1)} & 1 < \alpha < 2\\
                                  \bLisoBAR \ln \Mean{\bzNObs}                       & \alpha = 2\\
                                  \bLisoBAR                                          & \alpha > 2
                                  \end{cases} .
\end{equation}
The median value of $\bzAggLisoObs$ is $\sim \Mean{\bzNObs} \zMeanObs{\bLisoREGzObs}$. In particular, when $\alpha = 3/2$, typical values of $\bzAggLisoObs$ are usually around $\sim \Mean{\bzNObs}^2 \bLisoBAR$.

\subsubsection{Calculation of the distribution}
I use a bottom-up divide-and-conquer algorithm to calculate the PDF and the characteristic function, in terms of $X = \bLiso / \bLisoBAR$. I start with a PDF with $\Delta x_0 = 1 / 64$. Analogous to the degenerate luminosity distribution case, I start by calculating the characteristic function for $\bLiso / \bLisoBAR$. This is done simply by calculating the probability mass for a truncated power-law distribution in each $x$ bin to get a PMF and then taking a FFT. The PMF is normalized to have probability $1$.

As demonstrated in Appendix~\ref{sec:DivideAndConquer}, doubling the value of $\bzNObs$ can lead to the effective mean increasing by a much larger factor. Using the regularized mean as a guide, I set $\Delta x_k = \max(2^k, 2^{\lceil k/(\alpha - 1) \rceil}) \Delta x_0$.

If we had an infinitely long array, the PDF of $S_k$ would have a tail extending to infinitely large $x$:
\begin{equation}
\label{eqn:ParetoTailPDF}
\lim_{x \to \infty} \PDF{S_{k-1}} (x) = b \cdot 2^{k-1} \cdot \PDF{X}(x),
\end{equation}
with a PMF that can be easily calculated from this. This tail represents cases when the aggregate luminosity is dominated by a single extreme outlier broadcast with extraordinary brightness. Because of the finite length of the array, however, we only approximate $S_k$ as a truncated power law. So, after squaring the characteristic function of $S_{k-1}$ as described in Appendix~\ref{sec:DivideAndConquer}, I convert back to a PMF, upscale, then add a truncated power law tail that follows equation~\ref{eqn:ParetoTailPDF}. This tail is truncated at $i = \iFont{N}_R/2$, so that for the next doubling step, there are no wrapping effects leading to unphysical ramps in the probability distribution. The PMF of $S_{k}$, with this added tail, is then renormalized to have probability $1$. When I reach $k = a$, the calculated probability distribution of $\Mean{\bzNObs}$ itself,  I extend the tail in the PMF all the way to the end of the array.

Now, the resulting PMF is still for a truncated Pareto $\bLiso$ distribution. During the source count calculation, I extend the distribution by appending a power-law tail so that all values of the flux are considered (equation~\ref{eqn:ParetoTailPDF}).

\bibliographystyle{aasjournalv7}
\bibliography{ETIPopulations}

@ARTICLE{Ade11-Intro,
       author = {{Planck Collaboration} and {Ade}, P.~A.~R. and {Aghanim}, N. and {Arnaud}, M. and {Ashdown}, M. and {Aumont}, J. and {Baccigalupi}, C. and {Baker}, M. and {Balbi}, A. and {Banday}, A.~J. and {Barreiro}, R.~B. and {Bartlett}, J.~G. and {Battaner}, E. and {Benabed}, K. and {Bennett}, K. and {Beno{\^\i}t}, A. and {Bernard}, J. -P. and {Bersanelli}, M. and {Bhatia}, R. and {Bock}, J.~J. and {Bonaldi}, A. and {Bond}, J.~R. and {Borrill}, J. and {Bouchet}, F.~R. and {Bradshaw}, T. and {Bremer}, M. and {Bucher}, M. and {Burigana}, C. and {Butler}, R.~C. and {Cabella}, P. and {Cantalupo}, C.~M. and {Cappellini}, B. and {Cardoso}, J. -F. and {Carr}, R. and {Casale}, M. and {Catalano}, A. and {Cay{\'o}n}, L. and {Challinor}, A. and {Chamballu}, A. and {Charra}, J. and {Chary}, R. -R. and {Chiang}, L. -Y. and {Chiang}, C. and {Christensen}, P.~R. and {Clements}, D.~L. and {Colombi}, S. and {Couchot}, F. and {Coulais}, A. and {Crill}, B.~P. and {Crone}, G. and {Crook}, M. and {Cuttaia}, F. and {Danese}, L. and {D'Arcangelo}, O. and {Davies}, R.~D. and {Davis}, R.~J. and {de Bernardis}, P. and {de Bruin}, J. and {de Gasperis}, G. and {de Rosa}, A. and {de Zotti}, G. and {Delabrouille}, J. and {Delouis}, J. -M. and {D{\'e}sert}, F. -X. and {Dick}, J. and {Dickinson}, C. and {Dolag}, K. and {Dole}, H. and {Donzelli}, S. and {Dor{\'e}}, O. and {D{\"o}rl}, U. and {Douspis}, M. and {Dupac}, X. and {Efstathiou}, G. and {En{\ss}lin}, T.~A. and {Eriksen}, H.~K. and {Finelli}, F. and {Foley}, S. and {Forni}, O. and {Fosalba}, P. and {Frailis}, M. and {Franceschi}, E. and {Freschi}, M. and {Gaier}, T.~C. and {Galeotta}, S. and {Gallegos}, J. and {Gandolfo}, B. and {Ganga}, K. and {Giard}, M. and {Giardino}, G. and {Gienger}, G. and {Giraud-H{\'e}raud}, Y. and {Gonz{\'a}lez}, J. and {Gonz{\'a}lez-Nuevo}, J. and {G{\'o}rski}, K.~M. and {Gratton}, S. and {Gregorio}, A. and {Gruppuso}, A. and {Guyot}, G. and {Haissinski}, J. and {Hansen}, F.~K. and {Harrison}, D. and {Helou}, G. and {Henrot-Versill{\'e}}, S. and {Hern{\'a}ndez-Monteagudo}, C. and {Herranz}, D. and {Hildebrandt}, S.~R. and {Hivon}, E. and {Hobson}, M. and {Holmes}, W.~A. and {Hornstrup}, A. and {Hovest}, W. and {Hoyland}, R.~J. and {Huffenberger}, K.~M. and {Jaffe}, A.~H. and {Jagemann}, T. and {Jones}, W.~C. and {Juillet}, J.~J. and {Juvela}, M. and {Kangaslahti}, P. and {Keih{\"a}nen}, E. and {Keskitalo}, R. and {Kisner}, T.~S. and {Kneissl}, R. and {Knox}, L. and {Krassenburg}, M. and {Kurki-Suonio}, H. and {Lagache}, G. and {L{\"a}hteenm{\"a}ki}, A. and {Lamarre}, J. -M. and {Lange}, A.~E. and {Lasenby}, A. and {Laureijs}, R.~J. and {Lawrence}, C.~R. and {Leach}, S. and {Leahy}, J.~P. and {Leonardi}, R. and {Leroy}, C. and {Lilje}, P.~B. and {Linden-V{\o}rnle}, M. and {L{\'o}pez-Caniego}, M. and {Lowe}, S. and {Lubin}, P.~M. and {Mac{\'\i}as-P{\'e}rez}, J.~F. and {Maciaszek}, T. and {MacTavish}, C.~J. and {Maffei}, B. and {Maino}, D. and {Mandolesi}, N. and {Mann}, R. and {Maris}, M. and {Mart{\'\i}nez-Gonz{\'a}lez}, E. and {Masi}, S. and {Massardi}, M. and {Matarrese}, S. and {Matthai}, F. and {Mazzotta}, P. and {McDonald}, A. and {McGehee}, P. and {Meinhold}, P.~R. and {Melchiorri}, A. and {Melin}, J. -B. and {Mendes}, L. and {Mennella}, A. and {Mevi}, C. and {Miniscalco}, R. and {Mitra}, S. and {Miville-Desch{\^e}nes}, M. -A. and {Moneti}, A. and {Montier}, L. and {Morgante}, G. and {Morisset}, N. and {Mortlock}, D. and {Munshi}, D. and {Murphy}, A. and {Naselsky}, P. and {Natoli}, P. and {Netterfield}, C.~B. and {N{\o}rgaard-Nielsen}, H.~U. and {Noviello}, F. and {Novikov}, D. and {Novikov}, I. and {O'Dwyer}, I.~J. and {Ortiz}, I. and {Osborne}, S. and {Osuna}, P. and {Oxborrow}, C.~A. and {Pajot}, F. and {Paladini}, R. and {Partridge}, B. and {Pasian}, F. and {Passvogel}, T. and {Patanchon}, G. and {Pearson}, D. and {Pearson}, T.~J. and {Perdereau}, O. and {Perotto}, L. and {Perrotta}, F. and {Piacentini}, F. and {Piat}, M. and {Pierpaoli}, E. and {Plaszczynski}, S. and {Platania}, P. and {Pointecouteau}, E. and {Polenta}, G. and {Ponthieu}, N. and {Popa}, L. and {Poutanen}, T. and {Pr{\'e}zeau}, G. and {Prunet}, S. and {Puget}, J. -L. and {Rachen}, J.~P. and {Reach}, W.~T. and {Rebolo}, R. and {Reinecke}, M. and {Reix}, J. -M. and {Renault}, C. and {Ricciardi}, S. and {Riller}, T. and {Ristorcelli}, I. and {Rocha}, G. and {Rosset}, C. and {Rowan-Robinson}, M. and {Rubi{\~n}o-Mart{\'\i}n}, J.~A. and {Rusholme}, B. and {Salerno}, E. and {Sandri}, M. and {Santos}, D. and {Savini}, G. and {Schaefer}, B.~M. and {Scott}, D. and {Seiffert}, M.~D. and {Shellard}, P. and {Simonetto}, A. and {Smoot}, G.~F. and {Sozzi}, C. and {Starck}, J. -L. and {Sternberg}, J. and {Stivoli}, F. and {Stolyarov}, V. and {Stompor}, R. and {Stringhetti}, L. and {Sudiwala}, R. and {Sunyaev}, R. and {Sygnet}, J. -F. and {Tapiador}, D. and {Tauber}, J.~A. and {Tavagnacco}, D. and {Taylor}, D. and {Terenzi}, L. and {Texier}, D. and {Toffolatti}, L. and {Tomasi}, M. and {Torre}, J. -P. and {Tristram}, M. and {Tuovinen}, J. and {T{\"u}rler}, M. and {Tuttlebee}, M. and {Umana}, G. and {Valenziano}, L. and {Valiviita}, J. and {Varis}, J. and {Vibert}, L. and {Vielva}, P. and {Villa}, F. and {Vittorio}, N. and {Wade}, L.~A. and {Wandelt}, B.~D. and {Watson}, C. and {White}, S.~D.~M. and {White}, M. and {Wilkinson}, A. and {Yvon}, D. and {Zacchei}, A. and {Zonca}, A.},
        title = "{Planck early results. I. The Planck mission}",
      journal = {\aap},
         year = 2011,
        month = dec,
       volume = {536},
          eid = {A1},
        pages = {A1},
          doi = {10.1051/0004-6361/201116464},
archivePrefix = {arXiv},
       eprint = {1101.2022},
 primaryClass = {astro-ph.IM},
       adsurl = {https://ui.adsabs.harvard.edu/abs/2011A&A...536A...1P}
}

@ARTICLE{Ade11-EarlyCSC,
       author = {{Planck Collaboration} and {Ade}, P.~A.~R. and {Aghanim}, N. and {Arnaud}, M. and {Ashdown}, M. and {Aumont}, J. and {Baccigalupi}, C. and {Balbi}, A. and {Banday}, A.~J. and {Barreiro}, R.~B. and {Bartlett}, J.~G. and {Battaner}, E. and {Benabed}, K. and {Beno{\^\i}t}, A. and {Bernard}, J. -P. and {Bersanelli}, M. and {Bhatia}, R. and {Bonaldi}, A. and {Bonavera}, L. and {Bond}, J.~R. and {Borrill}, J. and {Bouchet}, F.~R. and {Bucher}, M. and {Burigana}, C. and {Butler}, R.~C. and {Cabella}, P. and {Cantalupo}, C.~M. and {Cappellini}, B. and {Cardoso}, J. -F. and {Carvalho}, P. and {Catalano}, A. and {Cay{\'o}n}, L. and {Challinor}, A. and {Chamballu}, A. and {Chary}, R. -R. and {Chen}, X. and {Chiang}, L. -Y. and {Chiang}, C. and {Christensen}, P.~R. and {Clements}, D.~L. and {Colombi}, S. and {Couchot}, F. and {Coulais}, A. and {Crill}, B.~P. and {Cuttaia}, F. and {Danese}, L. and {Davis}, R.~J. and {de Bernardis}, P. and {de Rosa}, A. and {de Zotti}, G. and {Delabrouille}, J. and {Delouis}, J. -M. and {D{\'e}sert}, F. -X. and {Dickinson}, C. and {Diego}, J.~M. and {Dolag}, K. and {Dole}, H. and {Donzelli}, S. and {Dor{\'e}}, O. and {D{\"o}rl}, U. and {Douspis}, M. and {Dupac}, X. and {Efstathiou}, G. and {En{\ss}lin}, T.~A. and {Eriksen}, H.~K. and {Finelli}, F. and {Forni}, O. and {Fosalba}, P. and {Frailis}, M. and {Franceschi}, E. and {Galeotta}, S. and {Ganga}, K. and {Giard}, M. and {Giraud-H{\'e}raud}, Y. and {Gonz{\'a}lez-Nuevo}, J. and {G{\'o}rski}, K.~M. and {Gratton}, S. and {Gregorio}, A. and {Gruppuso}, A. and {Haissinski}, J. and {Hansen}, F.~K. and {Harrison}, D. and {Helou}, G. and {Henrot-Versill{\'e}}, S. and {Hern{\'a}ndez-Monteagudo}, C. and {Herranz}, D. and {Hildebrandt}, S.~R. and {Hivon}, E. and {Hobson}, M. and {Holmes}, W.~A. and {Hornstrup}, A. and {Hovest}, W. and {Hoyland}, R.~J. and {Huffenberger}, K.~M. and {Huynh}, M. and {Jaffe}, A.~H. and {Jones}, W.~C. and {Juvela}, M. and {Keih{\"a}nen}, E. and {Keskitalo}, R. and {Kisner}, T.~S. and {Kneissl}, R. and {Knox}, L. and {Kurki-Suonio}, H. and {Lagache}, G. and {L{\"a}hteenm{\"a}ki}, A. and {Lamarre}, J. -M. and {Lasenby}, A. and {Laureijs}, R.~J. and {Lawrence}, C.~R. and {Leach}, S. and {Leahy}, J.~P. and {Leonardi}, R. and {Le{\'o}n-Tavares}, J. and {Leroy}, C. and {Lilje}, P.~B. and {Linden-V{\o}rnle}, M. and {L{\'o}pez-Caniego}, M. and {Lubin}, P.~M. and {Mac{\'\i}as-P{\'e}rez}, J.~F. and {MacTavish}, C.~J. and {Maffei}, B. and {Maggio}, G. and {Maino}, D. and {Mandolesi}, N. and {Mann}, R. and {Maris}, M. and {Marleau}, F. and {Marshall}, D.~J. and {Mart{\'\i}nez-Gonz{\'a}lez}, E. and {Masi}, S. and {Massardi}, M. and {Matarrese}, S. and {Matthai}, F. and {Mazzotta}, P. and {McGehee}, P. and {Meinhold}, P.~R. and {Melchiorri}, A. and {Melin}, J. -B. and {Mendes}, L. and {Mennella}, A. and {Mitra}, S. and {Miville-Desch{\^e}nes}, M. -A. and {Moneti}, A. and {Montier}, L. and {Morgante}, G. and {Mortlock}, D. and {Munshi}, D. and {Murphy}, A. and {Naselsky}, P. and {Natoli}, P. and {Netterfield}, C.~B. and {N{\o}rgaard-Nielsen}, H.~U. and {Noviello}, F. and {Novikov}, D. and {Novikov}, I. and {O'Dwyer}, I.~J. and {Osborne}, S. and {Pajot}, F. and {Paladini}, R. and {Partridge}, B. and {Pasian}, F. and {Patanchon}, G. and {Pearson}, T.~J. and {Perdereau}, O. and {Perotto}, L. and {Perrotta}, F. and {Piacentini}, F. and {Piat}, M. and {Piffaretti}, R. and {Plaszczynski}, S. and {Platania}, P. and {Pointecouteau}, E. and {Polenta}, G. and {Ponthieu}, N. and {Poutanen}, T. and {Pratt}, G.~W. and {Pr{\'e}zeau}, G. and {Prunet}, S. and {Puget}, J. -L. and {Rachen}, J.~P. and {Reach}, W.~T. and {Rebolo}, R. and {Reinecke}, M. and {Renault}, C. and {Ricciardi}, S. and {Riller}, T. and {Ristorcelli}, I. and {Rocha}, G. and {Rosset}, C. and {Rowan-Robinson}, M. and {Rubi{\~n}o-Mart{\'\i}n}, J.~A. and {Rusholme}, B. and {Sajina}, A. and {Sandri}, M. and {Santos}, D. and {Savini}, G. and {Schaefer}, B.~M. and {Scott}, D. and {Seiffert}, M.~D. and {Shellard}, P. and {Smoot}, G.~F. and {Starck}, J. -L. and {Stivoli}, F. and {Stolyarov}, V. and {Sudiwala}, R. and {Sunyaev}, R. and {Sygnet}, J. -F. and {Tauber}, J.~A. and {Tavagnacco}, D. and {Terenzi}, L. and {Toffolatti}, L. and {Tomasi}, M. and {Torre}, J. -P. and {Tristram}, M. and {Tuovinen}, J. and {T{\"u}rler}, M. and {Umana}, G. and {Valenziano}, L. and {Valiviita}, J. and {Varis}, J. and {Vielva}, P. and {Villa}, F. and {Vittorio}, N. and {Wade}, L.~A. and {Wandelt}, B.~D. and {White}, S.~D.~M. and {Wilkinson}, A. and {Yvon}, D. and {Zacchei}, A. and {Zonca}, A.},
        title = "{Planck early results. VII. The Early Release Compact Source Catalogue}",
      journal = {\aap},
         year = 2011,
        month = dec,
       volume = {536},
          eid = {A7},
        pages = {A7},
          doi = {10.1051/0004-6361/201116474},
archivePrefix = {arXiv},
       eprint = {1101.2041},
 primaryClass = {astro-ph.CO},
       adsurl = {https://ui.adsabs.harvard.edu/abs/2011A&A...536A...7P}
}

@ARTICLE{Ade13-SourceCounts,
       author = {{Planck Collaboration} and {Ade}, P.~A.~R. and {Aghanim}, N. and {Arg{\"u}eso}, F. and {Arnaud}, M. and {Ashdown}, M. and {Atrio-Barandela}, F. and {Aumont}, J. and {Baccigalupi}, C. and {Balbi}, A. and {Banday}, A.~J. and {Barreiro}, R.~B. and {Battaner}, E. and {Benabed}, K. and {Beno{\^\i}t}, A. and {Bernard}, J. -P. and {Bersanelli}, M. and {Bethermin}, M. and {Bhatia}, R. and {Bonaldi}, A. and {Bond}, J.~R. and {Borrill}, J. and {Bouchet}, F.~R. and {Burigana}, C. and {Cabella}, P. and {Cardoso}, J. -F. and {Catalano}, A. and {Cay{\'o}n}, L. and {Chamballu}, A. and {Chary}, R. -R. and {Chen}, X. and {Chiang}, L. -Y. and {Christensen}, P.~R. and {Clements}, D.~L. and {Colafrancesco}, S. and {Colombi}, S. and {Colombo}, L.~P.~L. and {Coulais}, A. and {Crill}, B.~P. and {Cuttaia}, F. and {Danese}, L. and {Davis}, R.~J. and {de Bernardis}, P. and {de Gasperis}, G. and {de Zotti}, G. and {Delabrouille}, J. and {Dickinson}, C. and {Diego}, J.~M. and {Dole}, H. and {Donzelli}, S. and {Dor{\'e}}, O. and {D{\"o}rl}, U. and {Douspis}, M. and {Dupac}, X. and {Efstathiou}, G. and {En{\ss}lin}, T.~A. and {Eriksen}, H.~K. and {Finelli}, F. and {Forni}, O. and {Fosalba}, P. and {Frailis}, M. and {Franceschi}, E. and {Galeotta}, S. and {Ganga}, K. and {Giard}, M. and {Giardino}, G. and {Giraud-H{\'e}raud}, Y. and {Gonz{\'a}lez-Nuevo}, J. and {G{\'o}rski}, K.~M. and {Gregorio}, A. and {Gruppuso}, A. and {Hansen}, F.~K. and {Harrison}, D. and {Henrot-Versill{\'e}}, S. and {Hern{\'a}ndez-Monteagudo}, C. and {Herranz}, D. and {Hildebrandt}, S.~R. and {Hivon}, E. and {Hobson}, M. and {Holmes}, W.~A. and {Jaffe}, T.~R. and {Jaffe}, A.~H. and {Jagemann}, T. and {Jones}, W.~C. and {Juvela}, M. and {Keih{\"a}nen}, E. and {Kisner}, T.~S. and {Kneissl}, R. and {Knoche}, J. and {Knox}, L. and {Kunz}, M. and {Kurinsky}, N. and {Kurki-Suonio}, H. and {Lagache}, G. and {L{\"a}hteenm{\"a}ki}, A. and {Lamarre}, J. -M. and {Lasenby}, A. and {Lawrence}, C.~R. and {Leonardi}, R. and {Lilje}, P.~B. and {L{\'o}pez-Caniego}, M. and {Mac{\'\i}as-P{\'e}rez}, J.~F. and {Maino}, D. and {Mandolesi}, N. and {Maris}, M. and {Marshall}, D.~J. and {Mart{\'\i}nez-Gonz{\'a}lez}, E. and {Masi}, S. and {Massardi}, M. and {Matarrese}, S. and {Mazzotta}, P. and {Melchiorri}, A. and {Mendes}, L. and {Mennella}, A. and {Mitra}, S. and {Miville-Desch{\`e}nes}, M. -A. and {Moneti}, A. and {Montier}, L. and {Morgante}, G. and {Mortlock}, D. and {Munshi}, D. and {Murphy}, J.~A. and {Naselsky}, P. and {Nati}, F. and {Natoli}, P. and {N{\o}rgaard-Nielsen}, H.~U. and {Noviello}, F. and {Novikov}, D. and {Novikov}, I. and {Osborne}, S. and {Pajot}, F. and {Paladini}, R. and {Paoletti}, D. and {Partridge}, B. and {Pasian}, F. and {Patanchon}, G. and {Perdereau}, O. and {Perotto}, L. and {Perrotta}, F. and {Piacentini}, F. and {Piat}, M. and {Pierpaoli}, E. and {Plaszczynski}, S. and {Pointecouteau}, E. and {Polenta}, G. and {Ponthieu}, N. and {Popa}, L. and {Poutanen}, T. and {Pratt}, G.~W. and {Prunet}, S. and {Puget}, J. -L. and {Rachen}, J.~P. and {Reach}, W.~T. and {Rebolo}, R. and {Reinecke}, M. and {Renault}, C. and {Ricciardi}, S. and {Riller}, T. and {Ristorcelli}, I. and {Rocha}, G. and {Rosset}, C. and {Rowan-Robinson}, M. and {Rubi{\~n}o-Mart{\'\i}n}, J.~A. and {Rusholme}, B. and {Sajina}, A. and {Sandri}, M. and {Savini}, G. and {Scott}, D. and {Smoot}, G.~F. and {Starck}, J. -L. and {Sudiwala}, R. and {Suur-Uski}, A. -S. and {Sygnet}, J. -F. and {Tauber}, J.~A. and {Terenzi}, L. and {Toffolatti}, L. and {Tomasi}, M. and {Tristram}, M. and {Tucci}, M. and {T{\"u}rler}, M. and {Valenziano}, L. and {Van Tent}, B. and {Vielva}, P. and {Villa}, F. and {Vittorio}, N. and {Wade}, L.~A. and {Wandelt}, B.~D. and {White}, M. and {Yvon}, D. and {Zacchei}, A. and {Zonca}, A.},
        title = "{Planck intermediate results. VII. Statistical properties of infrared and radio extragalactic sources from the Planck Early Release Compact Source Catalogue at frequencies between 100 and 857 GHz}",
      journal = {\aap},
         year = 2013,
        month = feb,
       volume = {550},
          eid = {A133},
        pages = {A133},
          doi = {10.1051/0004-6361/201220053},
archivePrefix = {arXiv},
       eprint = {1207.4706},
 primaryClass = {astro-ph.CO},
       adsurl = {https://ui.adsabs.harvard.edu/abs/2013A&A...550A.133P}
}

@ARTICLE{Ade16-PCCS2,
       author = {{Planck Collaboration} and {Ade}, P.~A.~R. and {Aghanim}, N. and {Arg{\"u}eso}, F. and {Arnaud}, M. and {Ashdown}, M. and {Aumont}, J. and {Baccigalupi}, C. and {Banday}, A.~J. and {Barreiro}, R.~B. and {Bartolo}, N. and {Battaner}, E. and {Beichman}, C. and {Benabed}, K. and {Beno{\^\i}t}, A. and {Benoit-L{\'e}vy}, A. and {Bernard}, J. -P. and {Bersanelli}, M. and {Bielewicz}, P. and {Bock}, J.~J. and {B{\"o}hringer}, H. and {Bonaldi}, A. and {Bonavera}, L. and {Bond}, J.~R. and {Borrill}, J. and {Bouchet}, F.~R. and {Boulanger}, F. and {Bucher}, M. and {Burigana}, C. and {Butler}, R.~C. and {Calabrese}, E. and {Cardoso}, J. -F. and {Carvalho}, P. and {Catalano}, A. and {Challinor}, A. and {Chamballu}, A. and {Chary}, R. -R. and {Chiang}, H.~C. and {Christensen}, P.~R. and {Clemens}, M. and {Clements}, D.~L. and {Colombi}, S. and {Colombo}, L.~P.~L. and {Combet}, C. and {Couchot}, F. and {Coulais}, A. and {Crill}, B.~P. and {Curto}, A. and {Cuttaia}, F. and {Danese}, L. and {Davies}, R.~D. and {Davis}, R.~J. and {de Bernardis}, P. and {de Rosa}, A. and {de Zotti}, G. and {Delabrouille}, J. and {D{\'e}sert}, F. -X. and {Dickinson}, C. and {Diego}, J.~M. and {Dole}, H. and {Donzelli}, S. and {Dor{\'e}}, O. and {Douspis}, M. and {Ducout}, A. and {Dupac}, X. and {Efstathiou}, G. and {Elsner}, F. and {En{\ss}lin}, T.~A. and {Eriksen}, H.~K. and {Falgarone}, E. and {Fergusson}, J. and {Finelli}, F. and {Forni}, O. and {Frailis}, M. and {Fraisse}, A.~A. and {Franceschi}, E. and {Frejsel}, A. and {Galeotta}, S. and {Galli}, S. and {Ganga}, K. and {Giard}, M. and {Giraud-H{\'e}raud}, Y. and {Gjerl{\o}w}, E. and {Gonz{\'a}lez-Nuevo}, J. and {G{\'o}rski}, K.~M. and {Gratton}, S. and {Gregorio}, A. and {Gruppuso}, A. and {Gudmundsson}, J.~E. and {Hansen}, F.~K. and {Hanson}, D. and {Harrison}, D.~L. and {Helou}, G. and {Henrot-Versill{\'e}}, S. and {Hern{\'a}ndez-Monteagudo}, C. and {Herranz}, D. and {Hildebrandt}, S.~R. and {Hivon}, E. and {Hobson}, M. and {Holmes}, W.~A. and {Hornstrup}, A. and {Hovest}, W. and {Huffenberger}, K.~M. and {Hurier}, G. and {Jaffe}, A.~H. and {Jaffe}, T.~R. and {Jones}, W.~C. and {Juvela}, M. and {Keih{\"a}nen}, E. and {Keskitalo}, R. and {Kisner}, T.~S. and {Kneissl}, R. and {Knoche}, J. and {Kunz}, M. and {Kurki-Suonio}, H. and {Lagache}, G. and {L{\"a}hteenm{\"a}ki}, A. and {Lamarre}, J. -M. and {Lasenby}, A. and {Lattanzi}, M. and {Lawrence}, C.~R. and {Leahy}, J.~P. and {Leonardi}, R. and {Le{\'o}n-Tavares}, J. and {Lesgourgues}, J. and {Levrier}, F. and {Liguori}, M. and {Lilje}, P.~B. and {Linden-V{\o}rnle}, M. and {L{\'o}pez-Caniego}, M. and {Lubin}, P.~M. and {Mac{\'\i}as-P{\'e}rez}, J.~F. and {Maggio}, G. and {Maino}, D. and {Mandolesi}, N. and {Mangilli}, A. and {Maris}, M. and {Marshall}, D.~J. and {Martin}, P.~G. and {Mart{\'\i}nez-Gonz{\'a}lez}, E. and {Masi}, S. and {Matarrese}, S. and {McGehee}, P. and {Meinhold}, P.~R. and {Melchiorri}, A. and {Mendes}, L. and {Mennella}, A. and {Migliaccio}, M. and {Mitra}, S. and {Miville-Desch{\^e}nes}, M. -A. and {Moneti}, A. and {Montier}, L. and {Morgante}, G. and {Mortlock}, D. and {Moss}, A. and {Munshi}, D. and {Murphy}, J.~A. and {Naselsky}, P. and {Nati}, F. and {Natoli}, P. and {Negrello}, M. and {Netterfield}, C.~B. and {N{\o}rgaard-Nielsen}, H.~U. and {Noviello}, F. and {Novikov}, D. and {Novikov}, I. and {Oxborrow}, C.~A. and {Paci}, F. and {Pagano}, L. and {Pajot}, F. and {Paladini}, R. and {Paoletti}, D. and {Partridge}, B. and {Pasian}, F. and {Patanchon}, G. and {Pearson}, T.~J. and {Perdereau}, O. and {Perotto}, L. and {Perrotta}, F. and {Pettorino}, V. and {Piacentini}, F. and {Piat}, M. and {Pierpaoli}, E. and {Pietrobon}, D. and {Plaszczynski}, S. and {Pointecouteau}, E. and {Polenta}, G. and {Pratt}, G.~W. and {Pr{\'e}zeau}, G. and {Prunet}, S. and {Puget}, J. -L. and {Rachen}, J.~P. and {Reach}, W.~T. and {Rebolo}, R. and {Reinecke}, M. and {Remazeilles}, M. and {Renault}, C. and {Renzi}, A. and {Ristorcelli}, I. and {Rocha}, G. and {Rosset}, C. and {Rossetti}, M. and {Roudier}, G. and {Rowan-Robinson}, M. and {Rubi{\~n}o-Mart{\'\i}n}, J.~A. and {Rusholme}, B. and {Sandri}, M. and {Sanghera}, H.~S. and {Santos}, D. and {Savelainen}, M. and {Savini}, G. and {Scott}, D. and {Seiffert}, M.~D. and {Shellard}, E.~P.~S. and {Spencer}, L.~D. and {Stolyarov}, V. and {Sudiwala}, R. and {Sunyaev}, R. and {Sutton}, D. and {Suur-Uski}, A. -S. and {Sygnet}, J. -F. and {Tauber}, J.~A. and {Terenzi}, L. and {Toffolatti}, L. and {Tomasi}, M. and {Tornikoski}, M. and {Tristram}, M. and {Tucci}, M. and {Tuovinen}, J. and {T{\"u}rler}, M. and {Umana}, G. and {Valenziano}, L. and {Valiviita}, J. and {Van Tent}, B. and {Vielva}, P. and {Villa}, F. and {Wade}, L.~A. and {Walter}, B. and {Wandelt}, B.~D. and {Wehus}, I.~K. and {Yvon}, D. and {Zacchei}, A. and {Zonca}, A.},
        title = "{Planck 2015 results. XXVI. The Second Planck Catalogue of Compact Sources}",
      journal = {\aap},
         year = 2016,
        month = sep,
       volume = {594},
          eid = {A26},
        pages = {A26},
          doi = {10.1051/0004-6361/201526914},
archivePrefix = {arXiv},
       eprint = {1507.02058},
 primaryClass = {astro-ph.CO},
       adsurl = {https://ui.adsabs.harvard.edu/abs/2016A&A...594A..26P}
}

@ARTICLE{Adelson66,
        title = "{Compound poisson distributions}",
       author = {{Adelson}, R. M.},
      journal = {Journal of the Operational Research Society},
       volume = {17},
       number = {1},
        pages = {73--75},
         year = {1966},
    publisher = {Taylor \& Francis},
	      doi = {10.1057/jors.1966.8}
}

@ARTICLE{Aghanim20,
       author = {{Planck Collaboration} and {Aghanim}, N. and {Akrami}, Y. and {Ashdown}, M. and {Aumont}, J. and {Baccigalupi}, C. and {Ballardini}, M. and {Banday}, A.~J. and {Barreiro}, R.~B. and {Bartolo}, N. and {Basak}, S. and {Battye}, R. and {Benabed}, K. and {Bernard}, J. -P. and {Bersanelli}, M. and {Bielewicz}, P. and {Bock}, J.~J. and {Bond}, J.~R. and {Borrill}, J. and {Bouchet}, F.~R. and {Boulanger}, F. and {Bucher}, M. and {Burigana}, C. and {Butler}, R.~C. and {Calabrese}, E. and {Cardoso}, J. -F. and {Carron}, J. and {Challinor}, A. and {Chiang}, H.~C. and {Chluba}, J. and {Colombo}, L.~P.~L. and {Combet}, C. and {Contreras}, D. and {Crill}, B.~P. and {Cuttaia}, F. and {de Bernardis}, P. and {de Zotti}, G. and {Delabrouille}, J. and {Delouis}, J. -M. and {Di Valentino}, E. and {Diego}, J.~M. and {Dor{\'e}}, O. and {Douspis}, M. and {Ducout}, A. and {Dupac}, X. and {Dusini}, S. and {Efstathiou}, G. and {Elsner}, F. and {En{\ss}lin}, T.~A. and {Eriksen}, H.~K. and {Fantaye}, Y. and {Farhang}, M. and {Fergusson}, J. and {Fernandez-Cobos}, R. and {Finelli}, F. and {Forastieri}, F. and {Frailis}, M. and {Fraisse}, A.~A. and {Franceschi}, E. and {Frolov}, A. and {Galeotta}, S. and {Galli}, S. and {Ganga}, K. and {G{\'e}nova-Santos}, R.~T. and {Gerbino}, M. and {Ghosh}, T. and {Gonz{\'a}lez-Nuevo}, J. and {G{\'o}rski}, K.~M. and {Gratton}, S. and {Gruppuso}, A. and {Gudmundsson}, J.~E. and {Hamann}, J. and {Handley}, W. and {Hansen}, F.~K. and {Herranz}, D. and {Hildebrandt}, S.~R. and {Hivon}, E. and {Huang}, Z. and {Jaffe}, A.~H. and {Jones}, W.~C. and {Karakci}, A. and {Keih{\"a}nen}, E. and {Keskitalo}, R. and {Kiiveri}, K. and {Kim}, J. and {Kisner}, T.~S. and {Knox}, L. and {Krachmalnicoff}, N. and {Kunz}, M. and {Kurki-Suonio}, H. and {Lagache}, G. and {Lamarre}, J. -M. and {Lasenby}, A. and {Lattanzi}, M. and {Lawrence}, C.~R. and {Le Jeune}, M. and {Lemos}, P. and {Lesgourgues}, J. and {Levrier}, F. and {Lewis}, A. and {Liguori}, M. and {Lilje}, P.~B. and {Lilley}, M. and {Lindholm}, V. and {L{\'o}pez-Caniego}, M. and {Lubin}, P.~M. and {Ma}, Y. -Z. and {Mac{\'\i}as-P{\'e}rez}, J.~F. and {Maggio}, G. and {Maino}, D. and {Mandolesi}, N. and {Mangilli}, A. and {Marcos-Caballero}, A. and {Maris}, M. and {Martin}, P.~G. and {Martinelli}, M. and {Mart{\'\i}nez-Gonz{\'a}lez}, E. and {Matarrese}, S. and {Mauri}, N. and {McEwen}, J.~D. and {Meinhold}, P.~R. and {Melchiorri}, A. and {Mennella}, A. and {Migliaccio}, M. and {Millea}, M. and {Mitra}, S. and {Miville-Desch{\^e}nes}, M. -A. and {Molinari}, D. and {Montier}, L. and {Morgante}, G. and {Moss}, A. and {Natoli}, P. and {N{\o}rgaard-Nielsen}, H.~U. and {Pagano}, L. and {Paoletti}, D. and {Partridge}, B. and {Patanchon}, G. and {Peiris}, H.~V. and {Perrotta}, F. and {Pettorino}, V. and {Piacentini}, F. and {Polastri}, L. and {Polenta}, G. and {Puget}, J. -L. and {Rachen}, J.~P. and {Reinecke}, M. and {Remazeilles}, M. and {Renzi}, A. and {Rocha}, G. and {Rosset}, C. and {Roudier}, G. and {Rubi{\~n}o-Mart{\'\i}n}, J.~A. and {Ruiz-Granados}, B. and {Salvati}, L. and {Sandri}, M. and {Savelainen}, M. and {Scott}, D. and {Shellard}, E.~P.~S. and {Sirignano}, C. and {Sirri}, G. and {Spencer}, L.~D. and {Sunyaev}, R. and {Suur-Uski}, A. -S. and {Tauber}, J.~A. and {Tavagnacco}, D. and {Tenti}, M. and {Toffolatti}, L. and {Tomasi}, M. and {Trombetti}, T. and {Valenziano}, L. and {Valiviita}, J. and {Van Tent}, B. and {Vibert}, L. and {Vielva}, P. and {Villa}, F. and {Vittorio}, N. and {Wandelt}, B.~D. and {Wehus}, I.~K. and {White}, M. and {White}, S.~D.~M. and {Zacchei}, A. and {Zonca}, A.},
        title = "{Planck 2018 results. VI. Cosmological parameters}",
      journal = {\aap},
         year = 2020,
        month = sep,
       volume = {641},
          eid = {A6},
        pages = {A6},
          doi = {10.1051/0004-6361/201833910},
archivePrefix = {arXiv},
       eprint = {1807.06209},
 primaryClass = {astro-ph.CO},
       adsurl = {https://ui.adsabs.harvard.edu/abs/2020A&A...641A...6P}
}

@INCOLLECTION{Baddeley07,
       author = {Baddeley, Adrian},
        title = "{Spatial Point Processes and their Applications}",
    booktitle = {Stochastic Geometry},
	   editor = {{Weil}, Wolfgang},
         year = {2007},
		pages = {1-75},
    publisher = {Springer},
	  address = {Berlin},
	      doi = {10.1007/978-3-540-38175-4_1}
}

@ARTICLE{Bauer04,
       author = {{Bauer}, F.~E. and {Alexander}, D.~M. and {Brandt}, W.~N. and {Schneider}, D.~P. and {Treister}, E. and {Hornschemeier}, A.~E. and {Garmire}, G.~P.},
        title = "{The Fall of Active Galactic Nuclei and the Rise of Star-forming Galaxies: A Close Look at the Chandra Deep Field X-Ray Number Counts}",
      journal = {\aj},
         year = 2004,
        month = nov,
       volume = {128},
       number = {5},
        pages = {2048-2065},
          doi = {10.1086/424859},
archivePrefix = {arXiv},
       eprint = {astro-ph/0408001},
 primaryClass = {astro-ph},
       adsurl = {https://ui.adsabs.harvard.edu/abs/2004AJ....128.2048B}
}

@ARTICLE{Behroozi15,
       author = {{Behroozi}, Peter and {Peeples}, Molly S.},
        title = "{On the history and future of cosmic planet formation}",
      journal = {\mnras},
         year = 2015,
        month = dec,
       volume = {454},
       number = {2},
        pages = {1811-1817},
          doi = {10.1093/mnras/stv1817},
archivePrefix = {arXiv},
       eprint = {1508.01202},
 primaryClass = {astro-ph.GA},
       adsurl = {https://ui.adsabs.harvard.edu/abs/2015MNRAS.454.1811B}
}

@ARTICLE{Brzycki23,
       author = {{Brzycki}, Bryan and {Siemion}, Andrew P.~V. and {de Pater}, Imke and {Cordes}, James M. and {Gajjar}, Vishal and {Lacki}, Brian and {Sheikh}, Sofia},
        title = "{On Detecting Interstellar Scintillation in Narrowband Radio SETI}",
      journal = {\apj},
         year = 2023,
        month = jul,
       volume = {952},
       number = {1},
          eid = {46},
        pages = {46},
          doi = {10.3847/1538-4357/acdee0},
archivePrefix = {arXiv},
       eprint = {2307.08793},
 primaryClass = {astro-ph.IM},
       adsurl = {https://ui.adsabs.harvard.edu/abs/2023ApJ...952...46B}
}

@ARTICLE{Carlstrom11,
       author = {{Carlstrom}, J.~E. and {Ade}, P.~A.~R. and {Aird}, K.~A. and {Benson}, B.~A. and {Bleem}, L.~E. and {Busetti}, S. and {Chang}, C.~L. and {Chauvin}, E. and {Cho}, H. -M. and {Crawford}, T.~M. and {Crites}, A.~T. and {Dobbs}, M.~A. and {Halverson}, N.~W. and {Heimsath}, S. and {Holzapfel}, W.~L. and {Hrubes}, J.~D. and {Joy}, M. and {Keisler}, R. and {Lanting}, T.~M. and {Lee}, A.~T. and {Leitch}, E.~M. and {Leong}, J. and {Lu}, W. and {Lueker}, M. and {Luong-Van}, D. and {McMahon}, J.~J. and {Mehl}, J. and {Meyer}, S.~S. and {Mohr}, J.~J. and {Montroy}, T.~E. and {Padin}, S. and {Plagge}, T. and {Pryke}, C. and {Ruhl}, J.~E. and {Schaffer}, K.~K. and {Schwan}, D. and {Shirokoff}, E. and {Spieler}, H.~G. and {Staniszewski}, Z. and {Stark}, A.~A. and {Tucker}, C. and {Vanderlinde}, K. and {Vieira}, J.~D. and {Williamson}, R.},
        title = "{The 10 Meter South Pole Telescope}",
      journal = {\pasp},
         year = 2011,
        month = may,
       volume = {123},
       number = {903},
        pages = {568},
          doi = {10.1086/659879},
archivePrefix = {arXiv},
       eprint = {0907.4445},
 primaryClass = {astro-ph.IM},
       adsurl = {https://ui.adsabs.harvard.edu/abs/2011PASP..123..568C}
}

@ARTICLE{Carter83,
       author = {{Carter}, B.},
        title = "{The Anthropic Principle and its Implications for Biological Evolution}",
      journal = {Philosophical Transactions of the Royal Society of London Series A},
         year = 1983,
        month = dec,
       volume = {310},
       number = {1512},
        pages = {347-363},
          doi = {10.1098/rsta.1983.0096},
       adsurl = {https://ui.adsabs.harvard.edu/abs/1983RSPTA.310..347C}
}

@ARTICLE{Chabrier03,
       author = {{Chabrier}, Gilles},
        title = "{The Galactic Disk Mass Function: Reconciliation of the Hubble Space Telescope and Nearby Determinations}",
      journal = {\apjl},
         year = 2003,
        month = apr,
       volume = {586},
       number = {2},
        pages = {L133-L136},
          doi = {10.1086/374879},
archivePrefix = {arXiv},
       eprint = {astro-ph/0302511},
 primaryClass = {astro-ph},
       adsurl = {https://ui.adsabs.harvard.edu/abs/2003ApJ...586L.133C}
}

@BOOK{Chiu13,
	   author = {{Chiu}, Sung Nok and {Stoyan}, Dietrich and {Kendall}, Wilfrid S. and {Mecke}, Joseph},
	    title = "{Stochastic Geometry and its Applications: Third Edition}",
		 year = 2013,
	  address = {New York},
	publisher = {Wiley},
	      doi = {10.1002/9781118658222}
}

@ARTICLE{Choza24,
       author = {{Choza}, Carmen and {Bautista}, Daniel and {Croft}, Steve and {Siemion}, Andrew P.~V. and {Brzycki}, Bryan and {Bhattaram}, Krishnakumar and {Czech}, Daniel and {de Pater}, Imke and {Gajjar}, Vishal and {Isaacson}, Howard and {Lacker}, Kevin and {Lacki}, Brian and {Lebofsky}, Matthew and {MacMahon}, David H.~E. and {Price}, Danny and {Schoultz}, Sarah and {Sheikh}, Sofia and {Varghese}, Savin Shynu and {Morgan}, Lawrence and {Drew}, Jamie and {Worden}, S. Pete},
        title = "{The Breakthrough Listen Search for Intelligent Life: Technosignature Search of 97 Nearby Galaxies}",
      journal = {\aj},
         year = 2024,
        month = jan,
       volume = {167},
       number = {1},
          eid = {10},
        pages = {10},
          doi = {10.3847/1538-3881/acf576},
archivePrefix = {arXiv},
       eprint = {2312.03943},
 primaryClass = {astro-ph.IM},
       adsurl = {https://ui.adsabs.harvard.edu/abs/2024AJ....167...10C}
}

@ARTICLE{Cirkovic08,
       author = {{{\'C}irkovi{\'c}}, Milan M. and {Vukoti{\'c}}, Branislav},
        title = "{Astrobiological Phase Transition: Towards Resolution of Fermi's Paradox}",
      journal = {Origins of Life and Evolution of the Biosphere},
         year = 2008,
        month = dec,
       volume = {38},
       number = {6},
        pages = {535-547},
          doi = {10.1007/s11084-008-9149-y},
       adsurl = {https://ui.adsabs.harvard.edu/abs/2008OLEB...38..535C}
}

@ARTICLE{Cocconi59,
       author = {{Cocconi}, Giuseppe and {Morrison}, Philip},
        title = "{Searching for Interstellar Communications}",
      journal = {\nat},
         year = 1959,
        month = sep,
       volume = {184},
       number = {4690},
        pages = {844-846},
          doi = {10.1038/184844a0},
       adsurl = {https://ui.adsabs.harvard.edu/abs/1959Natur.184..844C}
}

@ARTICLE{Condon84-Evol,
       author = {{Condon}, J.~J.},
        title = "{Cosmological evolution of radio sources.}",
      journal = {\apj},
         year = 1984,
        month = dec,
       volume = {287},
        pages = {461-474},
          doi = {10.1086/162705},
       adsurl = {https://ui.adsabs.harvard.edu/abs/1984ApJ...287..461C}
}

@ARTICLE{Condon92,
       author = {{Condon}, J.~J.},
        title = "{Radio emission from normal galaxies.}",
      journal = {\araa},
         year = 1992,
        month = jan,
       volume = {30},
        pages = {575-611},
          doi = {10.1146/annurev.aa.30.090192.003043},
       adsurl = {https://ui.adsabs.harvard.edu/abs/1992ARA&A..30..575C}
}

@ARTICLE{Condon98,
       author = {{Condon}, J.~J. and {Cotton}, W.~D. and {Greisen}, E.~W. and {Yin}, Q.~F. and {Perley}, R.~A. and {Taylor}, G.~B. and {Broderick}, J.~J.},
        title = "{The NRAO VLA Sky Survey}",
      journal = {\aj},
         year = 1998,
        month = may,
       volume = {115},
       number = {5},
        pages = {1693-1716},
          doi = {10.1086/300337},
       adsurl = {https://ui.adsabs.harvard.edu/abs/1998AJ....115.1693C}
}

@ARTICLE{Condon02,
       author = {{Condon}, J.~J. and {Cotton}, W.~D. and {Broderick}, J.~J.},
        title = "{Radio Sources and Star Formation in the Local Universe}",
      journal = {\aj},
         year = 2002,
        month = aug,
       volume = {124},
       number = {2},
        pages = {675-689},
          doi = {10.1086/341650},
       adsurl = {https://ui.adsabs.harvard.edu/abs/2002AJ....124..675C}
}

@ARTICLE{Condon12,
       author = {{Condon}, J.~J. and {Cotton}, W.~D. and {Fomalont}, E.~B. and {Kellermann}, K.~I. and {Miller}, N. and {Perley}, R.~A. and {Scott}, D. and {Vernstrom}, T. and {Wall}, J.~V.},
        title = "{Resolving the Radio Source Background: Deeper Understanding through Confusion}",
      journal = {\apj},
         year = 2012,
        month = oct,
       volume = {758},
       number = {1},
          eid = {23},
        pages = {23},
          doi = {10.1088/0004-637X/758/1/23},
archivePrefix = {arXiv},
       eprint = {1207.2439},
 primaryClass = {astro-ph.CO},
       adsurl = {https://ui.adsabs.harvard.edu/abs/2012ApJ...758...23C}
}

@ARTICLE{Cordes97,
       author = {{Cordes}, James M. and {Lazio}, Joseph W. and {Sagan}, Carl},
        title = "{Scintillation-induced Intermittency in SETI}",
      journal = {\apj},
         year = 1997,
        month = oct,
       volume = {487},
       number = {2},
        pages = {782-808},
          doi = {10.1086/304620},
archivePrefix = {arXiv},
       eprint = {astro-ph/9707039},
 primaryClass = {astro-ph},
       adsurl = {https://ui.adsabs.harvard.edu/abs/1997ApJ...487..782C}
}

@BOOK{Daley03,
	   author = {{Daley}, Daryl J. and {Vere-Jones}, David},
	    title = "{An Introduction to the Theory of Point Processes. Volume I: Elementary Theory and Methods}",
		 year = 2003,
	  address = {New York},
	publisher = {Springer},
	      doi = {10.1007/b97277}
}

@BOOK{Daley08,
	   author = {{Daley}, Daryl J. and {Vere-Jones}, David},
	    title = "{An Introduction to the Theory of Point Processes. Volume II: General Theory and Structure}",
		 year = 2008,
	  address = {New York},
	publisher = {Springer},
	      doi = {10.1007/978-0-387-49835-5}
}

@ARTICLE{Davies11,
       author = {{AMI Consortium} and {Davies}, Matthew L. and {Franzen}, Thomas M.~O. and {Waldram}, Elizabeth M. and {Grainge}, Keith J.~B. and {Hobson}, Michael P. and {Hurley-Walker}, Natasha and {Lasenby}, Anthony and {Olamaie}, Malak and {Pooley}, Guy G. and {Riley}, Julia M. and {Rodr{\'\i}guez-Gonz{\'a}lvez}, Carmen and {Saunders}, Richard D.~E. and {Scaife}, Anna M.~M. and {Schammel}, Michel P. and {Scott}, Paul F. and {Shimwell}, Timothy W. and {Titterington}, David J. and {Zwart}, Jonathan T.~L.},
        title = "{10C survey of radio sources at 15.7 GHz - II. First results}",
      journal = {\mnras},
         year = 2011,
        month = aug,
       volume = {415},
       number = {3},
        pages = {2708-2722},
          doi = {10.1111/j.1365-2966.2011.18925.x},
archivePrefix = {arXiv},
       eprint = {1012.3659},
 primaryClass = {astro-ph.CO},
       adsurl = {https://ui.adsabs.harvard.edu/abs/2011MNRAS.415.2708A}
}

@INPROCEEDINGS{Dixon85,
       author = {{Dixon}, R.~S.},
        title = "{The Ohio SETI program - the first decade.}",
    booktitle = {The Search for Extraterrestrial Life: Recent Developments},
       series = {IAU Symposium},
         year = 1985,
       editor = {{Papagiannis}, M.~D.},
       volume = {112},
        month = jan,
        pages = {305-314},
	  address = {Dordrecht},
    publisher = {D. Reidel Publishing Co.},
	      doi = {10.1007/978-94-009-5462-5_39},
       adsurl = {https://ui.adsabs.harvard.edu/abs/1985IAUS..112..305D}
}

@ARTICLE{Driver22,
       author = {{Driver}, Simon P. and {Bellstedt}, Sabine and {Robotham}, Aaron S.~G. and {Baldry}, Ivan K. and {Davies}, Luke J. and {Liske}, Jochen and {Obreschkow}, Danail and {Taylor}, Edward N. and {Wright}, Angus H. and {Alpaslan}, Mehmet and {Bamford}, Steven P. and {Bauer}, Amanda E. and {Bland-Hawthorn}, Joss and {Bilicki}, Maciej and {Bravo}, Mat{\'\i}as and {Brough}, Sarah and {Casura}, Sarah and {Cluver}, Michelle E. and {Colless}, Matthew and {Conselice}, Christopher J. and {Croom}, Scott M. and {de Jong}, Jelte and {D'Eugenio}, Franceso and {De Propris}, Roberto and {Dogruel}, Burak and {Drinkwater}, Michael J. and {Dvornik}, Andrej and {Farrow}, Daniel J. and {Frenk}, Carlos S. and {Giblin}, Benjamin and {Graham}, Alister W. and {Grootes}, Meiert W. and {Gunawardhana}, Madusha L.~P. and {Hashemizadeh}, Abdolhosein and {H{\"a}u{\ss}ler}, Boris and {Heymans}, Catherine and {Hildebrandt}, Hendrik and {Holwerda}, Benne W. and {Hopkins}, Andrew M. and {Jarrett}, Tom H. and {Heath Jones}, D. and {Kelvin}, Lee S. and {Koushan}, Soheil and {Kuijken}, Konrad and {Lara-L{\'o}pez}, Maritza A. and {Lange}, Rebecca and {L{\'o}pez-S{\'a}nchez}, {\'A}ngel R. and {Loveday}, Jon and {Mahajan}, Smriti and {Meyer}, Martin and {Moffett}, Amanda J. and {Napolitano}, Nicola R. and {Norberg}, Peder and {Owers}, Matt S. and {Radovich}, Mario and {Raouf}, Mojtaba and {Peacock}, John A. and {Phillipps}, Steven and {Pimbblet}, Kevin A. and {Popescu}, Cristina and {Said}, Khaled and {Sansom}, Anne E. and {Seibert}, Mark and {Sutherland}, Will J. and {Thorne}, Jessica E. and {Tuffs}, Richard J. and {Turner}, Ryan and {van der Wel}, Arjen and {van Kampen}, Eelco and {Wilkins}, Steve M.},
        title = "{Galaxy And Mass Assembly (GAMA): Data Release 4 and the z < 0.1 total and z < 0.08 morphological galaxy stellar mass functions}",
      journal = {\mnras},
         year = 2022,
        month = jun,
       volume = {513},
       number = {1},
        pages = {439-467},
          doi = {10.1093/mnras/stac472},
archivePrefix = {arXiv},
       eprint = {2203.08539},
 primaryClass = {astro-ph.GA},
       adsurl = {https://ui.adsabs.harvard.edu/abs/2022MNRAS.513..439D}
}

@ARTICLE{Enriquez17,
       author = {{Enriquez}, J. Emilio and {Siemion}, Andrew and {Foster}, Griffin and {Gajjar}, Vishal and {Hellbourg}, Greg and {Hickish}, Jack and {Isaacson}, Howard and {Price}, Danny C. and {Croft}, Steve and {DeBoer}, David and {Lebofsky}, Matt and {MacMahon}, David H.~E. and {Werthimer}, Dan},
        title = "{The Breakthrough Listen Search for Intelligent Life: 1.1-1.9 GHz Observations of 692 Nearby Stars}",
      journal = {\apj},
         year = 2017,
        month = nov,
       volume = {849},
       number = {2},
          eid = {104},
        pages = {104},
          doi = {10.3847/1538-4357/aa8d1b},
archivePrefix = {arXiv},
       eprint = {1709.03491},
 primaryClass = {astro-ph.EP},
       adsurl = {https://ui.adsabs.harvard.edu/abs/2017ApJ...849..104E}
}

@ARTICLE{Everett20,
       author = {{Everett}, W.~B. and {Zhang}, L. and {Crawford}, T.~M. and {Vieira}, J.~D. and {Aravena}, M. and {Archipley}, M.~A. and {Austermann}, J.~E. and {Benson}, B.~A. and {Bleem}, L.~E. and {Carlstrom}, J.~E. and {Chang}, C.~L. and {Chapman}, S. and {Crites}, A.~T. and {de Haan}, T. and {Dobbs}, M.~A. and {George}, E.~M. and {Halverson}, N.~W. and {Harrington}, N. and {Holder}, G.~P. and {Holzapfel}, W.~L. and {Hrubes}, J.~D. and {Knox}, L. and {Lee}, A.~T. and {Luong-Van}, D. and {Mangian}, A.~C. and {Marrone}, D.~P. and {McMahon}, J.~J. and {Meyer}, S.~S. and {Mocanu}, L.~M. and {Mohr}, J.~J. and {Natoli}, T. and {Padin}, S. and {Pryke}, C. and {Reichardt}, C.~L. and {Reuter}, C.~A. and {Ruhl}, J.~E. and {Sayre}, J.~T. and {Schaffer}, K.~K. and {Shirokoff}, E. and {Spilker}, J.~S. and {Stalder}, B. and {Staniszewski}, Z. and {Stark}, A.~A. and {Story}, K.~T. and {Switzer}, E.~R. and {Vanderlinde}, K. and {Wei{\ss}}, A. and {Williamson}, R.},
        title = "{Millimeter-wave Point Sources from the 2500 Square Degree SPT-SZ Survey: Catalog and Population Statistics}",
      journal = {\apj},
         year = 2020,
        month = sep,
       volume = {900},
       number = {1},
          eid = {55},
        pages = {55},
          doi = {10.3847/1538-4357/ab9df7},
archivePrefix = {arXiv},
       eprint = {2003.03431},
 primaryClass = {astro-ph.IM},
       adsurl = {https://ui.adsabs.harvard.edu/abs/2020ApJ...900...55E}
}

@ARTICLE{Franco18,
       author = {{Franco}, M. and {Elbaz}, D. and {B{\'e}thermin}, M. and {Magnelli}, B. and {Schreiber}, C. and {Ciesla}, L. and {Dickinson}, M. and {Nagar}, N. and {Silverman}, J. and {Daddi}, E. and {Alexander}, D.~M. and {Wang}, T. and {Pannella}, M. and {Le Floc'h}, E. and {Pope}, A. and {Giavalisco}, M. and {Maury}, A.~J. and {Bournaud}, F. and {Chary}, R. and {Demarco}, R. and {Ferguson}, H. and {Finkelstein}, S.~L. and {Inami}, H. and {Iono}, D. and {Juneau}, S. and {Lagache}, G. and {Leiton}, R. and {Lin}, L. and {Magdis}, G. and {Messias}, H. and {Motohara}, K. and {Mullaney}, J. and {Okumura}, K. and {Papovich}, C. and {Pforr}, J. and {Rujopakarn}, W. and {Sargent}, M. and {Shu}, X. and {Zhou}, L.},
        title = "{GOODS-ALMA: 1.1 mm galaxy survey. I. Source catalog and optically dark galaxies}",
      journal = {\aap},
         year = 2018,
        month = dec,
       volume = {620},
          eid = {A152},
        pages = {A152},
          doi = {10.1051/0004-6361/201832928},
archivePrefix = {arXiv},
       eprint = {1803.00157},
 primaryClass = {astro-ph.GA},
       adsurl = {https://ui.adsabs.harvard.edu/abs/2018A&A...620A.152F}
}

@ARTICLE{Franzen19,
       author = {{Franzen}, T.~M.~O. and {Vernstrom}, T. and {Jackson}, C.~A. and {Hurley-Walker}, N. and {Ekers}, R.~D. and {Heald}, G. and {Seymour}, N. and {White}, S.~V.},
        title = "{Source counts and confusion at 72-231 MHz in the MWA GLEAM survey}",
      journal = {\pasa},
         year = 2019,
        month = feb,
       volume = {36},
          eid = {e004},
        pages = {e004},
          doi = {10.1017/pasa.2018.52},
archivePrefix = {arXiv},
       eprint = {1812.00666},
 primaryClass = {astro-ph.GA},
       adsurl = {https://ui.adsabs.harvard.edu/abs/2019PASA...36....4F}
}

@ARTICLE{Freitas85,
       author = {{Freitas}, R.~A., Jr. and {Valdes}, F.},
        title = "{The Search for Extraterrestrial Artifacts (SETA).}",
      journal = {Acta Astronautica},
         year = 1985,
        month = jan,
       volume = {12},
       number = {12},
        pages = {1027-1034},
          doi = {10.1016/0094-5765(85)90031-1},
       adsurl = {https://ui.adsabs.harvard.edu/abs/1985AcAau..12.1027F}
}

@ARTICLE{Fujimoto16,
       author = {{Fujimoto}, Seiji and {Ouchi}, Masami and {Ono}, Yoshiaki and {Shibuya}, Takatoshi and {Ishigaki}, Masafumi and {Nagai}, Hiroshi and {Momose}, Rieko},
        title = "{ALMA Census of Faint 1.2 mm Sources Down to \raisebox{-0.5ex}\textasciitilde 0.02 mJy: Extragalactic Background Light and Dust-poor, High-z Galaxies}",
      journal = {\apjs},
         year = 2016,
        month = jan,
       volume = {222},
       number = {1},
          eid = {1},
        pages = {1},
          doi = {10.3847/0067-0049/222/1/1},
archivePrefix = {arXiv},
       eprint = {1505.03523},
 primaryClass = {astro-ph.GA},
       adsurl = {https://ui.adsabs.harvard.edu/abs/2016ApJS..222....1F}
}

@ARTICLE{Gajjar21,
       author = {{Gajjar}, Vishal and {Perez}, Karen I. and {Siemion}, Andrew P.~V. and {Foster}, Griffin and {Brzycki}, Bryan and {Chatterjee}, Shami and {Chen}, Yuhong and {Cordes}, James M. and {Croft}, Steve and {Czech}, Daniel and {DeBoer}, David and {DeMarines}, Julia and {Drew}, Jamie and {Gowanlock}, Michael and {Isaacson}, Howard and {Lacki}, Brian C. and {Lebofsky}, Matt and {MacMahon}, David H.~E. and {Morrison}, Ian S. and {Ng}, Cherry and {de Pater}, Imke and {Price}, Danny C. and {Sheikh}, Sofia Z. and {Suresh}, Akshay and {Webb}, Claire and {Pete Worden}, S.},
        title = "{The Breakthrough Listen Search For Intelligent Life Near the Galactic Center. I.}",
      journal = {\aj},
         year = 2021,
        month = jul,
       volume = {162},
       number = {1},
          eid = {33},
        pages = {33},
          doi = {10.3847/1538-3881/abfd36},
archivePrefix = {arXiv},
       eprint = {2104.14148},
 primaryClass = {astro-ph.HE},
       adsurl = {https://ui.adsabs.harvard.edu/abs/2021AJ....162...33G}
}

@ARTICLE{Garrett23,
       author = {{Garrett}, M.~A. and {Siemion}, A.~P.~V.},
        title = "{Constraints on extragalactic transmitters via Breakthrough Listen observations of background sources}",
      journal = {\mnras},
         year = 2023,
        month = mar,
       volume = {519},
       number = {3},
        pages = {4581-4588},
          doi = {10.1093/mnras/stac2607},
       adsurl = {https://ui.adsabs.harvard.edu/abs/2023MNRAS.519.4581G}
}

@ARTICLE{Gowanlock16,
       author = {{Gowanlock}, Michael G.},
        title = "{Astrobiological Effects of Gamma-ray Bursts in the Milky Way Galaxy}",
      journal = {\apj},
         year = 2016,
        month = nov,
       volume = {832},
       number = {1},
          eid = {38},
        pages = {38},
          doi = {10.3847/0004-637X/832/1/38},
archivePrefix = {arXiv},
       eprint = {1609.09355},
 primaryClass = {astro-ph.EP},
       adsurl = {https://ui.adsabs.harvard.edu/abs/2016ApJ...832...38G}
}

@ARTICLE{Gray17,
       author = {{Gray}, Robert H. and {Mooley}, Kunal},
        title = "{A VLA Search for Radio Signals from M31 and M33}",
      journal = {\aj},
         year = 2017,
        month = mar,
       volume = {153},
       number = {3},
          eid = {110},
        pages = {110},
          doi = {10.3847/1538-3881/153/3/110},
archivePrefix = {arXiv},
       eprint = {1702.03301},
 primaryClass = {astro-ph.EP},
       adsurl = {https://ui.adsabs.harvard.edu/abs/2017AJ....153..110G}
}

@ARTICLE{Greve04,
       author = {{Greve}, T.~R. and {Ivison}, R.~J. and {Bertoldi}, F. and {Stevens}, J.~A. and {Dunlop}, J.~S. and {Lutz}, D. and {Carilli}, C.~L.},
        title = "{A 1200-{\ensuremath{\mu}}m MAMBO survey of ELAISN2 and the Lockman Hole - I. Maps, sources and number counts}",
      journal = {\mnras},
         year = 2004,
        month = nov,
       volume = {354},
       number = {3},
        pages = {779-797},
          doi = {10.1111/j.1365-2966.2004.08235.x},
archivePrefix = {arXiv},
       eprint = {astro-ph/0405361},
 primaryClass = {astro-ph},
       adsurl = {https://ui.adsabs.harvard.edu/abs/2004MNRAS.354..779G}
}

@BOOK{Haenggi13,
        title = {Stochastic Geometry for Wireless Networks},
       author = {{Haenggi}, Martin},
         year = {2013},
    publisher = {Cambridge University Press},
      address = {Cambridge},
	      doi = {10.1017/cbo9781139043816}
}

@ARTICLE{Hanson21,
       author = {{Hanson}, Robin and {Martin}, Daniel and {McCarter}, Calvin and {Paulson}, Jonathan},
        title = "{If Loud Aliens Explain Human Earliness, Quiet Aliens Are Also Rare}",
      journal = {\apj},
         year = 2021,
        month = dec,
       volume = {922},
       number = {2},
          eid = {182},
        pages = {182},
          doi = {10.3847/1538-4357/ac2369},
archivePrefix = {arXiv},
       eprint = {2102.01522},
 primaryClass = {q-bio.OT},
       adsurl = {https://ui.adsabs.harvard.edu/abs/2021ApJ...922..182H}
}

@ARTICLE{Harp16,
       author = {{Harp}, G.~R. and {Richards}, Jon and {Tarter}, Jill C. and {Dreher}, John and {Jordan}, Jane and {Shostak}, Seth and {Smolek}, Ken and {Kilsdonk}, Tom and {Wilcox}, Bethany R. and {Wimberly}, M.~K.~R. and {Ross}, John and {Barott}, W.~C. and {Ackermann}, R.~F. and {Blair}, Samantha},
        title = "{SETI Observations of Exoplanets with the Allen Telescope Array}",
      journal = {\aj},
         year = 2016,
        month = dec,
       volume = {152},
       number = {6},
          eid = {181},
        pages = {181},
          doi = {10.3847/0004-6256/152/6/181},
archivePrefix = {arXiv},
       eprint = {1607.04207},
 primaryClass = {astro-ph.EP},
       adsurl = {https://ui.adsabs.harvard.edu/abs/2016AJ....152..181H}
}

@ARTICLE{Hogg99,
       author = {{Hogg}, David W.},
        title = "{Distance measures in cosmology}",
      journal = {arXiv e-prints},
         year = 1999,
        month = may,
          eid = {astro-ph/9905116},
        pages = {astro-ph/9905116},
archivePrefix = {arXiv},
       eprint = {astro-ph/9905116},
 primaryClass = {astro-ph},
       adsurl = {https://ui.adsabs.harvard.edu/abs/1999astro.ph..5116H}
}

@ARTICLE{Horowitz93,
       author = {{Horowitz}, Paul and {Sagan}, Carl},
        title = "{Five Years of Project META: an All-Sky Narrow-Band Radio Search for Extraterrestrial Signals}",
      journal = {\apj},
         year = 1993,
        month = sep,
       volume = {415},
        pages = {218},
          doi = {10.1086/173157},
       adsurl = {https://ui.adsabs.harvard.edu/abs/1993ApJ...415..218H}
}

@ARTICLE{Isaacson17,
       author = {{Isaacson}, Howard and {Siemion}, Andrew P.~V. and {Marcy}, Geoffrey W. and {Lebofsky}, Matt and {Price}, Danny C. and {MacMahon}, David and {Croft}, Steve and {DeBoer}, David and {Hickish}, Jack and {Werthimer}, Dan and {Sheikh}, Sofia and {Hellbourg}, Greg and {Enriquez}, J. Emilio},
        title = "{The Breakthrough Listen Search for Intelligent Life: Target Selection of Nearby Stars and Galaxies}",
      journal = {\pasp},
         year = 2017,
        month = may,
       volume = {129},
       number = {975},
        pages = {054501},
          doi = {10.1088/1538-3873/aa5800},
archivePrefix = {arXiv},
       eprint = {1701.06227},
 primaryClass = {astro-ph.IM},
       adsurl = {https://ui.adsabs.harvard.edu/abs/2017PASP..129e4501I}
}

@INPROCEEDINGS{Jones91,
       author = {{Jones}, Michael E.},
        title = "{Microwave background interferometry in Cambridge}",
    booktitle = {IAU Colloq. 131: Radio Interferometry. Theory, Techniques, and Applications},
         year = 1991,
       editor = {{Cornwell}, T.~J. and {Perley}, R.~A.},
       series = {Astronomical Society of the Pacific Conference Series},
       volume = {19},
        month = jan,
        pages = {395-399},
       adsurl = {https://ui.adsabs.harvard.edu/abs/1991ASPC...19..395J}
}

@ARTICLE{Kardashev64,
       author = {{Kardashev}, N.~S.},
        title = "{Transmission of Information by Extraterrestrial Civilizations}",
      journal = {\sovast},
         year = 1964,
        month = oct,
       volume = {8},
        pages = {217},
       adsurl = {https://ui.adsabs.harvard.edu/abs/1964SvA.....8..217K}
}

@BOOK{Kingman93,
        title = "{Poisson Processes}",
       author = {{Kingman}, John Frank Charles},
         year = {1993},
	  address = {Oxford},
    publisher = {Clarendon Press},
	      doi = {10.1093/oso/9780198536932.001.0001}
}

@ARTICLE{Kipping24,
       author = {{Kipping}, David and {Lewis}, Geraint},
        title = "{Do SETI Optimists Have a Fine-Tuning Problem?}",
      journal = {arXiv e-prints},
         year = 2024,
        month = jun,
          eid = {arXiv:2407.07097},
        pages = {arXiv:2407.07097},
          doi = {10.48550/arXiv.2407.07097},
archivePrefix = {arXiv},
       eprint = {2407.07097},
 primaryClass = {astro-ph.IM},
       adsurl = {https://ui.adsabs.harvard.edu/abs/2024arXiv240707097K}
}

@ARTICLE{Kuiper77,
       author = {{Kuiper}, T.~B.~H. and {Morris}, M.},
        title = "{Searching for Extraterrestrial Civilizations}",
      journal = {Science},
         year = 1977,
        month = may,
       volume = {196},
       number = {4290},
        pages = {616-621},
          doi = {10.1126/science.196.4290.616},
       adsurl = {https://ui.adsabs.harvard.edu/abs/1977Sci...196..616K}
}

@ARTICLE{Lacki21-Traversability,
       author = {{Lacki}, Brian C.},
        title = "{Galactic traversability: a new concept for extragalactic SETI}",
      journal = {International Journal of Astrobiology},
         year = 2021,
        month = oct,
       volume = {20},
       number = {5},
        pages = {359-376},
          doi = {10.1017/S1473550421000252},
archivePrefix = {arXiv},
       eprint = {2106.07739},
 primaryClass = {astro-ph.GA},
       adsurl = {https://ui.adsabs.harvard.edu/abs/2021IJAsB..20..359L}
}

@ARTICLE{Lacki24-ETIPop1,
       author = {{Lacki}, Brian C.},
        title = "{Artificial Broadcasts as Galactic Populations. I. A Point Process Formalism for Extraterrestrial Intelligences and Their Broadcasts}",
      journal = {\apj},
         year = 2024,
        month = may,
       volume = {966},
       number = {2},
          eid = {182},
        pages = {182},
          doi = {10.3847/1538-4357/ad11f2},
archivePrefix = {arXiv},
       eprint = {2405.04646},
 primaryClass = {astro-ph.GA},
       adsurl = {https://ui.adsabs.harvard.edu/abs/2024ApJ...966..182L}
}

@ARTICLE{Lacki24-ETIPop2,
       author = {{Lacki}, Brian C.},
        title = "{Artificial Broadcasts as Galactic Populations. II. Comparing Individualist and Collective Bounds on Broadcast Populations in Single Galaxies}",
      journal = {\apj},
         year = 2024,
        month = may,
       volume = {966},
       number = {2},
          eid = {183},
        pages = {183},
          doi = {10.3847/1538-4357/ad11f1},
archivePrefix = {arXiv},
       eprint = {2405.04651},
 primaryClass = {astro-ph.GA},
       adsurl = {https://ui.adsabs.harvard.edu/abs/2024ApJ...966..183L}
}

@BOOK{Last17,
        title = "{Lectures on the Poisson Process}",
       author = {{Last}, G{\"u}nter and {Penrose}, Mathew},
         year = {2017},
	  address = {Cambridge},
    publisher = {Cambridge University Press},
	      doi = {10.1017/9781316104477}
}

@ARTICLE{Lehmer12,
       author = {{Lehmer}, B.~D. and {Xue}, Y.~Q. and {Brandt}, W.~N. and {Alexander}, D.~M. and {Bauer}, F.~E. and {Brusa}, M. and {Comastri}, A. and {Gilli}, R. and {Hornschemeier}, A.~E. and {Luo}, B. and {Paolillo}, M. and {Ptak}, A. and {Shemmer}, O. and {Schneider}, D.~P. and {Tozzi}, P. and {Vignali}, C.},
        title = "{The 4 Ms Chandra Deep Field-South Number Counts Apportioned by Source Class: Pervasive Active Galactic Nuclei and the Ascent of Normal Galaxies}",
      journal = {\apj},
         year = 2012,
        month = jun,
       volume = {752},
       number = {1},
          eid = {46},
        pages = {46},
          doi = {10.1088/0004-637X/752/1/46},
archivePrefix = {arXiv},
       eprint = {1204.1977},
 primaryClass = {astro-ph.CO},
       adsurl = {https://ui.adsabs.harvard.edu/abs/2012ApJ...752...46L}
}

@ARTICLE{Lin14,
       author = {{Lin}, Henry W. and {Gonzalez Abad}, Gonzalo and {Loeb}, Abraham},
        title = "{Detecting Industrial Pollution in the Atmospheres of Earth-like Exoplanets}",
      journal = {\apjl},
         year = 2014,
        month = sep,
       volume = {792},
       number = {1},
          eid = {L7},
        pages = {L7},
          doi = {10.1088/2041-8205/792/1/L7},
archivePrefix = {arXiv},
       eprint = {1406.3025},
 primaryClass = {astro-ph.EP},
       adsurl = {https://ui.adsabs.harvard.edu/abs/2014ApJ...792L...7L}
}

@ARTICLE{Lindner11,
       author = {{Lindner}, R.~R. and {Baker}, A.~J. and {Omont}, A. and {Beelen}, A. and {Owen}, F.~N. and {Bertoldi}, F. and {Dole}, H. and {Fiolet}, N. and {Harris}, A.~I. and {Ivison}, R.~J. and {Lonsdale}, C.~J. and {Lutz}, D. and {Polletta}, M.},
        title = "{A Deep 1.2 mm Map of the Lockman Hole North Field}",
      journal = {\apj},
         year = 2011,
        month = aug,
       volume = {737},
       number = {2},
          eid = {83},
        pages = {83},
          doi = {10.1088/0004-637X/737/2/83},
archivePrefix = {arXiv},
       eprint = {1106.0344},
 primaryClass = {astro-ph.CO},
       adsurl = {https://ui.adsabs.harvard.edu/abs/2011ApJ...737...83L}
}

@ARTICLE{Lineweaver04,
       author = {{Lineweaver}, Charles H. and {Fenner}, Yeshe and {Gibson}, Brad K.},
        title = "{The Galactic Habitable Zone and the Age Distribution of Complex Life in the Milky Way}",
      journal = {Science},
         year = 2004,
        month = jan,
       volume = {303},
       number = {5654},
        pages = {59-62},
          doi = {10.1126/science.1092322},
archivePrefix = {arXiv},
       eprint = {astro-ph/0401024},
 primaryClass = {astro-ph},
       adsurl = {https://ui.adsabs.harvard.edu/abs/2004Sci...303...59L}
}

@ARTICLE{Margot21,
       author = {{Margot}, Jean-Luc and {Pinchuk}, Pavlo and {Geil}, Robert and {Alexander}, Stephen and {Arora}, Sparsh and {Biswas}, Swagata and {Cebreros}, Jose and {Desai}, Sanjana Prabhu and {Duclos}, Benjamin and {Dunne}, Riley and {Lin Fu}, Kristy Kwan and {Goel}, Shashwat and {Gonzales}, Julia and {Gonzalez}, Alexander and {Jain}, Rishabh and {Lam}, Adrian and {Lewis}, Briley and {Lewis}, Rebecca and {Li}, Grace and {MacDougall}, Mason and {Makarem}, Christopher and {Manan}, Ivan and {Molina}, Eden and {Nagib}, Caroline and {Neville}, Kyle and {O'Toole}, Connor and {Rockwell}, Valerie and {Rokushima}, Yoichiro and {Romanek}, Griffin and {Schmidgall}, Carlyn and {Seth}, Samar and {Shah}, Rehan and {Shimane}, Yuri and {Singhal}, Myank and {Tokadjian}, Armen and {Villafana}, Lizvette and {Wang}, Zhixian and {Yun}, In and {Zhu}, Lujia and {Lynch}, Ryan S.},
        title = "{A Search for Technosignatures around 31 Sun-like Stars with the Green Bank Telescope at 1.15-1.73 GHz}",
      journal = {\aj},
         year = 2021,
        month = feb,
       volume = {161},
       number = {2},
          eid = {55},
        pages = {55},
          doi = {10.3847/1538-3881/abcc77},
archivePrefix = {arXiv},
       eprint = {2011.05265},
 primaryClass = {astro-ph.EP},
       adsurl = {https://ui.adsabs.harvard.edu/abs/2021AJ....161...55M}
}

@ARTICLE{Maselli16,
       author = {{Maselli}, A. and {Massaro}, F. and {Cusumano}, G. and {La Parola}, V. and {Harris}, D.~E. and {Paggi}, A. and {Liuzzo}, E. and {Tremblay}, G.~R. and {Baum}, S.~A. and {O'Dea}, C.~P.},
        title = "{Swift observations of unidentified radio sources in the revised Third Cambridge Catalogue}",
      journal = {\mnras},
         year = 2016,
        month = aug,
       volume = {460},
       number = {4},
        pages = {3829-3837},
          doi = {10.1093/mnras/stw1222},
archivePrefix = {arXiv},
       eprint = {1609.07484},
 primaryClass = {astro-ph.HE},
       adsurl = {https://ui.adsabs.harvard.edu/abs/2016MNRAS.460.3829M}
}

@ARTICLE{Massardi11,
       author = {{Massardi}, Marcella and {Ekers}, Ronald D. and {Murphy}, Tara and {Mahony}, Elizabeth and {Hancock}, Paul J. and {Chhetri}, Rajan and {de Zotti}, Gianfranco and {Sadler}, Elaine M. and {Burke-Spolaor}, Sarah and {Calabretta}, Mark and {Edwards}, Philip G. and {Ekers}, Jennifer A. and {Jackson}, Carole A. and {Kesteven}, Michael J. and {Newton-McGee}, Katherine and {Phillips}, Chris and {Ricci}, Roberto and {Roberts}, Paul and {Sault}, Robert J. and {Staveley-Smith}, Lister and {Subrahmanyan}, Ravi and {Walker}, Mark A. and {Wilson}, Warwick E.},
        title = "{The Australia Telescope 20 GHz (AT20G) Survey: analysis of the extragalactic source sample}",
      journal = {\mnras},
         year = 2011,
        month = mar,
       volume = {412},
       number = {1},
        pages = {318-330},
          doi = {10.1111/j.1365-2966.2010.17917.x},
archivePrefix = {arXiv},
       eprint = {1010.5942},
 primaryClass = {astro-ph.CO},
       adsurl = {https://ui.adsabs.harvard.edu/abs/2011MNRAS.412..318M}
}

@ARTICLE{Matthews21,
       author = {{Matthews}, A.~M. and {Condon}, J.~J. and {Cotton}, W.~D. and {Mauch}, T.},
        title = "{Source Counts Spanning Eight Decades of Flux Density at 1.4 GHz}",
      journal = {\apj},
         year = 2021,
        month = mar,
       volume = {909},
       number = {2},
          eid = {193},
        pages = {193},
          doi = {10.3847/1538-4357/abdd37},
archivePrefix = {arXiv},
       eprint = {2101.07827},
 primaryClass = {astro-ph.GA},
       adsurl = {https://ui.adsabs.harvard.edu/abs/2021ApJ...909..193M}
}

@ARTICLE{Murphy10,
       author = {{Murphy}, Tara and {Sadler}, Elaine M. and {Ekers}, Ronald D. and {Massardi}, Marcella and {Hancock}, Paul J. and {Mahony}, Elizabeth and {Ricci}, Roberto and {Burke-Spolaor}, Sarah and {Calabretta}, Mark and {Chhetri}, Rajan and {de Zotti}, Gianfranco and {Edwards}, Philip G. and {Ekers}, Jennifer A. and {Jackson}, Carole A. and {Kesteven}, Michael J. and {Lindley}, Emma and {Newton-McGee}, Katherine and {Phillips}, Chris and {Roberts}, Paul and {Sault}, Robert J. and {Staveley-Smith}, Lister and {Subrahmanyan}, Ravi and {Walker}, Mark A. and {Wilson}, Warwick E.},
        title = "{The Australia Telescope 20 GHz Survey: the source catalogue}",
      journal = {\mnras},
         year = 2010,
        month = mar,
       volume = {402},
       number = {4},
        pages = {2403-2423},
          doi = {10.1111/j.1365-2966.2009.15961.x},
archivePrefix = {arXiv},
       eprint = {0911.0002},
 primaryClass = {astro-ph.GA},
       adsurl = {https://ui.adsabs.harvard.edu/abs/2010MNRAS.402.2403M}
}

@ARTICLE{Perley17,
       author = {{Perley}, R.~A. and {Butler}, B.~J.},
        title = "{An Accurate Flux Density Scale from 50 MHz to 50 GHz}",
      journal = {\apjs},
         year = 2017,
        month = may,
       volume = {230},
       number = {1},
          eid = {7},
        pages = {7},
          doi = {10.3847/1538-4365/aa6df9},
archivePrefix = {arXiv},
       eprint = {1609.05940},
 primaryClass = {astro-ph.IM},
       adsurl = {https://ui.adsabs.harvard.edu/abs/2017ApJS..230....7P}
}

@ARTICLE{Price20,
       author = {{Price}, Danny C. and {Enriquez}, J. Emilio and {Brzycki}, Bryan and {Croft}, Steve and {Czech}, Daniel and {DeBoer}, David and {DeMarines}, Julia and {Foster}, Griffin and {Gajjar}, Vishal and {Gizani}, Nectaria and {Hellbourg}, Greg and {Isaacson}, Howard and {Lacki}, Brian and {Lebofsky}, Matt and {MacMahon}, David H.~E. and {Pater}, Imke de and {Siemion}, Andrew P.~V. and {Werthimer}, Dan and {Green}, James A. and {Kaczmarek}, Jane F. and {Maddalena}, Ronald J. and {Mader}, Stacy and {Drew}, Jamie and {Worden}, S. Pete},
        title = "{The Breakthrough Listen Search for Intelligent Life: Observations of 1327 Nearby Stars Over 1.10-3.45 GHz}",
      journal = {\aj},
         year = 2020,
        month = mar,
       volume = {159},
       number = {3},
          eid = {86},
        pages = {86},
          doi = {10.3847/1538-3881/ab65f1},
archivePrefix = {arXiv},
       eprint = {1906.07750},
 primaryClass = {astro-ph.EP},
       adsurl = {https://ui.adsabs.harvard.edu/abs/2020AJ....159...86P}
}

@ARTICLE{Sadler02,
       author = {{Sadler}, Elaine M. and {Jackson}, Carole A. and {Cannon}, Russell D. and {McIntyre}, Vincent J. and {Murphy}, Tara and {Bland-Hawthorn}, Joss and {Bridges}, Terry and {Cole}, Shaun and {Colless}, Matthew and {Collins}, Chris and {Couch}, Warrick and {Dalton}, Gavin and {De Propris}, Roberto and {Driver}, Simon P. and {Efstathiou}, George and {Ellis}, Richard S. and {Frenk}, Carlos S. and {Glazebrook}, Karl and {Lahav}, Ofer and {Lewis}, Ian and {Lumsden}, Stuart and {Maddox}, Steve and {Madgwick}, Darren and {Norberg}, Peder and {Peacock}, John A. and {Peterson}, Bruce A. and {Sutherland}, Will and {Taylor}, Keith},
        title = "{Radio sources in the 2dF Galaxy Redshift Survey - II. Local radio luminosity functions for AGN and star-forming galaxies at 1.4 GHz}",
      journal = {\mnras},
         year = 2002,
        month = jan,
       volume = {329},
       number = {1},
        pages = {227-245},
          doi = {10.1046/j.1365-8711.2002.04998.x},
archivePrefix = {arXiv},
       eprint = {astro-ph/0106173},
 primaryClass = {astro-ph},
       adsurl = {https://ui.adsabs.harvard.edu/abs/2002MNRAS.329..227S}
}

@ARTICLE{Shostak96,
       author = {{Shostak}, Seth and {Ekers}, Ron and {Vaile}, Roberta},
        title = "{A Search for Artificial Signals From the Small Magellanic Cloud}",
      journal = {\aj},
         year = 1996,
        month = jul,
       volume = {112},
        pages = {164},
          doi = {10.1086/117996},
       adsurl = {https://ui.adsabs.harvard.edu/abs/1996AJ....112..164S}
}

@INPROCEEDINGS{Tarter85,
       author = {{Tarter}, J.},
        title = "{SETI observations worldwide}",
    booktitle = {The Search for Extraterrestrial Life: Recent Developments},
         year = 1985,
       editor = {{Papagiannis}, M.~D.},
       series = {IAU Symposium},
       volume = {112},
        month = jan,
        pages = {271-290},
	  address = {Dordrecht},
    publisher = {D. Reidel Publishing Co.},
	      doi = {10.1007/978-94-009-5462-5_37},
       adsurl = {https://ui.adsabs.harvard.edu/abs/1985IAUS..112..271T}
}

@ARTICLE{Tarter01,
	   author = {{Tarter}, J.},
		title = "{The Search for Extraterrestrial Intelligence (SETI)}",
	  journal = {\araa},
		 year = 2001,
	   volume = 39,
		pages = {511-548},
		  doi = {10.1146/annurev.astro.39.1.511},
	   adsurl = {http://adsabs.harvard.edu/abs/2001ARA%26A..39..511T}
}

@ARTICLE{Tauber10,
       author = {{Tauber}, J.~A. and {Mandolesi}, N. and {Puget}, J. -L. and {Banos}, T. and {Bersanelli}, M. and {Bouchet}, F.~R. and {Butler}, R.~C. and {Charra}, J. and {Crone}, G. and {Dodsworth}, J. and {Efstathiou}, G. and {Gispert}, R. and {Guyot}, G. and {Gregorio}, A. and {Juillet}, J.~J. and {Lamarre}, J. -M. and {Laureijs}, R.~J. and {Lawrence}, C.~R. and {N{\o}rgaard-Nielsen}, H.~U. and {Passvogel}, T. and {Reix}, J.~M. and {Texier}, D. and {Vibert}, L. and {Zacchei}, A. and {Ade}, P.~A.~R. and {Aghanim}, N. and {Aja}, B. and {Alippi}, E. and {Aloy}, L. and {Armand}, P. and {Arnaud}, M. and {Arondel}, A. and {Arreola-Villanueva}, A. and {Artal}, E. and {Artina}, E. and {Arts}, A. and {Ashdown}, M. and {Aumont}, J. and {Azzaro}, M. and {Bacchetta}, A. and {Baccigalupi}, C. and {Baker}, M. and {Balasini}, M. and {Balbi}, A. and {Banday}, A.~J. and {Barbier}, G. and {Barreiro}, R.~B. and {Bartelmann}, M. and {Battaglia}, P. and {Battaner}, E. and {Benabed}, K. and {Beney}, J. -L. and {Beneyton}, R. and {Bennett}, K. and {Benoit}, A. and {Bernard}, J. -P. and {Bhandari}, P. and {Bhatia}, R. and {Biggi}, M. and {Biggins}, R. and {Billig}, G. and {Blanc}, Y. and {Blavot}, H. and {Bock}, J.~J. and {Bonaldi}, A. and {Bond}, R. and {Bonis}, J. and {Borders}, J. and {Borrill}, J. and {Boschini}, L. and {Boulanger}, F. and {Bouvier}, J. and {Bouzit}, M. and {Bowman}, R. and {Br{\'e}elle}, E. and {Bradshaw}, T. and {Braghin}, M. and {Bremer}, M. and {Brienza}, D. and {Broszkiewicz}, D. and {Burigana}, C. and {Burkhalter}, M. and {Cabella}, P. and {Cafferty}, T. and {Cairola}, M. and {Caminade}, S. and {Camus}, P. and {Cantalupo}, C.~M. and {Cappellini}, B. and {Cardoso}, J. -F. and {Carr}, R. and {Catalano}, A. and {Cay{\'o}n}, L. and {Cesa}, M. and {Chaigneau}, M. and {Challinor}, A. and {Chamballu}, A. and {Chambelland}, J.~P. and {Charra}, M. and {Chiang}, L. -Y. and {Chlewicki}, G. and {Christensen}, P.~R. and {Church}, S. and {Ciancietta}, E. and {Cibrario}, M. and {Cizeron}, R. and {Clements}, D. and {Collaudin}, B. and {Colley}, J. -M. and {Colombi}, S. and {Colombo}, A. and {Colombo}, F. and {Corre}, O. and {Couchot}, F. and {Cougrand}, B. and {Coulais}, A. and {Couzin}, P. and {Crane}, B. and {Crill}, B. and {Crook}, M. and {Crumb}, D. and {Cuttaia}, F. and {D{\"o}rl}, U. and {da Silva}, P. and {Daddato}, R. and {Damasio}, C. and {Danese}, L. and {D'Aquino}, G. and {D'Arcangelo}, O. and {Dassas}, K. and {Davies}, R.~D. and {Davies}, W. and {Davis}, R.~J. and {de Bernardis}, P. and {de Chambure}, D. and {de Gasperis}, G. and {de La Fuente}, M.~L. and {de Paco}, P. and {de Rosa}, A. and {de Troia}, G. and {de Zotti}, G. and {Dehamme}, M. and {Delabrouille}, J. and {Delouis}, J. -M. and {D{\'e}sert}, F. -X. and {di Girolamo}, G. and {Dickinson}, C. and {Doelling}, E. and {Dolag}, K. and {Domken}, I. and {Douspis}, M. and {Doyle}, D. and {Du}, S. and {Dubruel}, D. and {Dufour}, C. and {Dumesnil}, C. and {Dupac}, X. and {Duret}, P. and {Eder}, C. and {Elfving}, A. and {En{\ss}lin}, T.~A. and {Eng}, P. and {English}, K. and {Eriksen}, H.~K. and {Estaria}, P. and {Falvella}, M.~C. and {Ferrari}, F. and {Finelli}, F. and {Fishman}, A. and {Fogliani}, S. and {Foley}, S. and {Fonseca}, A. and {Forma}, G. and {Forni}, O. and {Fosalba}, P. and {Fourmond}, J. -J. and {Frailis}, M. and {Franceschet}, C. and {Franceschi}, E. and {Fran{\c{c}}ois}, S. and {Frerking}, M. and {G{\'o}mez-Re{\~n}asco}, M.~F. and {G{\'o}rski}, K.~M. and {Gaier}, T.~C. and {Galeotta}, S. and {Ganga}, K. and {Garc{\'\i}a L{\'a}zaro}, J. and {Garnica}, A. and {Gaspard}, M. and {Gavila}, E. and {Giard}, M. and {Giardino}, G. and {Gienger}, G. and {Giraud-Heraud}, Y. and {Glorian}, J. -M. and {Griffin}, M. and {Gruppuso}, A. and {Guglielmi}, L. and {Guichon}, D. and {Guillaume}, B. and {Guillouet}, P. and {Haissinski}, J. and {Hansen}, F.~K. and {Hardy}, J. and {Harrison}, D. and {Hazell}, A. and {Hechler}, M. and {Heckenauer}, V. and {Heinzer}, D. and {Hell}, R. and {Henrot-Versill{\'e}}, S. and {Hern{\'a}ndez-Monteagudo}, C. and {Herranz}, D. and {Herreros}, J.~M. and {Hervier}, V. and {Heske}, A. and {Heurtel}, A. and {Hildebrandt}, S.~R. and {Hills}, R. and {Hivon}, E. and {Hobson}, M. and {Hollert}, D. and {Holmes}, W. and {Hornstrup}, A. and {Hovest}, W. and {Hoyland}, R.~J. and {Huey}, G. and {Huffenberger}, K.~M. and {Hughes}, N. and {Israelsson}, U. and {Jackson}, B. and {Jaffe}, A. and {Jaffe}, T.~R. and {Jagemann}, T. and {Jessen}, N.~C. and {Jewell}, J. and {Jones}, W. and {Juvela}, M. and {Kaplan}, J. and {Karlman}, P. and {Keck}, F. and {Keih{\"a}nen}, E. and {King}, M. and {Kisner}, T.~S. and {Kletzkine}, P. and {Kneissl}, R. and {Knoche}, J. and {Knox}, L. and {Koch}, T. and {Krassenburg}, M. and {Kurki-Suonio}, H. and {L{\"a}hteenm{\"a}ki}, A. and {Lagache}, G. and {Lagorio}, E. and {Lami}, P. and {Lande}, J. and {Lange}, A. and {Langlet}, F. and {Lapini}, R. and {Lapolla}, M. and {Lasenby}, A. and {Le Jeune}, M. and {Leahy}, J.~P. and {Lefebvre}, M. and {Legrand}, F. and {Le Meur}, G. and {Leonardi}, R. and {Leriche}, B. and {Leroy}, C. and {Leutenegger}, P. and {Levin}, S.~M. and {Lilje}, P.~B. and {Lindensmith}, C. and {Linden-V{\o}rnle}, M. and {Loc}, A. and {Longval}, Y. and {Lubin}, P.~M. and {Luchik}, T. and {Luthold}, I. and {Macias-Perez}, J.~F. and {Maciaszek}, T. and {MacTavish}, C. and {Madden}, S. and {Maffei}, B. and {Magneville}, C. and {Maino}, D. and {Mambretti}, A. and {Mansoux}, B. and {Marchioro}, D. and {Maris}, M. and {Marliani}, F. and {Marrucho}, J. -C. and {Mart{\'\i}-Canales}, J. and {Mart{\'\i}nez-Gonz{\'a}lez}, E. and {Mart{\'\i}n-Polegre}, A. and {Martin}, P. and {Marty}, C. and {Marty}, W. and {Masi}, S. and {Massardi}, M. and {Matarrese}, S. and {Matthai}, F. and {Mazzotta}, P. and {McDonald}, A. and {McGrath}, P. and {Mediavilla}, A. and {Meinhold}, P.~R. and {M{\'e}lin}, J. -B. and {Melot}, F. and {Mendes}, L. and {Mennella}, A. and {Mervier}, C. and {Meslier}, L. and {Miccolis}, M. and {Miville-Deschenes}, M. -A. and {Moneti}, A. and {Montet}, D. and {Montier}, L. and {Mora}, J. and {Morgante}, G. and {Morigi}, G. and {Morinaud}, G. and {Morisset}, N. and {Mortlock}, D. and {Mottet}, S. and {Mulder}, J. and {Munshi}, D. and {Murphy}, A. and {Murphy}, P. and {Musi}, P. and {Narbonne}, J. and {Naselsky}, P. and {Nash}, A. and {Nati}, F. and {Natoli}, P. and {Netterfield}, B. and {Newell}, J. and {Nexon}, M. and {Nicolas}, C. and {Nielsen}, P.~H. and {Ninane}, N. and {Noviello}, F. and {Novikov}, D. and {Novikov}, I. and {O'Dwyer}, I.~J. and {Oldeman}, P. and {Olivier}, P. and {Ouchet}, L. and {Oxborrow}, C.~A. and {P{\'e}rez-Cuevas}, L. and {Pagan}, L. and {Paine}, C. and {Pajot}, F. and {Paladini}, R. and {Pancher}, F. and {Panh}, J. and {Parks}, G. and {Parnaudeau}, P. and {Partridge}, B. and {Parvin}, B. and {Pascual}, J.~P. and {Pasian}, F. and {Pearson}, D.~P. and {Pearson}, T. and {Pecora}, M. and {Perdereau}, O. and {Perotto}, L. and {Perrotta}, F. and {Piacentini}, F. and {Piat}, M. and {Pierpaoli}, E. and {Piersanti}, O. and {Plaige}, E. and {Plaszczynski}, S. and {Platania}, P. and {Pointecouteau}, E. and {Polenta}, G. and {Ponthieu}, N. and {Popa}, L. and {Poulleau}, G. and {Poutanen}, T. and {Pr{\'e}zeau}, G. and {Pradell}, L. and {Prina}, M. and {Prunet}, S. and {Rachen}, J.~P. and {Rambaud}, D. and {Rame}, F. and {Rasmussen}, I. and {Rautakoski}, J. and {Reach}, W.~T. and {Rebolo}, R. and {Reinecke}, M. and {Reiter}, J. and {Renault}, C. and {Ricciardi}, S. and {Rideau}, P. and {Riller}, T. and {Ristorcelli}, I. and {Riti}, J.~B. and {Rocha}, G. and {Roche}, Y. and {Pons}, R. and {Rohlfs}, R. and {Romero}, D. and {Roose}, S. and {Rosset}, C. and {Rouberol}, S. and {Rowan-Robinson}, M. and {Rubi{\~n}o-Mart{\'\i}n}, J.~A. and {Rusconi}, P. and {Rusholme}, B. and {Salama}, M. and {Salerno}, E. and {Sandri}, M. and {Santos}, D. and {Sanz}, J.~L. and {Sauter}, L. and {Sauvage}, F. and {Savini}, G. and {Schmelzel}, M. and {Schnorhk}, A. and {Schwarz}, W. and {Scott}, D. and {Seiffert}, M.~D. and {Shellard}, P. and {Shih}, C. and {Sias}, M. and {Silk}, J.~I. and {Silvestri}, R. and {Sippel}, R. and {Smoot}, G.~F. and {Starck}, J. -L. and {Stassi}, P. and {Sternberg}, J. and {Stivoli}, F. and {Stolyarov}, V. and {Stompor}, R. and {Stringhetti}, L. and {Strommen}, D. and {Stute}, T. and {Sudiwala}, R. and {Sugimura}, R. and {Sunyaev}, R. and {Sygnet}, J. -F. and {T{\"u}rler}, M. and {Taddei}, E. and {Tallon}, J. and {Tamiatto}, C. and {Taurigna}, M. and {Taylor}, D. and {Terenzi}, L. and {Thuerey}, S. and {Tillis}, J. and {Tofani}, G. and {Toffolatti}, L. and {Tommasi}, E. and {Tomasi}, M. and {Tonazzini}, E. and {Torre}, J. -P. and {Tosti}, S. and {Touze}, F. and {Tristram}, M. and {Tuovinen}, J. and {Tuttlebee}, M. and {Umana}, G. and {Valenziano}, L. and {Vall{\'e}e}, D. and {van der Vlis}, M. and {van Leeuwen}, F. and {Vanel}, J. -C. and {van-Tent}, B. and {Varis}, J. and {Vassallo}, E. and {Vescovi}, C. and {Vezzu}, F. and {Vibert}, D. and {Vielva}, P. and {Vierra}, J. and {Villa}, F. and {Vittorio}, N. and {Vuerli}, C. and {Wade}, L.~A. and {Walker}, A.~R. and {Wandelt}, B.~D. and {Watson}, C. and {Werner}, D. and {White}, M. and {White}, S.~D.~M. and {Wilkinson}, A. and {Wilson}, P. and {Woodcraft}, A. and {Yoffo}, B. and {Yun}, M. and {Yurchenko}, V. and {Yvon}, D. and {Zhang}, B. and {Zimmermann}, O. and {Zonca}, A. and {Zorita}, D.},
        title = "{Planck pre-launch status: The Planck mission}",
      journal = {\aap},
         year = 2010,
        month = sep,
       volume = {520},
          eid = {A1},
        pages = {A1},
          doi = {10.1051/0004-6361/200912983},
       adsurl = {https://ui.adsabs.harvard.edu/abs/2010A&A...520A...1T}
}

@ARTICLE{Thorne21,
       author = {{Thorne}, Jessica E. and {Robotham}, Aaron S.~G. and {Davies}, Luke J.~M. and {Bellstedt}, Sabine and {Driver}, Simon P. and {Bravo}, Mat{\'\i}as and {Bremer}, Malcolm N. and {Holwerda}, Benne W. and {Hopkins}, Andrew M. and {Lagos}, Claudia del P. and {Phillipps}, Steven and {Siudek}, Malgorzata and {Taylor}, Edward N. and {Wright}, Angus H.},
        title = "{Deep Extragalactic VIsible Legacy Survey (DEVILS): SED fitting in the D10-COSMOS field and the evolution of the stellar mass function and SFR-M$_{{\ensuremath{\star}}}$ relation}",
      journal = {\mnras},
         year = 2021,
        month = jul,
       volume = {505},
       number = {1},
        pages = {540-567},
          doi = {10.1093/mnras/stab1294},
archivePrefix = {arXiv},
       eprint = {2011.13605},
 primaryClass = {astro-ph.GA},
       adsurl = {https://ui.adsabs.harvard.edu/abs/2021MNRAS.505..540T}
}

@ARTICLE{Tremblay20,
       author = {{Tremblay}, C.~D. and {Tingay}, S.~J.},
        title = "{A SETI survey of the Vela region using the Murchison Widefield Array: Orders of magnitude expansion in search space}",
      journal = {\pasa},
         year = 2020,
        month = sep,
       volume = {37},
          eid = {e035},
        pages = {e035},
          doi = {10.1017/pasa.2020.27},
archivePrefix = {arXiv},
       eprint = {2009.03267},
 primaryClass = {astro-ph.IM},
       adsurl = {https://ui.adsabs.harvard.edu/abs/2020PASA...37...35T}
}

@ARTICLE{Tremblay22,
       author = {{Tremblay}, Chenoa D. and {Price}, Danny C. and {Tingay}, Steven J.},
        title = "{A search for technosignatures toward the Galactic Centre at 150 MHz}",
      journal = {\pasa},
         year = 2022,
        month = mar,
       volume = {39},
          eid = {e008},
        pages = {e008},
          doi = {10.1017/pasa.2022.5},
archivePrefix = {arXiv},
       eprint = {2202.03324},
 primaryClass = {astro-ph.GA},
       adsurl = {https://ui.adsabs.harvard.edu/abs/2022PASA...39....8T}
}

@ARTICLE{Tremblay24,
       author = {{Tremblay}, C.~D. and {Tingay}, S.~J.},
        title = "{An Extragalactic Widefield Search for Technosignatures with the Murchison Widefield Array}",
      journal = {\apj},
         year = 2024,
        month = sep,
       volume = {972},
       number = {1},
          eid = {76},
        pages = {76},
          doi = {10.3847/1538-4357/ad6b11},
archivePrefix = {arXiv},
       eprint = {2408.10372},
 primaryClass = {astro-ph.GA},
       adsurl = {https://ui.adsabs.harvard.edu/abs/2024ApJ...972...76T}
}

@ARTICLE{Uno23,
       author = {{Uno}, Yuri and {Hashimoto}, Tetsuya and {Goto}, Tomotsugu and {Ho}, Simon C. -C. and {Hsu}, Tzu-Yin and {Burns}, Ross},
        title = "{Upper limits on transmitter rate of extragalactic civilizations placed by Breakthrough Listen observations}",
      journal = {\mnras},
         year = 2023,
        month = jul,
       volume = {522},
       number = {3},
        pages = {4649-4653},
          doi = {10.1093/mnras/stad993},
archivePrefix = {arXiv},
       eprint = {2304.02756},
 primaryClass = {astro-ph.HE},
       adsurl = {https://ui.adsabs.harvard.edu/abs/2023MNRAS.522.4649U}
}

@ARTICLE{Vernstrom14,
       author = {{Vernstrom}, T. and {Scott}, Douglas and {Wall}, J.~V. and {Condon}, J.~J. and {Cotton}, W.~D. and {Fomalont}, E.~B. and {Kellermann}, K.~I. and {Miller}, N. and {Perley}, R.~A.},
        title = "{Deep 3 GHz number counts from a P(D) fluctuation analysis}",
      journal = {\mnras},
         year = 2014,
        month = may,
       volume = {440},
       number = {3},
        pages = {2791-2809},
          doi = {10.1093/mnras/stu470},
archivePrefix = {arXiv},
       eprint = {1311.7451},
 primaryClass = {astro-ph.CO},
       adsurl = {https://ui.adsabs.harvard.edu/abs/2014MNRAS.440.2791V}
}

@ARTICLE{Vieira10,
       author = {{Vieira}, J.~D. and {Crawford}, T.~M. and {Switzer}, E.~R. and {Ade}, P.~A.~R. and {Aird}, K.~A. and {Ashby}, M.~L.~N. and {Benson}, B.~A. and {Bleem}, L.~E. and {Brodwin}, M. and {Carlstrom}, J.~E. and {Chang}, C.~L. and {Cho}, H. -M. and {Crites}, A.~T. and {de Haan}, T. and {Dobbs}, M.~A. and {Everett}, W. and {George}, E.~M. and {Gladders}, M. and {Hall}, N.~R. and {Halverson}, N.~W. and {High}, F.~W. and {Holder}, G.~P. and {Holzapfel}, W.~L. and {Hrubes}, J.~D. and {Joy}, M. and {Keisler}, R. and {Knox}, L. and {Lee}, A.~T. and {Leitch}, E.~M. and {Lueker}, M. and {Marrone}, D.~P. and {McIntyre}, V. and {McMahon}, J.~J. and {Mehl}, J. and {Meyer}, S.~S. and {Mohr}, J.~J. and {Montroy}, T.~E. and {Padin}, S. and {Plagge}, T. and {Pryke}, C. and {Reichardt}, C.~L. and {Ruhl}, J.~E. and {Schaffer}, K.~K. and {Shaw}, L. and {Shirokoff}, E. and {Spieler}, H.~G. and {Stalder}, B. and {Staniszewski}, Z. and {Stark}, A.~A. and {Vanderlinde}, K. and {Walsh}, W. and {Williamson}, R. and {Yang}, Y. and {Zahn}, O. and {Zenteno}, A.},
        title = "{Extragalactic Millimeter-wave Sources in South Pole Telescope Survey Data: Source Counts, Catalog, and Statistics for an 87 Square-degree Field}",
      journal = {\apj},
         year = 2010,
        month = aug,
       volume = {719},
       number = {1},
        pages = {763-783},
          doi = {10.1088/0004-637X/719/1/763},
archivePrefix = {arXiv},
       eprint = {0912.2338},
 primaryClass = {astro-ph.CO},
       adsurl = {https://ui.adsabs.harvard.edu/abs/2010ApJ...719..763V}
}

@ARTICLE{Waldram10,
       author = {{Waldram}, E.~M. and {Pooley}, G.~G. and {Davies}, M.~L. and {Grainge}, K.~J.~B. and {Scott}, P.~F.},
        title = "{9C continued: results from a deeper radio-source survey at 15 GHz}",
      journal = {\mnras},
         year = 2010,
        month = may,
       volume = {404},
       number = {2},
        pages = {1005-1017},
          doi = {10.1111/j.1365-2966.2010.16333.x},
archivePrefix = {arXiv},
       eprint = {0908.0066},
 primaryClass = {astro-ph.CO},
       adsurl = {https://ui.adsabs.harvard.edu/abs/2010MNRAS.404.1005W}
}

@ARTICLE{Wesson90,
       author = {{Wesson}, Paul S.},
        title = "{Cosmology, Extraterrestrial Intelligence, and a Resolution of the Fermi-Hart Paradox}",
      journal = {\qjras},
         year = 1990,
        month = jun,
       volume = {31},
        pages = {161},
       adsurl = {https://ui.adsabs.harvard.edu/abs/1990QJRAS..31..161W}
}

@ARTICLE{Whittam16,
       author = {{Whittam}, I.~H. and {Riley}, J.~M. and {Green}, D.~A. and {Davies}, M.~L. and {Franzen}, T.~M.~O. and {Rumsey}, C. and {Schammel}, M.~P. and {Waldram}, E.~M.},
        title = "{10C continued: a deeper radio survey at 15.7 GHz}",
      journal = {\mnras},
         year = 2016,
        month = apr,
       volume = {457},
       number = {2},
        pages = {1496-1506},
          doi = {10.1093/mnras/stv2960},
archivePrefix = {arXiv},
       eprint = {1601.00282},
 primaryClass = {astro-ph.GA},
       adsurl = {https://ui.adsabs.harvard.edu/abs/2016MNRAS.457.1496W}
}

@ARTICLE{Williams16,
       author = {{Williams}, W.~L. and {van Weeren}, R.~J. and {R{\"o}ttgering}, H.~J.~A. and {Best}, P. and {Dijkema}, T.~J. and {de Gasperin}, F. and {Hardcastle}, M.~J. and {Heald}, G. and {Prandoni}, I. and {Sabater}, J. and {Shimwell}, T.~W. and {Tasse}, C. and {van Bemmel}, I.~M. and {Br{\"u}ggen}, M. and {Brunetti}, G. and {Conway}, J.~E. and {En{\ss}lin}, T. and {Engels}, D. and {Falcke}, H. and {Ferrari}, C. and {Haverkorn}, M. and {Jackson}, N. and {Jarvis}, M.~J. and {Kapi{\'n}ska}, A.~D. and {Mahony}, E.~K. and {Miley}, G.~K. and {Morabito}, L.~K. and {Morganti}, R. and {Orr{\'u}}, E. and {Retana-Montenegro}, E. and {Sridhar}, S.~S. and {Toribio}, M.~C. and {White}, G.~J. and {Wise}, M.~W. and {Zwart}, J.~T.~L.},
        title = "{LOFAR 150-MHz observations of the Bo{\"o}tes field: catalogue and source counts}",
      journal = {\mnras},
         year = 2016,
        month = aug,
       volume = {460},
       number = {3},
        pages = {2385-2412},
          doi = {10.1093/mnras/stw1056},
archivePrefix = {arXiv},
       eprint = {1605.01531},
 primaryClass = {astro-ph.CO},
       adsurl = {https://ui.adsabs.harvard.edu/abs/2016MNRAS.460.2385W}
}

@ARTICLE{WlodarczykSroka20,
       author = {{Wlodarczyk-Sroka}, B.~S. and {Garrett}, M.~A. and {Siemion}, A.~P.~V.},
        title = "{Extending the Breakthrough Listen nearby star survey to other stellar objects in the field}",
      journal = {\mnras},
         year = 2020,
        month = nov,
       volume = {498},
       number = {4},
        pages = {5720-5729},
          doi = {10.1093/mnras/staa2672},
archivePrefix = {arXiv},
       eprint = {2006.09756},
 primaryClass = {astro-ph.IM},
       adsurl = {https://ui.adsabs.harvard.edu/abs/2020MNRAS.498.5720W}
}

@ARTICLE{Worden17,
	   author = {{Worden}, S.~P. and {Drew}, J. and {Siemion}, A. and {Werthimer}, D. and 
		{DeBoer}, D. and {Croft}, S. and {MacMahon}, D. and {Lebofsky}, M. and 
		{Isaacson}, H. and {Hickish}, J. and {Price}, D. and {Gajjar}, V. and 
		{Wright}, J.~T.},
		title = "{Breakthrough Listen - A new search for life in the universe}",
	  journal = {Acta Astronautica},
		 year = 2017,
		month = oct,
	   volume = 139,
		pages = {98-101},
		  doi = {10.1016/j.actaastro.2017.06.008},
	   adsurl = {http://adsabs.harvard.edu/abs/2017AcAau.139...98W}
}

@ARTICLE{Wright18,
       author = {{Wright}, Jason T. and {Kanodia}, Shubham and {Lubar}, Emily},
        title = "{How Much SETI Has Been Done? Finding Needles in the n-dimensional Cosmic Haystack}",
      journal = {\aj},
         year = 2018,
        month = dec,
       volume = {156},
       number = {6},
          eid = {260},
        pages = {260},
          doi = {10.3847/1538-3881/aae099},
archivePrefix = {arXiv},
       eprint = {1809.07252},
 primaryClass = {astro-ph.IM},
       adsurl = {https://ui.adsabs.harvard.edu/abs/2018AJ....156..260W}
}

@ARTICLE{Wright20,
       author = {{Wright}, Jason T.},
        title = "{Dyson Spheres}",
      journal = {Serbian Astronomical Journal},
         year = 2020,
        month = jun,
       volume = {200},
        pages = {1-18},
          doi = {10.2298/SAJ2000001W},
archivePrefix = {arXiv},
       eprint = {2006.16734},
 primaryClass = {astro-ph.EP},
       adsurl = {https://ui.adsabs.harvard.edu/abs/2020SerAJ.200....1W}
}

@ARTICLE{Yun01,
       author = {{Yun}, Min S. and {Reddy}, Naveen A. and {Condon}, J.~J.},
        title = "{Radio Properties of Infrared-selected Galaxies in the IRAS 2 Jy Sample}",
      journal = {\apj},
         year = 2001,
        month = jun,
       volume = {554},
       number = {2},
        pages = {803-822},
          doi = {10.1086/323145},
archivePrefix = {arXiv},
       eprint = {astro-ph/0102154},
 primaryClass = {astro-ph},
       adsurl = {https://ui.adsabs.harvard.edu/abs/2001ApJ...554..803Y}
}

@ARTICLE{Zackrisson16,
       author = {{Zackrisson}, Erik and {Calissendorff}, Per and {Gonz{\'a}lez}, Juan and {Benson}, Andrew and {Johansen}, Anders and {Janson}, Markus},
        title = "{Terrestrial Planets across Space and Time}",
      journal = {\apj},
         year = 2016,
        month = dec,
       volume = {833},
       number = {2},
          eid = {214},
        pages = {214},
          doi = {10.3847/1538-4357/833/2/214},
archivePrefix = {arXiv},
       eprint = {1602.00690},
 primaryClass = {astro-ph.GA},
       adsurl = {https://ui.adsabs.harvard.edu/abs/2016ApJ...833..214Z}
}

\end{document}